\documentclass[reprint,aps,prc,twocolumn,superscriptaddress,floatfix,10pt]{revtex4-1}

\usepackage{bm}
\usepackage{dcolumn}
\usepackage{amsthm}
\usepackage{amssymb}
\usepackage{amsmath}
\usepackage{graphicx}
\usepackage{xcolor}
\usepackage{color}
\usepackage{fix-cm}
\usepackage{mathptmx} 
\usepackage[T1]{fontenc}
\usepackage[colorlinks,allcolors=blue]{hyperref}
\makeatletter
\def\NAT@def@citea{\def\@citea{\NAT@separator}}
\makeatother

\begin{document}

\title{Microscopic predictions for production of neutron rich nuclei in the reaction $^\mathbf{176}\mathbf{Yb}+{}^\mathbf{176}\mathbf{Yb}$}

\author{K. Godbey}\email{kyle.s.godbey@vanderbilt.edu}
\affiliation{Department of Physics and Astronomy, Vanderbilt University, Nashville, Tennessee 37235, USA}
\author{C. Simenel}\email{cedric.simenel@anu.edu.au}
\affiliation{Department of Theoretical Physics and Department of Nuclear Physics, Research School of Physics, The Australian National University, Canberra ACT 2601, Australia}
\author{A. S. Umar}\email{umar@compsci.cas.vanderbilt.edu}
\affiliation{Department of Physics and Astronomy, Vanderbilt University, Nashville, Tennessee 37235, USA}

\date{\today}


\begin{abstract}
\edef\oldrightskip{\the\rightskip}
\begin{description}
\rightskip\oldrightskip\relax
\setlength{\parskip}{0pt} 
\item[Background]
Production of neutron-rich nuclei is of vital importance to both understanding nuclear structure far from stability and to informing astrophysical models of the rapid neutron capture process (r-process). Multinucleon transfer (MNT) in heavy-ion collisions offers a possibility to produce neutron-rich nuclei far from stability.
\item[Purpose]
The $^{176}\mathrm{Yb}+{}^{176}\mathrm{Yb}$ reaction has been suggested as a potential candidate to explore the neutron-rich region surrounding the principal fragments. The current study has been conducted with the goal of providing guidance for future experiments wishing to study this (or similar) system.
\item[Methods]
Time-dependent Hartree-Fock (TDHF) and its time-dependent random-phase approximation (TDRPA) extension are used to examine both scattering and MNT characteristics in $^{176}\mathrm{Yb}+{}^{176}\mathrm{Yb}$. TDRPA calculations are performed to compute fluctuations and correlations of the neutron and proton numbers, allowing for estimates of primary fragment production probabilities. 
\item[Results]
Both scattering results from TDHF and transfer results from the TDRPA are presented for different energies, orientations, and impact parameters.
In addition to fragment composition, scattering angles and total kinetic energies, as well as correlations between these observables are presented.
\item[Conclusions]
$^{176}\mathrm{Yb}+{}^{176}\mathrm{Yb}$ appears to be an interesting probe for the mid-mass neutron-rich region of the chart of nuclides. The predictions of both TDHF and TDRPA are speculative, and will benefit from future experimental results to test the validity of this approach to studying MNT in heavy, symmetric collisions.
\end{description}

\end{abstract}
\maketitle

\section{Introduction}

The synthesis of neutron-rich nuclei is one of the most exciting and challenging tasks in both experimental and theoretical nuclear physics.
From the lightest systems to the superheavy regime, knowledge about the nuclei at the extremes of the chart of nuclides is vital to understanding physical phenomena at multiple scales.
At the foremost, neutron-rich nuclei are at the literal and figurative center of the rapid neutron capture process (r-process).
Attempts at modeling the r-process utilize input from nuclear models to inform threshold energies for the reaction types that characterize this process~\cite{cowan2019}.
Thus, strong theoretical understanding of both the static and dynamic properties of nuclei far from stability can give vital insight into the formation of stable heavy nuclei.

The production of neutron-rich nuclei is also of interest for studying nuclear structure, where exploring this region of the nuclear landscape clearly probes the edges of our current understanding of how finite nuclei form and are composed~\cite{otsuka2018}. 
This includes studies of neutron-rich nuclei of all masses, ranging from oxygen~\cite{desouza2013} up to the superheavy element (SHE) region.
SHEs are of particular note, as the formation and static properties of said nuclei have been the focus of many experimental~\cite{hofmann2002,munzenberg2015,morita2015,oganessian2015,roberto2015} and theoretical~\cite{bender1999,nazarewicz2002, cwiok2005,pei2009a,stone2019} studies.

Over the years, many theoretical approaches to studying neutron-rich nuclei formation have been pursued for various reaction types.
One such technique is to use models to study neutron enrichment via multinucleon transfer (MNT) in deep-inelastic collisions (DIC) and  quasifission reactions~\cite{adamian2003,zagrebaev2007,umar2008a,golabek2009,aritomo2009,kedziora2010,zhao2016,sekizawa2017a,wu2019}.
While quasifission occurs at a much shorter time-scale than fusion-fission~\cite{toke1985,durietz2011} and is the primary reaction mechanism that limits the formation of superheavy nuclei, the fragments produced may still be neutron-rich.

Quasifission reactions are often studied in asymmetric systems with, e.g., an actinide target \cite{toke1985,hinde1992,hinde1995,itkis2004,wakhle2014}.
However, quasifission can also be present in symmetric systems. In fact, the extreme case of quasifission in
 actinide-actinide collisions has been suggested as a possible reaction mechanism to obtain neutron-rich isotopes of high $Z$ nuclei in particular as well as a possible means to search for SHE~\cite{majka2018,wuenschel2018}.
Theoretically, the investigation of actinide-actinide collisions has a rich history with various approaches, including the dinuclear system (DNS) model~\cite{feng2009a}, relativistic mean-field (RMF) and Skyrme HF studies~\cite{gupta2007b}, reduced density-matrix formalism~\cite{sargsyan2009}, quantum molecular dynamics (QMD)~\cite{zhao2009}, and improved quantum molecular dynamics (ImQMD)~\cite{tian2008,zhao2016} calculations, as well as time-dependent Hartree-Fock (TDHF) studies~\cite{cusson1980,golabek2009,kedziora2010}.
Over recent years, TDHF has proved to be a tool of choice to investigate fragment properties produced in various reactions, such as DIC~\cite{umar2017,wu2019}, quasifission~\cite{wakhle2014,oberacker2014,hammerton2015,umar2015c,umar2016,sekizawa2017a,godbey2019}, and fission~\cite{simenel2014a,scamps2015a,goddard2015,tanimura2015,goddard2016,bulgac2016,tanimura2017,scamps2018,bulgac2018,scamps2019}.
Recent reviews~\cite{simenel2018,sekizawa2019} succinctly summarize the current state of TDHF (and its extensions) as it has been applied to various MNT reactions.

In this work, we present a study of the $^{176}\mathrm{Yb}+{}^{176}\mathrm{Yb}$ system using TDHF and the time-dependent random phase approximation (TDRPA) extension that considers the effect of one-body fluctuations around the TDHF trajectory.
As discussed before, microscopic approaches such as TDHF and its extensions are commonly used in heavy-ion collision studies in different regions of the nuclear chart, positioning TDHF and TDRPA as tools of choice for the current investigation.
Symmetric $^{176}$Yb reactions were chosen because they are considered as a potential candidate to  explore the neutron-rich region around the mass region $A\sim 170-180$ of the nuclear chart.
Specifically, an experimental investigation of this reaction are being considered in Dubna by Oganessian \textit{et al.} and the work presented here was undertaken at their suggestion~\cite{priv_oganessian}.
The base theory (TDHF) and the primary extension (TDRPA) are briefly described in Section~\ref{sec:TDHF}.
Results for both scattering characteristics and transfer characteristics are discussed in Section~\ref{sec:scat} and Section~\ref{sec:tran} respectively.
A summary and outlook are then presented in Section~\ref{sec:conclusions}.



\section{Formalism: TDHF and TDRPA} \label{sec:TDHF}

The TDHF theory provides a microscopic approach with which one may investigate a wide range of phenomena observed in low energy nuclear physics~\cite{negele1982,simenel2012,simenel2018,sekizawa2019}.
Specifically, TDHF provides a dynamic quantum many-body description of nuclear reactions in the vicinity of the Coulomb barrier, such as fusion~\cite{bonche1978,flocard1978,simenel2001,umar2006d,washiyama2008,umar2010a,umar2009a,guo2012,keser2012,simenel2013a,oberacker2012,oberacker2010,umar2012a,simenel2013b,umar2014a,jiang2014} 
and transfer reactions~\cite{koonin1977,simenel2010,simenel2011,umar2008a,sekizawa2013,scamps2013a,sekizawa2014,bourgin2016,umar2017,sekizawa2019}.

The TDHF equations for the single-particle wave functions
\begin{equation}
h(\{\phi_{\mu}\}) \ \phi_{\lambda} (r,t) = i \hbar \frac{\partial}{\partial t} \phi_{\lambda} (r,t)
\ \ \ \ (\lambda = 1,...,A) \ ,
\label{eq:TDHF}
\end{equation}
can be derived from a variational principle.
The principal approximation in TDHF is that the many-body wave function $\Phi(t)$  is assumed to be a single time-dependent Slater determinant at all times.
It describes the time-evolution of the single-particle wave functions in a mean-field corresponding to the dominant reaction channel.
During the past decade it has become numerically feasible to perform TDHF calculations on a 3D Cartesian grid without any symmetry restrictions and with much more accurate numerical methods~\cite{bottcher1989,umar2006c,sekizawa2013,maruhn2014}.

The main limitation in the TDHF theory when studying features like particle transfer, however, is that it is optimized for the prediction of expectation values of one-body observables~\cite{balian1981} and will under-predict fluctuations of those observables~\cite{dasso1979}. 
This is due to the fact that the fluctuation of one-body operators (such as the particle number operator) includes the expectation value of the square of a one-body operator,
\begin{equation}
\sigma_{XX}=\sqrt{\langle\hat{X}^2\rangle - \langle\hat{X}\rangle^2},
\end{equation}
that is outside the variational space of TDHF \cite{balian1981}.

To obtain such quantities one needs to go beyond standard TDHF and consider the fluctuations around the TDHF mean-field trajectory using techniques like the stochastic mean-field theory~\cite{ayik2008,lacroix2014} or TDRPA~\cite{balian1984}, which has recently been used to investigate DIC~\cite{broomfield2009,simenel2011,williams2018}.
In this work we follow a similar approach as in~\cite{simenel2011,williams2018} to obtain particle number fluctuations and distributions about the outgoing fragments.

The procedure to obtain the desired correlations involves first transforming the states after the collision as
\begin{equation}
\tilde{\phi}^X_\alpha(r,t_f)=\exp[-i\epsilon N_X\Theta_V({r})]\phi_\alpha(r,t_f),\label{eq:phi_x}
\end{equation}
where $X$ stands for neutron ($N$), proton ($Z$), or total nucleon number ($A$).
The operator $N_X$ ensures that the transformation acts only on nucleons with the correct isospin, with $N_A=1$, $N_Z=\frac{1-\tau_3}{2}$, and $N_N=\frac{1+\tau_3}{2}$.  
The operator $\Theta_V(\hat{r})$ is a step function that is either $1$ or $0$ depending on whether $r$ is within a volume of space, $V$, delimiting the fragment of interest.
Finally, $\epsilon$ is a small number that is varied to achieve convergence.

These transformed states are then propagated backwards in time from the final time $t_f$ to the initial time $t_i$.
The correlations can then be calculated by evaluating
\begin{equation}\label{eq:sigma}
\sigma_{XY} = \sqrt{\lim_{\epsilon\rightarrow0}\frac{\eta_{00}+\eta_{XY}-\eta_{0X}-\eta_{0Y}}{2\epsilon^2}},
\end{equation}
with $\eta_{XY}$ describing the overlap between the states at  time $t=t_i$,
\begin{equation}
\eta_{XY}=\sum_{\alpha \beta}\left|\langle\phi_\alpha^X(t_i)|\phi_\beta^Y(t_i)\rangle\right|^2.
\end{equation}
In the case of $X,Y=0$, this refers to states obtained with $\epsilon=0$ in Eq.~(\ref{eq:phi_x}).
In principle, one should recover exactly the initial state as the evolution is unitary.
However, using states that have been evolved forward and then backward in time with $\varepsilon=0$ minimizes systematic errors from numerical inaccuracies \cite{bonche1985,broomfield2009}. 

\begin{figure}[!htb]
	\includegraphics*[width=8.6cm]{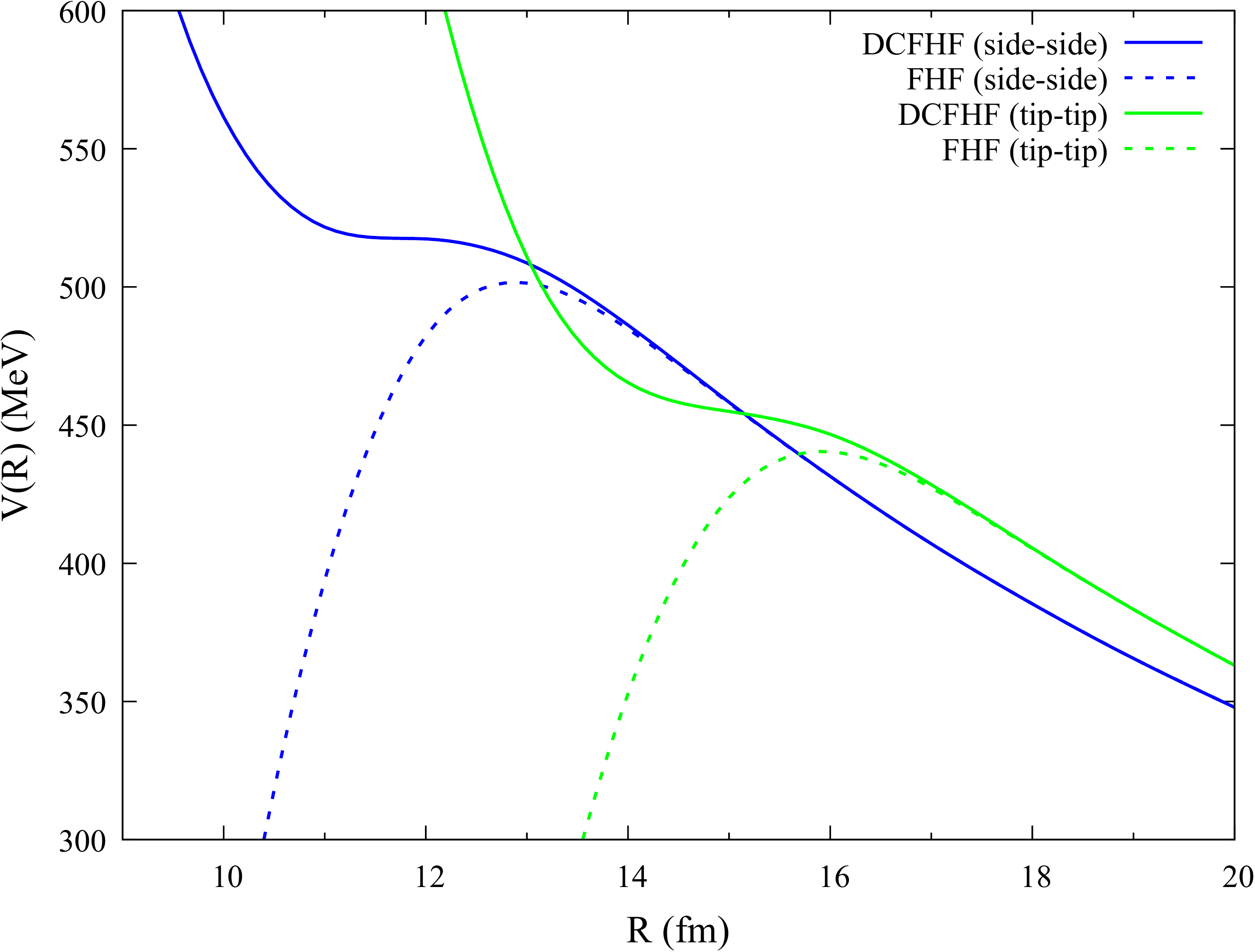}
	\caption{\protect(Color online). Static nuclear potentials for $^{176}\mathrm{Yb}+{}^{176}\mathrm{Yb}$ in the side-side (blue (dark) lines) and tip-tip (green (light) lines) orientations from FHF and DCFHF.}
	\label{fig:pot}
\end{figure}

The SLy4$d$ parametrization of the Skyrme functional is used~\cite{kim1997} and
all calculations were performed in a numerical box with $66 \times 66$ points in the reaction plane, and 36 points along the axis perpendicular to the reaction plane.
The grid spacing used was a standard $1.0$~fm which provides an excellent numerical representation of spatial quantities using the basis spline collocation method~\cite{umar1991a}.
For the TDRPA calculations, each initial orientation, energy, and impact parameter resulted in three additional TDHF evolutions (one for each $X$) for the time reversed evolution at one value of $\epsilon=2\times10^{-3}$ in addition to occasionally scanning $\epsilon$ to ensure convergence of Eq.~(\ref{eq:sigma}).
In total, $200$ full TDHF evolutions were required for the results presented in this work with each taking on the order of $10\sim55$~hours of wall time due to the large, three-dimensional box size chosen.
This corresponds to roughly $250$ days of computation time split among multiple nodes for the $^{176}$Yb HF ground state configuration with a prolate deformation. 

The proton and neutron numbers correlations and fluctuations computed with TDRPA are used to estimate probabilities for the formation of a given nuclide using 
 Gaussian bivariate normal  distributions of the form
\begin{equation}
\mathcal{P}(n,z) = \mathcal{P}(0,0)\exp\left[ -\frac{1}{1-\rho^2} \left( \frac{n^2}{\sigma_{NN}^2}+\frac{z^2}{\sigma_{ZZ}^2} - \frac{2\rho nz}{\sigma_{NN}\sigma_{ZZ}}\right) \right], 
\label{eq:Pnz}
\end{equation}
where $n$ and $z$ are the number of transferred neutrons and protons, respectively.  
The correlations between $N$ and $Z$ are quantified by the parameter
\begin{equation}
\rho = \mbox{sign}(\sigma_{NZ})\frac{\sigma_{NZ}^2}{\sigma_{NN}\sigma_{ZZ}}=\frac{\langle nz\rangle}{\sqrt{\langle n^2\rangle\langle z^2\rangle}}.
\end{equation}
In principle, $n$ and $z$ could be very large and lead to unphysical predictions with fragments having, e.g., a negative number of protons and neutrons, or more nucleons than available. 
In practice, such spurious results could only happen for the most violent collisions where the fluctuations are large.  
To avoid such spurious effects, the probabilities are shifted so that $\mathcal{P}$ is zero when one fragment has all (or more) protons or neutrons. 
The resulting distribution is then normalized. 

Although the $^{176}$Yb nuclide is in a region where shape coexistence is often found~\cite{fu2018,nomura2011,robledo2009,sarriguren2008,xu2011}, TDHF calculations can only be performed with one well-defined deformation (and orientation) of each collision partners in the entrance channel. 
In our calculations, the ground state is found to have a prolate deformation with $\beta_2\simeq 0.33$ in its HF ground state.
A higher energy oblate solution is also found with a difference of around $5$~MeV in total binding energy.
A set of calculations were also performed for the oblate solution, though the overall transfer behavior was found to be similar for both deformations despite the oblate one resulting in slightly lower fluctuations.
In the following, we thus only show results for the prolate ground state. 

This deformation allows for possible choices of the orientation of the nuclei. 
Extreme orientations are called ``side'' (``tip'') when the deformation axis is initially perpendicular (parallel) to the collision axis. 
Although various intermediate orientations could be considered~\cite{godbey2019}, we limit our study to tip-tip and side-side orientations where the initial orientations of both nuclei are identical. 
In addition to saving computational time, this restriction is necessary to ensure fully symmetric collisions and to avoid unphysical results in TDRPA~\cite{williams2018}.

Figure~\ref{fig:pot} shows the nucleus-nucleus potentials computed using the frozen Hartree-Fock (FHF)~\cite{simenel2008,washiyama2008} and density-constrained frozen Hartree-Fock (DCFHF)~\cite{simenel2017} methods, 
respectively neglecting and including the Pauli exclusion principle between the nucleons of different nuclei.
Due to Pauli repulsion in DCFHF, the inner pocket potential is very shallow in the side-side configuration, and disappears in the tip-tip one. 
In this work, the effect of the orientation is studied by comparing tip-tip and side-side configurations at a center of mass energy 
$E_\mathrm{c.m.}=660$~MeV. 
In addition, calculations are also performed at $E_\mathrm{c.m.}=880$~MeV for both orientations to investigate the role of the energy on the reaction outcome. 


\section{Results}\label{sec:results}

In this section we present the results of TDHF and TDRPA studies of $^{176}\mathrm{Yb}+{}^{176}\mathrm{Yb}$ reactions at different center of mass energies and initial orientations  for a range of impact parameters.
Both scattering features and particle number fluctuation derived quantities were calculated and are shown below.


\subsection{Scattering Characteristics}\label{sec:scat}

The following section presents scattering results from the standard TDHF calculations of $^{176}\mathrm{Yb}+{}^{176}\mathrm{Yb}$ collisions.
The TDRPA extension to TDHF is not needed for these results, though this means the points can only be interpreted as the most likely outcome for each initial condition.

\begin{figure}[!htb]
	\includegraphics*[width=8.6cm]{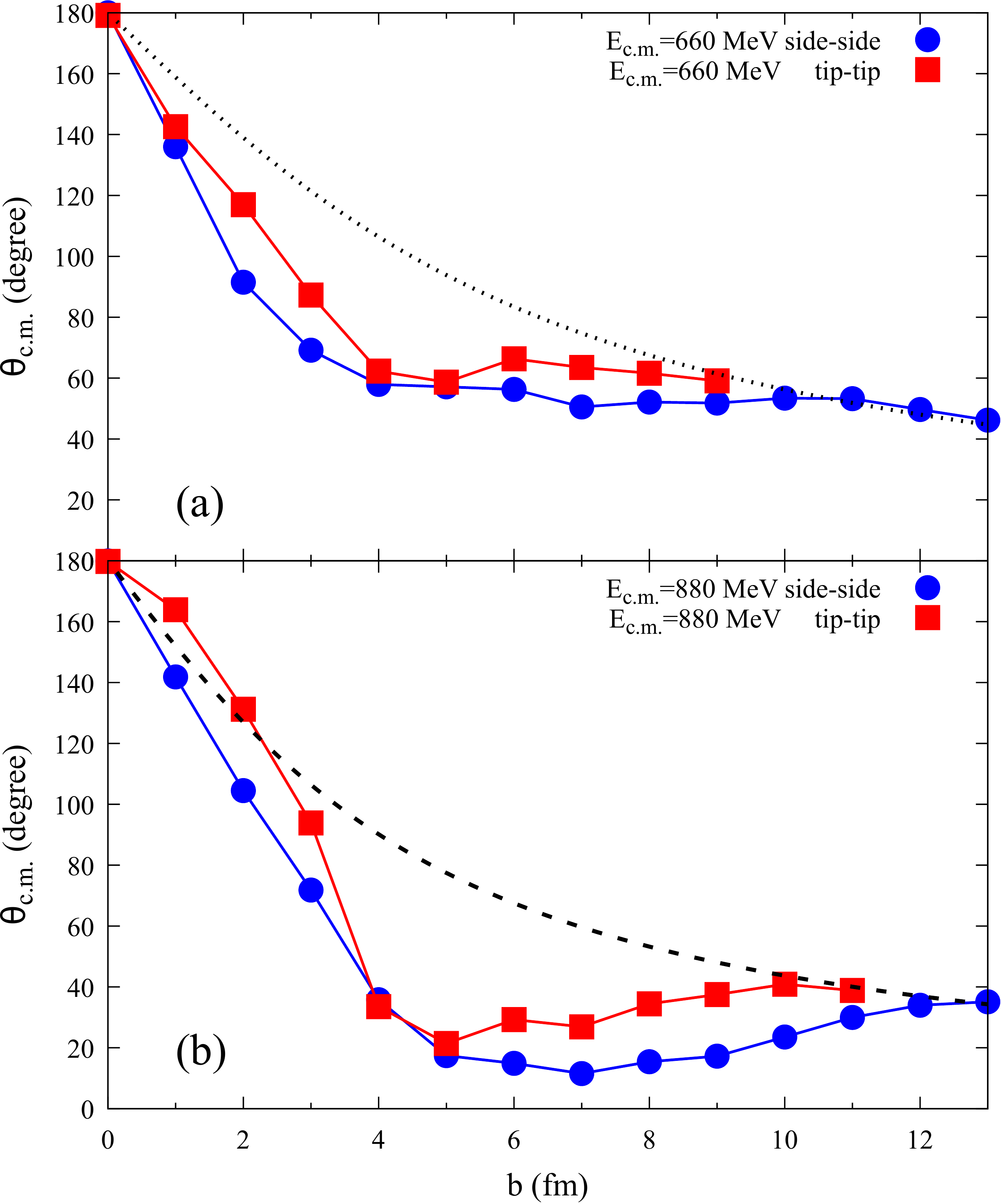}
	\caption{\protect(Color online). Scattering angles for $^{176}\mathrm{Yb}+{}^{176}\mathrm{Yb}$ collisions at center of mass energies (a) $\mathrm{E_{c.m.}}=660$~MeV and (b) $\mathrm{E_{c.m.}}=880$~MeV in the side-side (circles) and tip-tip (squares) orientations. The  dotted (dashed) line plots the Rutherford scattering angle for $\mathrm{E_{c.m.}}=660$~MeV ($880$~MeV).}
	\label{fig:bvstheta}
\end{figure}

Scattering angles for the $^{176}\mathrm{Yb}+{}^{176}\mathrm{Yb}$ system for both orientations are presented in Fig.~\ref{fig:bvstheta}. 
A similar deviation from Rutherford scattering is observed at impact parameters $b\le8$~fm for both orientations. 
These deviations are due to nuclear deflection and partial orbiting of the system.
Note that no fusion is observed.
The relatively flat shape of the curve around $50-60^\circ$ at 660~MeV and $20-40^\circ$ at 880~MeV implies a large number of events in these particular angular ranges.

\begin{figure}[!htb]
	\includegraphics*[width=8.6cm]{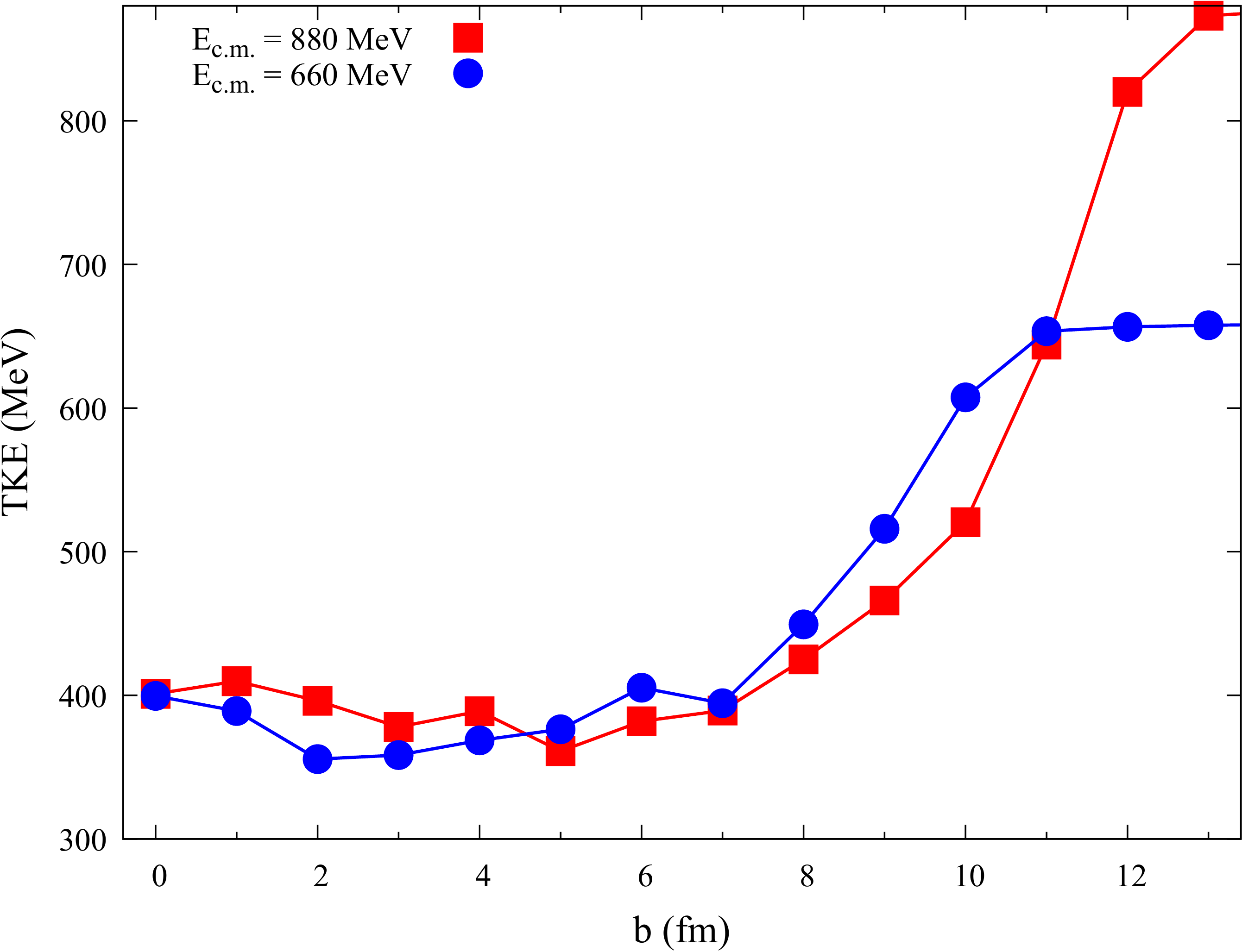}
	\caption{\protect(Color online). Total kinetic energies of the outgoing fragments in $^{176}\mathrm{Yb}+{}^{176}\mathrm{Yb}$ collisions at center of mass energies $\mathrm{E_{c.m.}}=660$~MeV (blue circles) and $\mathrm{E_{c.m.}}=880$~MeV (red squares) in the side-side orientation.}
	\label{fig:tke}
\end{figure}

The TKE of the outgoing fragments is plotted in Fig.~\ref{fig:tke} 
as a function of the impact parameter $b$ for side-side collisions at the two center of mass energies.
Although dissipation occurs at different impact parameter ranges ($b<10$~fm at $E_{c.m.}=660$~MeV and $b<12$~fm at $E_{c.m.}=880$~MeV), both curves exhibit similar behavior.
In particular, the TKEs saturate at roughly the same energy ($\sim350-400$~MeV) indicating full damping of the initial TKE for the most central collisions.

Among the mechanisms responsible for energy dissipation, nucleon transfer is expected to play an important role. 
Of course, in symmetric collisions the average number of nucleons in the fragments does not change.
Nevertheless, multinucleon transfer is possible thanks to fluctuations, leading to finite widths in the fragment particle number distributions. 
These fluctuations are explored in the following section.

\subsection{Transfer Characteristics}\label{sec:tran}

This section focuses on the results obtained by extending TDHF to recover particle number fluctuations and correlations with the TDRPA.

\begin{figure}[!htb]
	\includegraphics*[width=8.6cm]{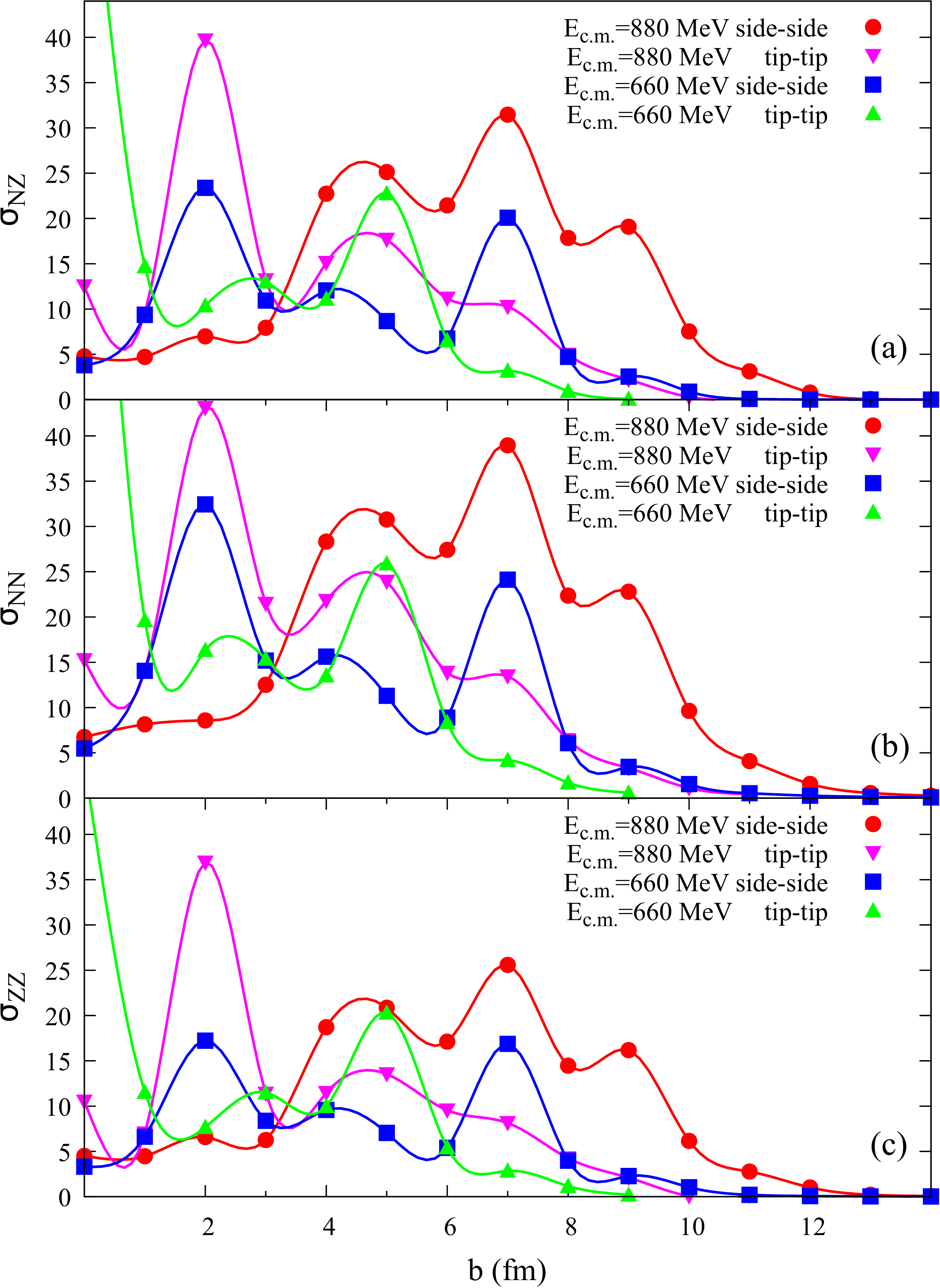}
	\caption{\protect(Color online). TDRPA predictions of correlations $\sigma_{NZ}$ (a) and fluctuations $\sigma_{NN}$ (b) and $\sigma_{ZZ}$ (c) for $^{176}\mathrm{Yb}+{}^{176}\mathrm{Yb}$ collisions for four initial configurations over a range of impact parameters.}
	\label{fig:fluc}
\end{figure}

Particle number fluctuations ($\sigma_{ZZ}$ and $\sigma_{NN}$) and correlations ($\sigma_{NZ}$) calculated from Eq.~(\ref{eq:sigma}) are shown in Fig.~\ref{fig:fluc} as a function of  impact parameters for different initial conditions.
The fluctuations are greater in general at the smaller impact parameters, though they do not converge to a single value.
Similar variations in fluctuations were already observed in earlier TDRPA studies of deep inelastic collisions in lighter systems~\cite{simenel2011,williams2018}. 
Particularly large values are sometimes obtained, such as at 660~MeV in tip-tip central ($b=0$) collisions, indicating approximately flat distributions around the TDHF average.

\begin{figure}[!htb]
	\includegraphics*[width=8.6cm]{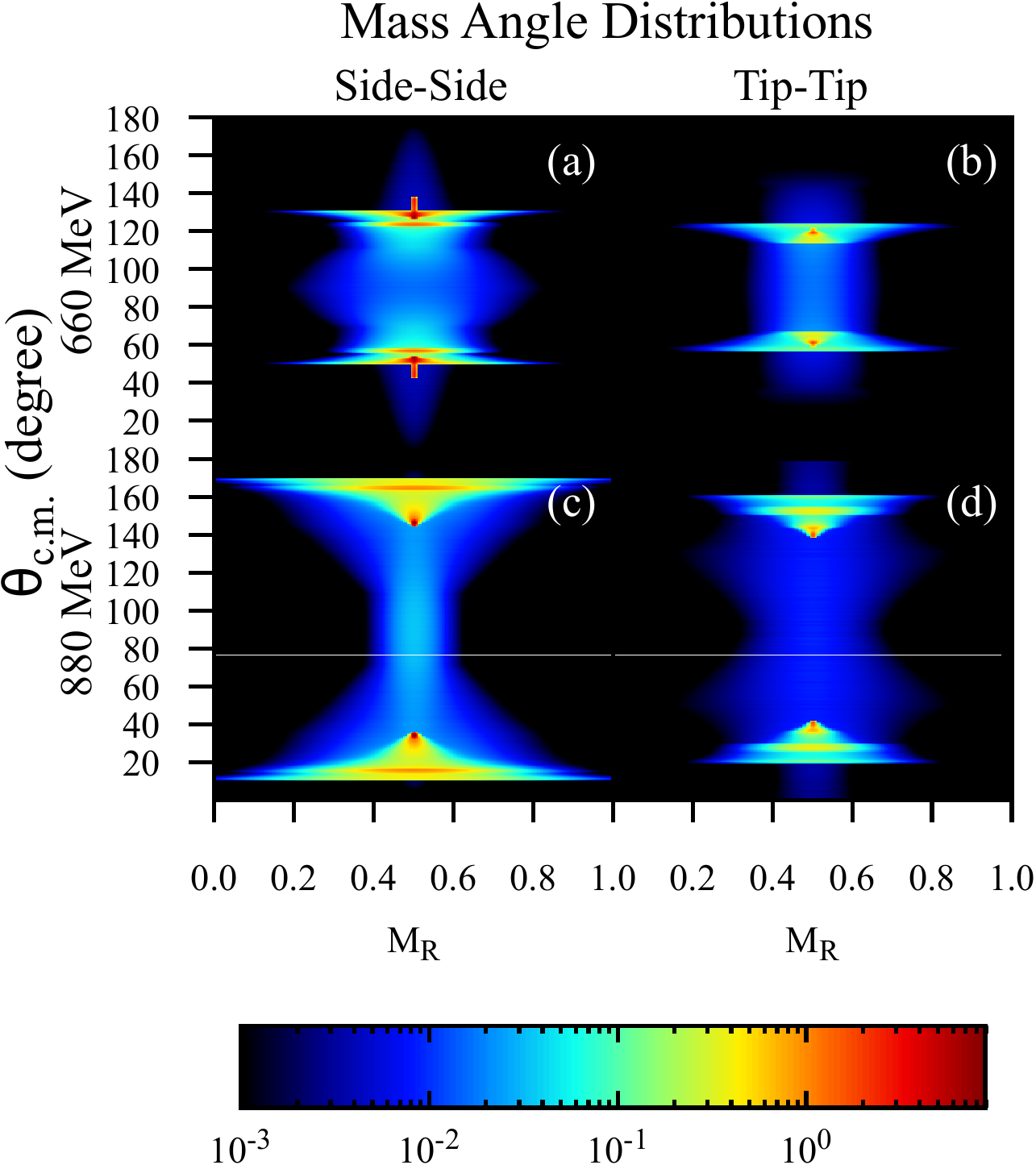}
	\caption{\protect(Color online). Mass angle distributions for $^{176}\mathrm{Yb}+{}^{176}\mathrm{Yb}$ collisions at (a) $\mathrm{E_{c.m.}}=660$~MeV in the side-side orientation, (b) $\mathrm{E_{c.m.}}=660$~MeV in the tip-tip orientation, (c) $\mathrm{E_{c.m.}}=880$~MeV in the side-side orientation, and (d) $\mathrm{E_{c.m.}}=880$~MeV in the tip-tip orientation. The colorbar represents cross sections in millibarns per bin of mass ratio and degree.}
	\label{fig:mad}
\end{figure}

\begin{figure}[!htb]
	\includegraphics*[width=8.6cm]{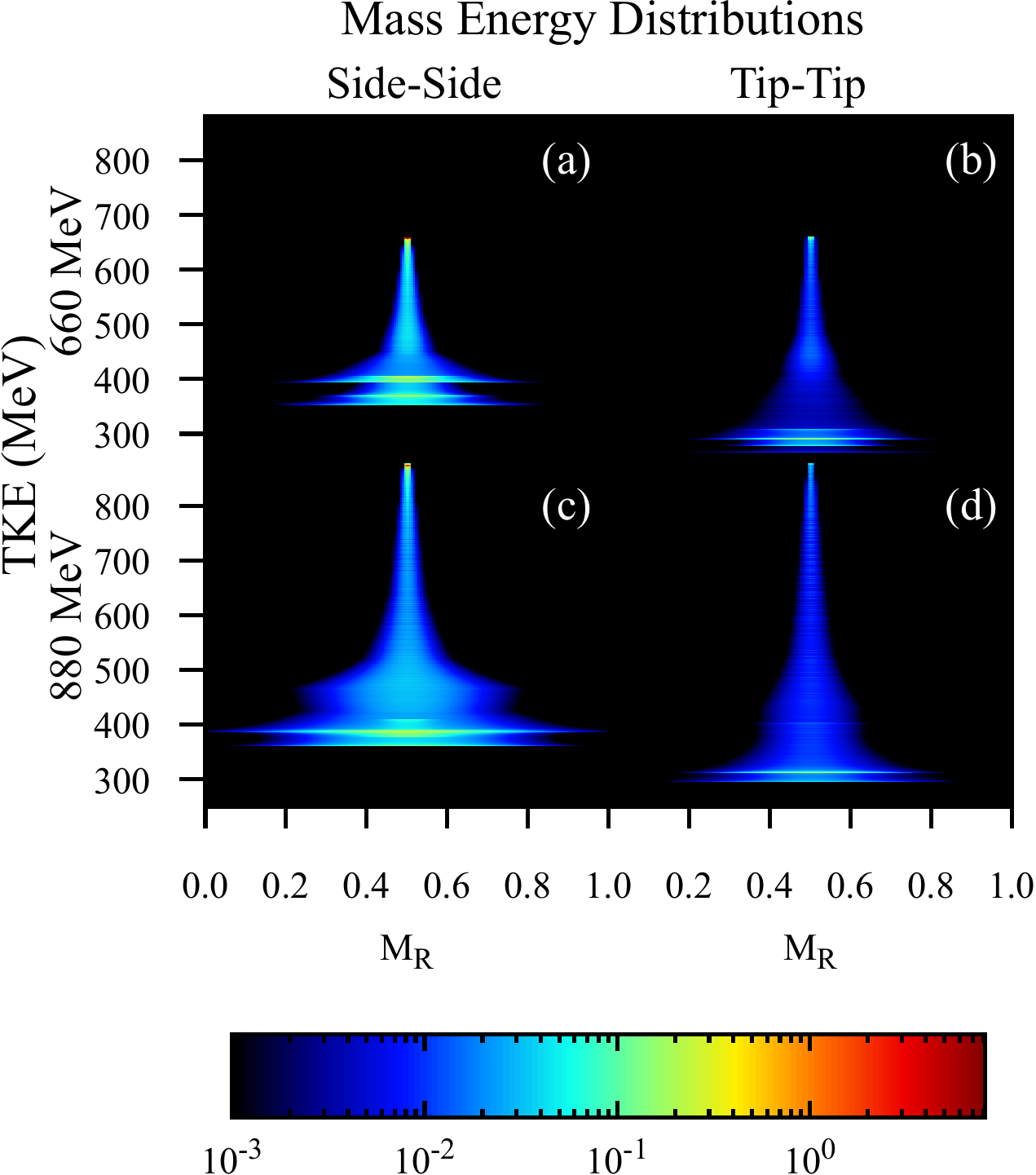}
	\caption{\protect(Color online). Mass energy distributions for $^{176}\mathrm{Yb}+{}^{176}\mathrm{Yb}$ collisions at (a) $\mathrm{E_{c.m.}}=660$~MeV in the side-side orientation, (b) $\mathrm{E_{c.m.}}=660$~MeV in the tip-tip orientation, (c) $\mathrm{E_{c.m.}}=880$~MeV in the side-side orientation, and (d) $\mathrm{E_{c.m.}}=880$~MeV in the tip-tip orientation. The colorbar represents cross sections in millibarns per bin of mass ratio and MeV.}
	\label{fig:med}
\end{figure}

Fragment mass-angle distributions (MADs) are a standard tool used experimentally to interpret the dynamics of heavy-ion collisions \cite{toke1985,shen1987,hinde2008,simenel2012b,durietz2013,wakhle2014,hammerton2015,morjean2017,mohanto2018,hinde2018}.
Although TDHF has been used to help interpret theoretically these distributions \cite{wakhle2014,hammerton2015,umar2016,sekizawa2016}, 
these earlier calculations only incorporate fluctuations coming from the distribution of initial conditions (e.g., different orientations). 
Here, we go beyond the mean-field prediction by including the fragment mass fluctuations from TDRPA.
Note that we only include mass fluctuations, not fluctuations in scattering angle which are still determined solely by TDHF. 
Calculating quantum fluctuations of scattering angles is beyond the scope of this work, although they might be necessary for a more detailed comparison with experimental MADs.

The resulting MADs for $^{176}\mathrm{Yb}+{}^{176}\mathrm{Yb}$ reactions are shown in Fig.~\ref{fig:mad}.
The mass ratio $M_R$ is defined as the ratio of the fragment mass over the total mass of the system. 
The distributions of mass ratios are determined assuming Gaussian distributions with standard deviation $\sigma_{M_R}=\sigma_{AA}/A$, limited and normalized to the physical region $0\le M_R\le1$ (see section~\ref{sec:TDHF}). 
There is then an $M_R$ distribution per initial condition (defined by $E_{c.m.}$, $b$, and the orientations), but only a single scattering angle $\theta_{c.m.}$.
To obtain a continuous representation of the scattering angle, $\theta_{c.m.}$ is discretized into bins of $\Delta\theta=1$ degree and interpolated between the values obtained by TDHF.


The figures are symmetric about 90$^{\circ}$ as both outgoing fragments are identically the same and will then travel outwards at complimentary angles.
Specific orientations such as side-side and tip-tip will not be accessible in an experimental setting of course.
Interestingly, when investigating initial energy dependence of the MAD (compare panels (a) and (c), (b) and (d) in Fig.~\ref{fig:mad}), 
it can be seen that different outgoing angles are preferred depending on the incoming center of mass energy with back (and forward) scattering events being more prevalent in the higher energy regime.

This  agrees well with what is seen in Fig.~\ref{fig:bvstheta}, where many impact parameters result in  scattering angles around $50-60$~degrees at $E_{c.m.}=660$~MeV and around $20-40$~degrees at 880~MeV.
This is the case for both tip-tip and side-side orientations, though the tip-tip results tend further towards the intermediate angles than side-side at the same energy.

While the predictive capability of this method needs to be compared with experimental results and tested, this suggests a strong energy dependence and that detection of fragment production will greatly benefit from large angle detectors.
The energy dependence seen in the MAD is not intuitive, and may prove to be useful for informing experimental setups.

\begin{figure*}[htb]
	\includegraphics*[width=12.9cm]{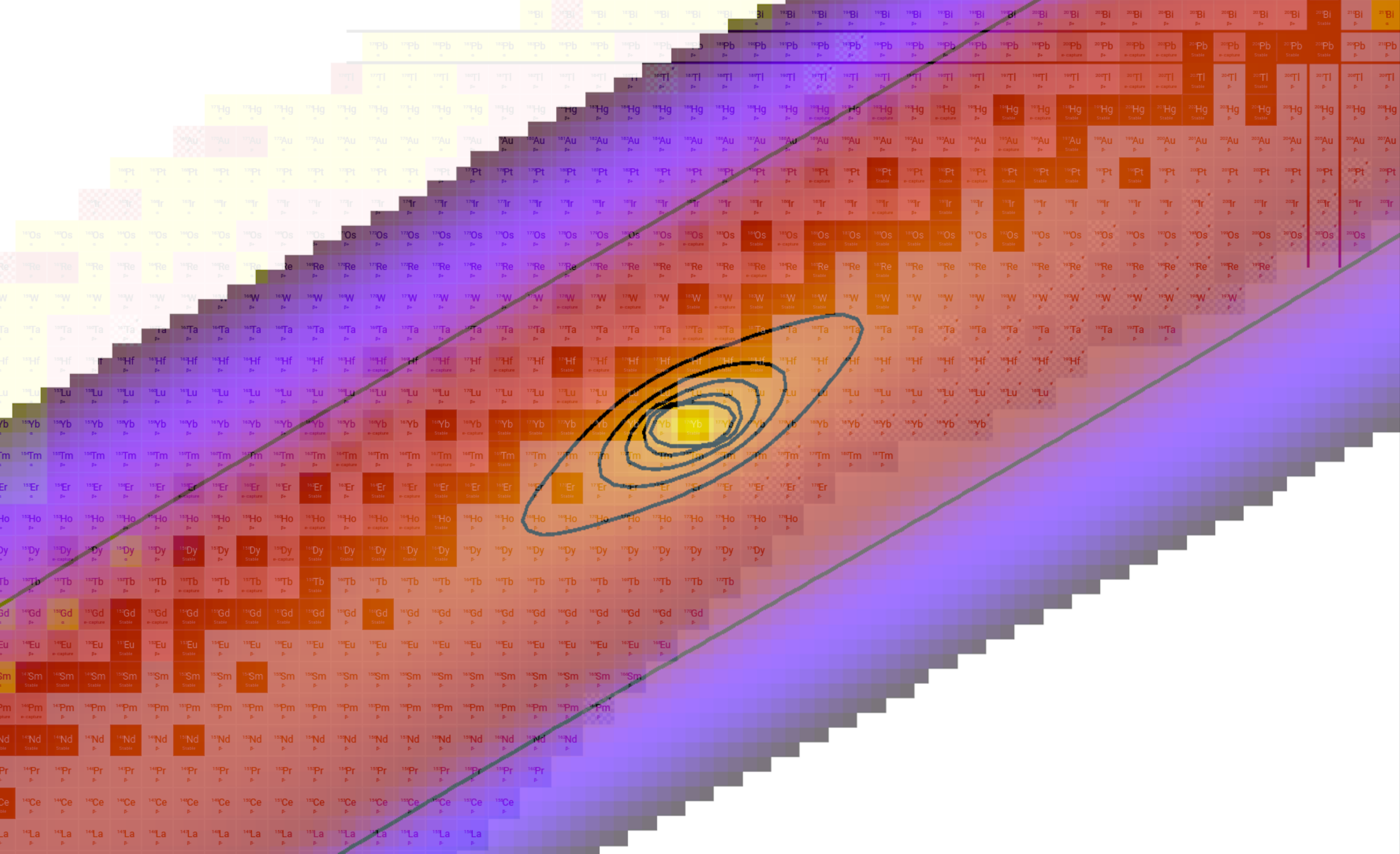}
	\caption{\protect(Color online). Primary fragments production cross sections for $^{176}\mathrm{Yb}+{}^{176}\mathrm{Yb}$ collisions at $\mathrm{E_{c.m.}}=660$~MeV in the side-side orientation overlaid onto the chart of nuclides. 
	The innermost contour corresponds to a cross section of 1~millibarn, with subsequent contours drawn every 0.2~mb.
	Finally, we also plot a boundary contour drawn at the microbarn level. Chart from~\cite{anu_chart}.}
	\label{fig:chart}
\end{figure*}

Useful information can also be obtained from the correlations between fragment mass and kinetic energy \cite{itkis2004,itkis2011,itkis2015,kozulin2019,banerjee2019}.
Figure~\ref{fig:med} presents mass energy distributions (MED) that detail the predicted TKE of outgoing fragments.
It should be noted here that, while the theory provides particle number fluctuations, the values for TKE are single points (as in the case of $\theta_{c.m.}$) as predicted by TDHF alone.
That is, widths of the TKE distributions are currently unknown with the method used here.
This would make for an excellent extension to the theory, bringing it more in line with what can be experimentally observed.

The MEDs exhibit a continuous broadening of the mass distribution with increasing energy dissipation. 
The saturation of TKE lies around $350-400$~MeV for side-side collisions (see also Fig.~\ref{fig:tke}) and around $250-300$~MeV for tip-tip. 
This difference between orientations is interesting as it indicates a larger kinetic energy dissipation with less compact configurations. 
 A possible explanation is that the nuclei overlap at a larger distance in the tip-tip configuration, thus producing energy dissipation earlier in the collision process than in the side-side orientation. 
 
In general, the MEDs show peaks around the elastic and fully damped regions which results from the large range of impact parameters contributing to both mechanisms.

\subsection{Primary fragments production}

Using the  correlations and fluctuations shown in Fig.~\ref{fig:fluc}, a map of probabilities can be made in the $N$--$Z$ plane assuming a modified Gaussian bivariate normal distribution (See section~\ref{sec:TDHF} and Eq.~(\ref{eq:Pnz})).
This choice of using a Gaussian is the primary assumption when calculating probabilities and related quantities and may not accurately describe the true distribution far from the center.

These probability distributions at multiple impact parameters can then be integrated over to produce a map of primary fragment production cross sections which is presented in Fig.~\ref{fig:chart} overlaid atop a section of the chart of nuclides in the region surrounding $^{176}$Yb~\cite{anu_chart}.
As the probability distributions for each impact parameter will be centered around the $^{176}$Yb ($Z=70$, $N=106$) nuclide, the resulting cross sections are also symmetric about $^{176}$Yb.
The inclusion of correlations between protons and neutrons via $\sigma_{NZ}$ more or less aligns the distribution parallel to the valley of stability  due to the symmetry energy.

Subsequent decay of the fragments would inevitably bring the final products closer to the valley of stability.
Here, our focus is on primary fragment productions and the prediction of evaporation residue cross-sections are beyond the scope of this work.
In fact, experimental measurements of mass-angle distributions using time of flight techniques are for primary fragments as they assume two-body kinematics \cite{thomas2008}. 
To estimate the evaporation residue cross-sections would require to first compute the excitation energy of the fragments and then predict their decay with a statistical model \cite{umar2017,sekizawa2017}. 

One way to minimize evaporation is to consider less violent collisions. 
In terms of primary fragment productions, 660 and 880~MeV center of mass energies are quite similar (this can be seen by the relatively similar particle number fluctuations in Fig.~\ref{fig:fluc}). 
However, the higher energy will lead to more neutron evaporation and thus to less exotic evaporation residues. 
Use of relatively neutron-rich $^{176}$Yb nuclei in symmetric collisions may then allow for this reaction to act as a probe of the neutron-rich region surrounding the principal outgoing fragment.


\section{Summary and Discussion}\label{sec:conclusions}

Multiple TDHF and TDRPA calculations have been performed for the $^{176}\mathrm{Yb}+{}^{176}\mathrm{Yb}$ system with various initial orientations, energies, and impact parameters.
Standard TDHF allows for the classification of general scattering characteristics, while the TDRPA technique extends the approach to include correlations and fluctuations of particle numbers of the reaction fragments.
This extension provides a theoretical framework that more closely resembles what will be seen in experimental investigations of this (and similar) systems.

In examining figures such as the mass-angle distributions in Fig.~\ref{fig:mad}, information regarding the angular distribution of fragments can be gleaned and suggest large acceptance detectors to maximize measurement capability.
Mass-energy distributions shown in Fig.~\ref{fig:med} are also useful to investigate, e.g., the interplay between dissipation and fluctuations. 
In both cases, however, fluctuations of $\theta_{c.m.}$ and of TKE are not predicted in the present study.
The latter would require new implementations of the TDRPA to these observables, or the use of alternative approaches such as the stochastic mean-field theory \cite{tanimura2017} or an extension of the Langevin equation \cite{bulgac2019}. 
Both methods have been recently used to investigate kinetic energy distributions in fission fragments. 
In order to benchmark our theoretical methods as applied to symmetric heavy nuclei, all predictions presented in this study would greatly benefit from experimental verification.

The methods used here provide a very powerful tool for investigating symmetric systems, though an important caveat should be discussed regarding the interpretation of these results.
TDRPA produces only correlations and fluctuations, not the actual distributions themselves, which are then taken to be of a Gaussian nature.
This assumption may break down when far from the center of the distribution or if the shape at the center itself is too flat and deviates sufficiently from a Gaussian behavior.
It is then extremely important to compare with observations made in experimental studies such that we may better understand how to interpret the results coming from these methods.

Regardless, the $^{176}\mathrm{Yb}+{}^{176}\mathrm{Yb}$ system presents itself as a viable candidate for studies of MNT processes and production of neutron rich nuclei in the region around $A\sim176$.
The map of possible primary fragments loosely painted in Fig.~\ref{fig:chart} presents an exciting range of previously inaccessible nuclei, with the above caveat applying the further one goes from the center of the distribution.
Another caveat is that the predicted distribution is for primary fragments only and that statistical decay should be included in order to predict fragment produced after evaporation, e.g., following \cite{sekizawa2017,umar2017,wu2019}.


\begin{acknowledgments}
We thank Yu. Ts. Oganessian and D. J. Hinde for stimulating discussions.
This work has been supported by the U.S. Department of Energy under grant No.
DE-SC0013847 with Vanderbilt University and by the
Australian Research Councils Grant No. DP190100256.
\end{acknowledgments}


\bibliography{VU_bibtex_master}

\begin{thebibliography}{116}%
\makeatletter
\providecommand \@ifxundefined [1]{%
 \@ifx{#1\undefined}
}%
\providecommand \@ifnum [1]{%
 \ifnum #1\expandafter \@firstoftwo
 \else \expandafter \@secondoftwo
 \fi
}%
\providecommand \@ifx [1]{%
 \ifx #1\expandafter \@firstoftwo
 \else \expandafter \@secondoftwo
 \fi
}%
\providecommand \natexlab [1]{#1}%
\providecommand \enquote  [1]{``#1''}%
\providecommand \bibnamefont  [1]{#1}%
\providecommand \bibfnamefont [1]{#1}%
\providecommand \citenamefont [1]{#1}%
\providecommand \href@noop [0]{\@secondoftwo}%
\providecommand \href [0]{\begingroup \@sanitize@url \@href}%
\providecommand \@href[1]{\@@startlink{#1}\@@href}%
\providecommand \@@href[1]{\endgroup#1\@@endlink}%
\providecommand \@sanitize@url [0]{\catcode `\\12\catcode `\$12\catcode
  `\&12\catcode `\#12\catcode `\^12\catcode `\_12\catcode `\%12\relax}%
\providecommand \@@startlink[1]{}%
\providecommand \@@endlink[0]{}%
\providecommand \url  [0]{\begingroup\@sanitize@url \@url }%
\providecommand \@url [1]{\endgroup\@href {#1}{\urlprefix }}%
\providecommand \urlprefix  [0]{URL }%
\providecommand \Eprint [0]{\href }%
\providecommand \doibase [0]{https://doi.org/}%
\providecommand \selectlanguage [0]{\@gobble}%
\providecommand \bibinfo  [0]{\@secondoftwo}%
\providecommand \bibfield  [0]{\@secondoftwo}%
\providecommand \translation [1]{[#1]}%
\providecommand \BibitemOpen [0]{}%
\providecommand \bibitemStop [0]{}%
\providecommand \bibitemNoStop [0]{.\EOS\space}%
\providecommand \EOS [0]{\spacefactor3000\relax}%
\providecommand \BibitemShut  [1]{\csname bibitem#1\endcsname}%
\let\auto@bib@innerbib\@empty
\bibitem [{\citenamefont {Cowan}\ \emph {et~al.}(2019)\citenamefont {Cowan},
  \citenamefont {Sneden}, \citenamefont {Lawler}, \citenamefont {Aprahamian},
  \citenamefont {Wiescher}, \citenamefont {Langanke}, \citenamefont
  {Mart\'{\i}nez-Pinedo},\ and\ \citenamefont {Thielemann}}]{cowan2019}%
  \BibitemOpen
  \bibfield  {author} {\bibinfo {author} {\bibfnamefont {J.~J.}\ \bibnamefont
  {Cowan}}, \bibinfo {author} {\bibfnamefont {C.}~\bibnamefont {Sneden}},
  \bibinfo {author} {\bibfnamefont {J.~E.}\ \bibnamefont {Lawler}}, \bibinfo
  {author} {\bibfnamefont {A.}~\bibnamefont {Aprahamian}}, \bibinfo {author}
  {\bibfnamefont {M.}~\bibnamefont {Wiescher}}, \bibinfo {author}
  {\bibfnamefont {K.}~\bibnamefont {Langanke}}, \bibinfo {author}
  {\bibfnamefont {G.}~\bibnamefont {Mart\'{\i}nez-Pinedo}},\ and\ \bibinfo
  {author} {\bibfnamefont {F.-K.}\ \bibnamefont {Thielemann}},\ }\bibfield
  {title} {\bibinfo {title} {{Making the Heaviest Elements in the Universe: A
  Review of the Rapid Neutron Capture Process}},\ }\href
  {https://arxiv.org/abs/1901.01410} {\bibfield  {journal} {\bibinfo  {journal}
  {Arxiv:1901.01410}\ } (\bibinfo {year} {2019})}\BibitemShut {NoStop}%
\bibitem [{\citenamefont {Otsuka}\ \emph {et~al.}(2018)\citenamefont {Otsuka},
  \citenamefont {Gade}, \citenamefont {Sorlin}, \citenamefont {Suzuki},\ and\
  \citenamefont {Utsuno}}]{otsuka2018}%
  \BibitemOpen
  \bibfield  {author} {\bibinfo {author} {\bibfnamefont {T.}~\bibnamefont
  {Otsuka}}, \bibinfo {author} {\bibfnamefont {A.}~\bibnamefont {Gade}},
  \bibinfo {author} {\bibfnamefont {O.}~\bibnamefont {Sorlin}}, \bibinfo
  {author} {\bibfnamefont {T.}~\bibnamefont {Suzuki}},\ and\ \bibinfo {author}
  {\bibfnamefont {Y.}~\bibnamefont {Utsuno}},\ }\bibfield  {title} {\bibinfo
  {title} {{Evolution of nuclear structure in exotic nuclei driven by nuclear
  forces}},\ }\href {https://arxiv.org/abs/1805.06501} {\bibfield  {journal}
  {\bibinfo  {journal} {Arxiv:1805.06501}\ } (\bibinfo {year} {2018})},\
  \bibinfo {note} {accepted in Rev. Mod. Phys.}\BibitemShut {Stop}%
\bibitem [{\citenamefont {deSouza}\ \emph {et~al.}(2013)\citenamefont
  {deSouza}, \citenamefont {Hudan}, \citenamefont {Oberacker},\ and\
  \citenamefont {Umar}}]{desouza2013}%
  \BibitemOpen
  \bibfield  {author} {\bibinfo {author} {\bibfnamefont {R.~T.}\ \bibnamefont
  {deSouza}}, \bibinfo {author} {\bibfnamefont {S.}~\bibnamefont {Hudan}},
  \bibinfo {author} {\bibfnamefont {V.~E.}\ \bibnamefont {Oberacker}},\ and\
  \bibinfo {author} {\bibfnamefont {A.~S.}\ \bibnamefont {Umar}},\ }\bibfield
  {title} {\bibinfo {title} {{C}onfronting measured near- and sub-barrier
  fusion cross sections for $^{20}\mathrm{O}+{}^{12}\mathrm{C}$ with a
  microscopic method},\ }\href {https://doi.org/10.1103/PhysRevC.88.014602}
  {\bibfield  {journal} {\bibinfo  {journal} {Phys. Rev. C}\ }\textbf {\bibinfo
  {volume} {88}},\ \bibinfo {pages} {014602} (\bibinfo {year}
  {2013})}\BibitemShut {NoStop}%
\bibitem [{\citenamefont {Hofmann}\ \emph {et~al.}(2002)\citenamefont
  {Hofmann}, \citenamefont {He\ss{}berger}, \citenamefont {Ackermann},
  \citenamefont {M\"unzenberg}, \citenamefont {Antalic}, \citenamefont
  {Cagarda}, \citenamefont {Kindler}, \citenamefont {Kojouharova},
  \citenamefont {Leino}, \citenamefont {Lommel}, \citenamefont {Mann},
  \citenamefont {Popeko}, \citenamefont {Reshitko}, \citenamefont {\'Saro},
  \citenamefont {Uusitalo},\ and\ \citenamefont {Yeremin}}]{hofmann2002}%
  \BibitemOpen
  \bibfield  {author} {\bibinfo {author} {\bibfnamefont {S.}~\bibnamefont
  {Hofmann}}, \bibinfo {author} {\bibfnamefont {F.~P.}\ \bibnamefont
  {He\ss{}berger}}, \bibinfo {author} {\bibfnamefont {D.}~\bibnamefont
  {Ackermann}}, \bibinfo {author} {\bibfnamefont {G.}~\bibnamefont
  {M\"unzenberg}}, \bibinfo {author} {\bibfnamefont {S.}~\bibnamefont
  {Antalic}}, \bibinfo {author} {\bibfnamefont {P.}~\bibnamefont {Cagarda}},
  \bibinfo {author} {\bibfnamefont {B.}~\bibnamefont {Kindler}}, \bibinfo
  {author} {\bibfnamefont {J.}~\bibnamefont {Kojouharova}}, \bibinfo {author}
  {\bibfnamefont {M.}~\bibnamefont {Leino}}, \bibinfo {author} {\bibfnamefont
  {B.}~\bibnamefont {Lommel}}, \bibinfo {author} {\bibfnamefont
  {R.}~\bibnamefont {Mann}}, \bibinfo {author} {\bibfnamefont {A.~G.}\
  \bibnamefont {Popeko}}, \bibinfo {author} {\bibfnamefont {S.}~\bibnamefont
  {Reshitko}}, \bibinfo {author} {\bibfnamefont {S.}~\bibnamefont {\'Saro}},
  \bibinfo {author} {\bibfnamefont {J.}~\bibnamefont {Uusitalo}},\ and\
  \bibinfo {author} {\bibfnamefont {A.~V.}\ \bibnamefont {Yeremin}},\
  }\bibfield  {title} {\bibinfo {title} {{N}ew results on elements 111 and
  112},\ }\href {https://doi.org/10.1140/epja/i2001-10119-x} {\bibfield
  {journal} {\bibinfo  {journal} {Eur. Phys. J. A}\ }\textbf {\bibinfo {volume}
  {14}},\ \bibinfo {pages} {147} (\bibinfo {year} {2002})}\BibitemShut
  {NoStop}%
\bibitem [{\citenamefont {M\"unzenberg}\ and\ \citenamefont
  {Morita}(2015)}]{munzenberg2015}%
  \BibitemOpen
  \bibfield  {author} {\bibinfo {author} {\bibfnamefont {G.}~\bibnamefont
  {M\"unzenberg}}\ and\ \bibinfo {author} {\bibfnamefont {K.}~\bibnamefont
  {Morita}},\ }\bibfield  {title} {\bibinfo {title} {Synthesis of the heaviest
  nuclei in cold fusion reactions},\ }\href
  {https://doi.org/10.1016/j.nuclphysa.2015.06.007} {\bibfield  {journal}
  {\bibinfo  {journal} {Nucl. Phys. A}\ }\textbf {\bibinfo {volume} {944}},\
  \bibinfo {pages} {3} (\bibinfo {year} {2015})}\BibitemShut {NoStop}%
\bibitem [{\citenamefont {Morita}(2015)}]{morita2015}%
  \BibitemOpen
  \bibfield  {author} {\bibinfo {author} {\bibfnamefont {K.}~\bibnamefont
  {Morita}},\ }\bibfield  {title} {\bibinfo {title} {{SHE} research at
  {RIKEN/GARIS}},\ }\href {https://doi.org/10.1016/j.nuclphysa.2015.10.007}
  {\bibfield  {journal} {\bibinfo  {journal} {Nucl. Phys. A}\ }\textbf
  {\bibinfo {volume} {944}},\ \bibinfo {pages} {30} (\bibinfo {year}
  {2015})}\BibitemShut {NoStop}%
\bibitem [{\citenamefont {{Yu. Ts. Oganessian}}\ and\ \citenamefont
  {Utyonkov}(2015)}]{oganessian2015}%
  \BibitemOpen
  \bibfield  {author} {\bibinfo {author} {\bibnamefont {{Yu. Ts. Oganessian}}}\
  and\ \bibinfo {author} {\bibfnamefont {V.~K.}\ \bibnamefont {Utyonkov}},\
  }\bibfield  {title} {\bibinfo {title} {Superheavy nuclei from
  $^{48}${C}a-induced reactions},\ }\href
  {https://doi.org/10.1016/j.nuclphysa.2015.07.003} {\bibfield  {journal}
  {\bibinfo  {journal} {Nucl. Phys. A}\ }\textbf {\bibinfo {volume} {944}},\
  \bibinfo {pages} {62} (\bibinfo {year} {2015})}\BibitemShut {NoStop}%
\bibitem [{\citenamefont {Roberto}\ \emph {et~al.}(2015)\citenamefont
  {Roberto}, \citenamefont {Alexander}, \citenamefont {Boll}, \citenamefont
  {Burns}, \citenamefont {Ezold}, \citenamefont {Felker}, \citenamefont
  {Hogle},\ and\ \citenamefont {Rykaczewski}}]{roberto2015}%
  \BibitemOpen
  \bibfield  {author} {\bibinfo {author} {\bibfnamefont {J.~B.}\ \bibnamefont
  {Roberto}}, \bibinfo {author} {\bibfnamefont {C.~W.}\ \bibnamefont
  {Alexander}}, \bibinfo {author} {\bibfnamefont {R.~A.}\ \bibnamefont {Boll}},
  \bibinfo {author} {\bibfnamefont {J.~D.}\ \bibnamefont {Burns}}, \bibinfo
  {author} {\bibfnamefont {J.~G.}\ \bibnamefont {Ezold}}, \bibinfo {author}
  {\bibfnamefont {L.~K.}\ \bibnamefont {Felker}}, \bibinfo {author}
  {\bibfnamefont {S.~L.}\ \bibnamefont {Hogle}},\ and\ \bibinfo {author}
  {\bibfnamefont {K.~P.}\ \bibnamefont {Rykaczewski}},\ }\bibfield  {title}
  {\bibinfo {title} {Actinide targets for the synthesis of super-heavy
  elements},\ }\href {https://doi.org/10.1016/j.nuclphysa.2015.06.009}
  {\bibfield  {journal} {\bibinfo  {journal} {Nucl. Phys. A}\ }\textbf
  {\bibinfo {volume} {944}},\ \bibinfo {pages} {99} (\bibinfo {year}
  {2015})}\BibitemShut {NoStop}%
\bibitem [{\citenamefont {Bender}\ \emph {et~al.}(1999)\citenamefont {Bender},
  \citenamefont {Rutz}, \citenamefont {Reinhard}, \citenamefont {Maruhn},\ and\
  \citenamefont {Greiner}}]{bender1999}%
  \BibitemOpen
  \bibfield  {author} {\bibinfo {author} {\bibfnamefont {M.}~\bibnamefont
  {Bender}}, \bibinfo {author} {\bibfnamefont {K.}~\bibnamefont {Rutz}},
  \bibinfo {author} {\bibfnamefont {P.-G.}\ \bibnamefont {Reinhard}}, \bibinfo
  {author} {\bibfnamefont {J.~A.}\ \bibnamefont {Maruhn}},\ and\ \bibinfo
  {author} {\bibfnamefont {W.}~\bibnamefont {Greiner}},\ }\bibfield  {title}
  {\bibinfo {title} {{S}hell structure of superheavy nuclei in self-consistent
  mean-field models},\ }\href {https://doi.org/10.1103/PhysRevC.60.034304}
  {\bibfield  {journal} {\bibinfo  {journal} {Phys. Rev. C}\ }\textbf {\bibinfo
  {volume} {60}},\ \bibinfo {pages} {034304} (\bibinfo {year}
  {1999})}\BibitemShut {NoStop}%
\bibitem [{\citenamefont {Nazarewicz}\ \emph {et~al.}(2002)\citenamefont
  {Nazarewicz}, \citenamefont {Bender}, \citenamefont {\'{C}wiok},
  \citenamefont {Heenen}, \citenamefont {Kruppa}, \citenamefont {Reinhard},\
  and\ \citenamefont {Vertse}}]{nazarewicz2002}%
  \BibitemOpen
  \bibfield  {author} {\bibinfo {author} {\bibfnamefont {W.}~\bibnamefont
  {Nazarewicz}}, \bibinfo {author} {\bibfnamefont {M.}~\bibnamefont {Bender}},
  \bibinfo {author} {\bibfnamefont {S.}~\bibnamefont {\'{C}wiok}}, \bibinfo
  {author} {\bibfnamefont {P.~H.}\ \bibnamefont {Heenen}}, \bibinfo {author}
  {\bibfnamefont {A.~T.}\ \bibnamefont {Kruppa}}, \bibinfo {author}
  {\bibfnamefont {P.-G.}\ \bibnamefont {Reinhard}},\ and\ \bibinfo {author}
  {\bibfnamefont {T.}~\bibnamefont {Vertse}},\ }\bibfield  {title} {\bibinfo
  {title} {{T}heoretical description of superheavy nuclei},\ }\href
  {https://doi.org/10.1016/S0375-9474(01)01567-6} {\bibfield  {journal}
  {\bibinfo  {journal} {Nucl. Phys. A}\ }\textbf {\bibinfo {volume} {701}},\
  \bibinfo {pages} {165} (\bibinfo {year} {2002})}\BibitemShut {NoStop}%
\bibitem [{\citenamefont {\'{C}wiok}\ \emph {et~al.}(2005)\citenamefont
  {\'{C}wiok}, \citenamefont {Heenen},\ and\ \citenamefont
  {Nazarewicz}}]{cwiok2005}%
  \BibitemOpen
  \bibfield  {author} {\bibinfo {author} {\bibfnamefont {S.}~\bibnamefont
  {\'{C}wiok}}, \bibinfo {author} {\bibfnamefont {P.-H.}\ \bibnamefont
  {Heenen}},\ and\ \bibinfo {author} {\bibfnamefont {W.}~\bibnamefont
  {Nazarewicz}},\ }\bibfield  {title} {\bibinfo {title} {{S}hape coexistence
  and triaxiality in the superheavy nuclei},\ }\href
  {https://doi.org/10.1038/nature03336} {\bibfield  {journal} {\bibinfo
  {journal} {Nature}\ }\textbf {\bibinfo {volume} {433}},\ \bibinfo {pages}
  {705} (\bibinfo {year} {2005})}\BibitemShut {NoStop}%
\bibitem [{\citenamefont {Pei}\ \emph {et~al.}(2009)\citenamefont {Pei},
  \citenamefont {Nazarewicz}, \citenamefont {Sheikh},\ and\ \citenamefont
  {Kerman}}]{pei2009a}%
  \BibitemOpen
  \bibfield  {author} {\bibinfo {author} {\bibfnamefont {J.~C.}\ \bibnamefont
  {Pei}}, \bibinfo {author} {\bibfnamefont {W.}~\bibnamefont {Nazarewicz}},
  \bibinfo {author} {\bibfnamefont {J.~A.}\ \bibnamefont {Sheikh}},\ and\
  \bibinfo {author} {\bibfnamefont {A.~K.}\ \bibnamefont {Kerman}},\ }\bibfield
   {title} {\bibinfo {title} {{F}ission {B}arriers of {C}ompound {S}uperheavy
  {N}uclei},\ }\href {https://doi.org/10.1103/PhysRevLett.102.192501}
  {\bibfield  {journal} {\bibinfo  {journal} {Phys. Rev. Lett.}\ }\textbf
  {\bibinfo {volume} {102}},\ \bibinfo {pages} {192501} (\bibinfo {year}
  {2009})}\BibitemShut {NoStop}%
\bibitem [{\citenamefont {Stone}\ \emph {et~al.}(2019)\citenamefont {Stone},
  \citenamefont {Morita}, \citenamefont {Guichon},\ and\ \citenamefont
  {Thomas}}]{stone2019}%
  \BibitemOpen
  \bibfield  {author} {\bibinfo {author} {\bibfnamefont {J.~R.}\ \bibnamefont
  {Stone}}, \bibinfo {author} {\bibfnamefont {K.}~\bibnamefont {Morita}},
  \bibinfo {author} {\bibfnamefont {P.~A.~M.}\ \bibnamefont {Guichon}},\ and\
  \bibinfo {author} {\bibfnamefont {A.~W.}\ \bibnamefont {Thomas}},\ }\bibfield
   {title} {\bibinfo {title} {Physics of even-even superheavy nuclei with
  $96<${Z}$<110$ in the quark-meson-coupling model},\ }\href
  {https://doi.org/10.1103/PhysRevC.100.044302} {\bibfield  {journal} {\bibinfo
   {journal} {Phys. Rev. C}\ }\textbf {\bibinfo {volume} {100}},\ \bibinfo
  {pages} {044302} (\bibinfo {year} {2019})}\BibitemShut {NoStop}%
\bibitem [{\citenamefont {Adamian}\ \emph {et~al.}(2003)\citenamefont
  {Adamian}, \citenamefont {Antonenko},\ and\ \citenamefont
  {Scheid}}]{adamian2003}%
  \BibitemOpen
  \bibfield  {author} {\bibinfo {author} {\bibfnamefont {G.~G.}\ \bibnamefont
  {Adamian}}, \bibinfo {author} {\bibfnamefont {N.~V.}\ \bibnamefont
  {Antonenko}},\ and\ \bibinfo {author} {\bibfnamefont {W.}~\bibnamefont
  {Scheid}},\ }\bibfield  {title} {\bibinfo {title} {{C}haracteristics of
  quasifission products within the dinuclear system model},\ }\href
  {https://doi.org/10.1103/PhysRevC.68.034601} {\bibfield  {journal} {\bibinfo
  {journal} {Phys. Rev. C}\ }\textbf {\bibinfo {volume} {68}},\ \bibinfo
  {pages} {034601} (\bibinfo {year} {2003})}\BibitemShut {NoStop}%
\bibitem [{\citenamefont {{Valery Zagrebaev}}\ and\ \citenamefont {{Walter
  Greiner}}(2007)}]{zagrebaev2007}%
  \BibitemOpen
  \bibfield  {author} {\bibinfo {author} {\bibnamefont {{Valery Zagrebaev}}}\
  and\ \bibinfo {author} {\bibnamefont {{Walter Greiner}}},\ }\bibfield
  {title} {\bibinfo {title} {{S}hell effects in damped collisions: a new way to
  superheavies},\ }\href {https://doi.org/10.1088/0954-3899/34/11/004}
  {\bibfield  {journal} {\bibinfo  {journal} {J. Phys. G}\ }\textbf {\bibinfo
  {volume} {34}},\ \bibinfo {pages} {2265} (\bibinfo {year}
  {2007})}\BibitemShut {NoStop}%
\bibitem [{\citenamefont {Umar}\ \emph {et~al.}(2008)\citenamefont {Umar},
  \citenamefont {Oberacker},\ and\ \citenamefont {Maruhn}}]{umar2008a}%
  \BibitemOpen
  \bibfield  {author} {\bibinfo {author} {\bibfnamefont {A.~S.}\ \bibnamefont
  {Umar}}, \bibinfo {author} {\bibfnamefont {V.~E.}\ \bibnamefont
  {Oberacker}},\ and\ \bibinfo {author} {\bibfnamefont {J.~A.}\ \bibnamefont
  {Maruhn}},\ }\bibfield  {title} {\bibinfo {title} {{N}eutron transfer
  dynamics and doorway to fusion in time-dependent {H}artree-{F}ock theory},\
  }\href {https://doi.org/10.1140/epja/i2008-10614-6} {\bibfield  {journal}
  {\bibinfo  {journal} {Eur. Phys. J. A}\ }\textbf {\bibinfo {volume} {37}},\
  \bibinfo {pages} {245} (\bibinfo {year} {2008})}\BibitemShut {NoStop}%
\bibitem [{\citenamefont {{C\'edric Golabek}}\ and\ \citenamefont {{C\'edric
  Simenel}}(2009)}]{golabek2009}%
  \BibitemOpen
  \bibfield  {author} {\bibinfo {author} {\bibnamefont {{C\'edric Golabek}}}\
  and\ \bibinfo {author} {\bibnamefont {{C\'edric Simenel}}},\ }\bibfield
  {title} {\bibinfo {title} {{C}ollision {D}ynamics of {T}wo $^{238}${U A}tomic
  {N}uclei},\ }\href {https://doi.org/10.1103/PhysRevLett.103.042701}
  {\bibfield  {journal} {\bibinfo  {journal} {Phys. Rev. Lett.}\ }\textbf
  {\bibinfo {volume} {103}},\ \bibinfo {pages} {042701} (\bibinfo {year}
  {2009})}\BibitemShut {NoStop}%
\bibitem [{\citenamefont {Aritomo}(2009)}]{aritomo2009}%
  \BibitemOpen
  \bibfield  {author} {\bibinfo {author} {\bibfnamefont {Y.}~\bibnamefont
  {Aritomo}},\ }\bibfield  {title} {\bibinfo {title} {{A}nalysis of dynamical
  processes using the mass distribution of fission fragments in heavy-ion
  reactions},\ }\href {https://doi.org/10.1103/PhysRevC.80.064604} {\bibfield
  {journal} {\bibinfo  {journal} {Phys. Rev. C}\ }\textbf {\bibinfo {volume}
  {80}},\ \bibinfo {pages} {064604} (\bibinfo {year} {2009})}\BibitemShut
  {NoStop}%
\bibitem [{\citenamefont {{David J. Kedziora}}\ and\ \citenamefont {{C\'edric
  Simenel}}(2010)}]{kedziora2010}%
  \BibitemOpen
  \bibfield  {author} {\bibinfo {author} {\bibnamefont {{David J. Kedziora}}}\
  and\ \bibinfo {author} {\bibnamefont {{C\'edric Simenel}}},\ }\bibfield
  {title} {\bibinfo {title} {{N}ew inverse quasifission mechanism to produce
  neutron-rich transfermium nuclei},\ }\href
  {https://doi.org/10.1103/PhysRevC.81.044613} {\bibfield  {journal} {\bibinfo
  {journal} {Phys. Rev. C}\ }\textbf {\bibinfo {volume} {81}},\ \bibinfo
  {pages} {044613} (\bibinfo {year} {2010})}\BibitemShut {NoStop}%
\bibitem [{\citenamefont {Zhao}\ \emph {et~al.}(2016)\citenamefont {Zhao},
  \citenamefont {Li}, \citenamefont {Zhang}, \citenamefont {Wang},
  \citenamefont {Li}, \citenamefont {Shen}, \citenamefont {Wang},\ and\
  \citenamefont {Wu}}]{zhao2016}%
  \BibitemOpen
  \bibfield  {author} {\bibinfo {author} {\bibfnamefont {K.}~\bibnamefont
  {Zhao}}, \bibinfo {author} {\bibfnamefont {Z.}~\bibnamefont {Li}}, \bibinfo
  {author} {\bibfnamefont {Y.}~\bibnamefont {Zhang}}, \bibinfo {author}
  {\bibfnamefont {N.}~\bibnamefont {Wang}}, \bibinfo {author} {\bibfnamefont
  {Q.}~\bibnamefont {Li}}, \bibinfo {author} {\bibfnamefont {C.}~\bibnamefont
  {Shen}}, \bibinfo {author} {\bibfnamefont {Y.}~\bibnamefont {Wang}},\ and\
  \bibinfo {author} {\bibfnamefont {X.}~\bibnamefont {Wu}},\ }\bibfield
  {title} {\bibinfo {title} {Production of unknown neutron--rich isotopes in
  $^{238}\mathrm{U}+{}^{238}\mathrm{U}$ collisions at near--barrier energy},\
  }\href {https://doi.org/10.1103/PhysRevC.94.024601} {\bibfield  {journal}
  {\bibinfo  {journal} {Phys. Rev. C}\ }\textbf {\bibinfo {volume} {94}},\
  \bibinfo {pages} {024601} (\bibinfo {year} {2016})}\BibitemShut {NoStop}%
\bibitem [{\citenamefont {Sekizawa}(2017{\natexlab{a}})}]{sekizawa2017a}%
  \BibitemOpen
  \bibfield  {author} {\bibinfo {author} {\bibfnamefont {K.}~\bibnamefont
  {Sekizawa}},\ }\bibfield  {title} {\bibinfo {title} {Enhanced nucleon
  transfer in tip collisions of $^{238}\mathrm{U}+{}^{124}\mathrm{Sn}$},\
  }\href {https://doi.org/10.1103/PhysRevC.96.041601} {\bibfield  {journal}
  {\bibinfo  {journal} {Phys. Rev. C}\ }\textbf {\bibinfo {volume} {96}},\
  \bibinfo {pages} {041601} (\bibinfo {year}
  {2017}{\natexlab{a}})}\BibitemShut {NoStop}%
\bibitem [{\citenamefont {Wu}\ and\ \citenamefont {Guo}(2019)}]{wu2019}%
  \BibitemOpen
  \bibfield  {author} {\bibinfo {author} {\bibfnamefont {Z.}~\bibnamefont
  {Wu}}\ and\ \bibinfo {author} {\bibfnamefont {L.}~\bibnamefont {Guo}},\
  }\bibfield  {title} {\bibinfo {title} {Microscopic studies of production
  cross sections in multinucleon transfer reaction
  $^{58}\mathrm{Ni}+^{124}\mathrm{Sn}$},\ }\href
  {https://doi.org/10.1103/PhysRevC.100.014612} {\bibfield  {journal} {\bibinfo
   {journal} {Phys. Rev. C}\ }\textbf {\bibinfo {volume} {100}},\ \bibinfo
  {pages} {014612} (\bibinfo {year} {2019})}\BibitemShut {NoStop}%
\bibitem [{\citenamefont {T{\~{o}}ke}\ \emph {et~al.}(1985)\citenamefont
  {T{\~{o}}ke}, \citenamefont {Bock}, \citenamefont {Dai}, \citenamefont
  {Gobbi}, \citenamefont {Gralla}, \citenamefont {Hildenbrand}, \citenamefont
  {Kuzminski}, \citenamefont {M\"uller}, \citenamefont {Olmi}, \citenamefont
  {Stelzer}, \citenamefont {Back},\ and\ \citenamefont
  {Bj\o{}rnholm}}]{toke1985}%
  \BibitemOpen
  \bibfield  {author} {\bibinfo {author} {\bibfnamefont {J.}~\bibnamefont
  {T{\~{o}}ke}}, \bibinfo {author} {\bibfnamefont {R.}~\bibnamefont {Bock}},
  \bibinfo {author} {\bibfnamefont {G.~X.}\ \bibnamefont {Dai}}, \bibinfo
  {author} {\bibfnamefont {A.}~\bibnamefont {Gobbi}}, \bibinfo {author}
  {\bibfnamefont {S.}~\bibnamefont {Gralla}}, \bibinfo {author} {\bibfnamefont
  {K.~D.}\ \bibnamefont {Hildenbrand}}, \bibinfo {author} {\bibfnamefont
  {J.}~\bibnamefont {Kuzminski}}, \bibinfo {author} {\bibfnamefont {W.~F.~J.}\
  \bibnamefont {M\"uller}}, \bibinfo {author} {\bibfnamefont {A.}~\bibnamefont
  {Olmi}}, \bibinfo {author} {\bibfnamefont {H.}~\bibnamefont {Stelzer}},
  \bibinfo {author} {\bibfnamefont {B.~B.}\ \bibnamefont {Back}},\ and\
  \bibinfo {author} {\bibfnamefont {S.}~\bibnamefont {Bj\o{}rnholm}},\
  }\bibfield  {title} {\bibinfo {title} {{Q}uasi-fission: {T}he mass-drift mode
  in heavy-ion reactions},\ }\href
  {https://doi.org/10.1016/0375-9474(85)90344-6} {\bibfield  {journal}
  {\bibinfo  {journal} {Nucl. Phys. A}\ }\textbf {\bibinfo {volume} {440}},\
  \bibinfo {pages} {327} (\bibinfo {year} {1985})}\BibitemShut {NoStop}%
\bibitem [{\citenamefont {{du Rietz}}\ \emph {et~al.}(2011)\citenamefont {{du
  Rietz}}, \citenamefont {Hinde}, \citenamefont {Dasgupta}, \citenamefont
  {Thomas}, \citenamefont {Gasques}, \citenamefont {Evers}, \citenamefont
  {Lobanov},\ and\ \citenamefont {Wakhle}}]{durietz2011}%
  \BibitemOpen
  \bibfield  {author} {\bibinfo {author} {\bibfnamefont {R.}~\bibnamefont {{du
  Rietz}}}, \bibinfo {author} {\bibfnamefont {D.~J.}\ \bibnamefont {Hinde}},
  \bibinfo {author} {\bibfnamefont {M.}~\bibnamefont {Dasgupta}}, \bibinfo
  {author} {\bibfnamefont {R.~G.}\ \bibnamefont {Thomas}}, \bibinfo {author}
  {\bibfnamefont {L.~R.}\ \bibnamefont {Gasques}}, \bibinfo {author}
  {\bibfnamefont {M.}~\bibnamefont {Evers}}, \bibinfo {author} {\bibfnamefont
  {N.}~\bibnamefont {Lobanov}},\ and\ \bibinfo {author} {\bibfnamefont
  {A.}~\bibnamefont {Wakhle}},\ }\bibfield  {title} {\bibinfo {title}
  {Predominant {T}ime {S}cales in {F}ission {P}rocesses in {R}eactions of {S},
  {T}i and {N}i with {W}: {Z}eptosecond versus {A}ttosecond},\ }\href
  {https://doi.org/10.1103/PhysRevLett.106.052701} {\bibfield  {journal}
  {\bibinfo  {journal} {Phys. Rev. Lett.}\ }\textbf {\bibinfo {volume} {106}},\
  \bibinfo {pages} {052701} (\bibinfo {year} {2011})}\BibitemShut {NoStop}%
\bibitem [{\citenamefont {Hinde}\ \emph {et~al.}(1992)\citenamefont {Hinde},
  \citenamefont {Hilscher}, \citenamefont {Rossner}, \citenamefont {Gebauer},
  \citenamefont {Lehmann},\ and\ \citenamefont {Wilpert}}]{hinde1992}%
  \BibitemOpen
  \bibfield  {author} {\bibinfo {author} {\bibfnamefont {D.~J.}\ \bibnamefont
  {Hinde}}, \bibinfo {author} {\bibfnamefont {D.}~\bibnamefont {Hilscher}},
  \bibinfo {author} {\bibfnamefont {H.}~\bibnamefont {Rossner}}, \bibinfo
  {author} {\bibfnamefont {B.}~\bibnamefont {Gebauer}}, \bibinfo {author}
  {\bibfnamefont {M.}~\bibnamefont {Lehmann}},\ and\ \bibinfo {author}
  {\bibfnamefont {M.}~\bibnamefont {Wilpert}},\ }\bibfield  {title} {\bibinfo
  {title} {{N}eutron emission as a probe of fusion-fission and quasi-fission
  dynamics},\ }\href {https://doi.org/10.1103/PhysRevC.45.1229} {\bibfield
  {journal} {\bibinfo  {journal} {Phys. Rev. C}\ }\textbf {\bibinfo {volume}
  {45}},\ \bibinfo {pages} {1229} (\bibinfo {year} {1992})}\BibitemShut
  {NoStop}%
\bibitem [{\citenamefont {Hinde}\ \emph {et~al.}(1995)\citenamefont {Hinde},
  \citenamefont {Dasgupta}, \citenamefont {Leigh}, \citenamefont {Lestone},
  \citenamefont {Mein}, \citenamefont {Morton}, \citenamefont {Newton},\ and\
  \citenamefont {Timmers}}]{hinde1995}%
  \BibitemOpen
  \bibfield  {author} {\bibinfo {author} {\bibfnamefont {D.~J.}\ \bibnamefont
  {Hinde}}, \bibinfo {author} {\bibfnamefont {M.}~\bibnamefont {Dasgupta}},
  \bibinfo {author} {\bibfnamefont {J.~R.}\ \bibnamefont {Leigh}}, \bibinfo
  {author} {\bibfnamefont {J.~P.}\ \bibnamefont {Lestone}}, \bibinfo {author}
  {\bibfnamefont {J.~C.}\ \bibnamefont {Mein}}, \bibinfo {author}
  {\bibfnamefont {C.~R.}\ \bibnamefont {Morton}}, \bibinfo {author}
  {\bibfnamefont {J.~O.}\ \bibnamefont {Newton}},\ and\ \bibinfo {author}
  {\bibfnamefont {H.}~\bibnamefont {Timmers}},\ }\bibfield  {title} {\bibinfo
  {title} {{F}usion-{F}ission versus {Q}uasifission: {E}ffect of {N}uclear
  {O}rientation},\ }\href {https://doi.org/10.1103/PhysRevLett.74.1295}
  {\bibfield  {journal} {\bibinfo  {journal} {Phys. Rev. Lett.}\ }\textbf
  {\bibinfo {volume} {74}},\ \bibinfo {pages} {1295} (\bibinfo {year}
  {1995})}\BibitemShut {NoStop}%
\bibitem [{\citenamefont {Itkis}\ \emph {et~al.}(2004)\citenamefont {Itkis},
  \citenamefont {\"Ayst\"o}, \citenamefont {Beghini}, \citenamefont {Bogachev},
  \citenamefont {Corradi}, \citenamefont {Dorvaux}, \citenamefont {Gadea},
  \citenamefont {Giardina}, \citenamefont {Hanappe}, \citenamefont {Itkis},
  \citenamefont {Jandel}, \citenamefont {Kliman}, \citenamefont {Khlebnikov},
  \citenamefont {Kniajeva}, \citenamefont {Kondratiev}, \citenamefont
  {Kozulin}, \citenamefont {Krupa}, \citenamefont {Latina}, \citenamefont
  {Materna}, \citenamefont {Montagnoli}, \citenamefont {Oganessian},
  \citenamefont {Pokrovsky}, \citenamefont {Prokhorova}, \citenamefont
  {Rowley}, \citenamefont {Rubchenya}, \citenamefont {Rusanov}, \citenamefont
  {Sagaidak}, \citenamefont {Scarlassara}, \citenamefont {Stefanini},
  \citenamefont {Stuttge}, \citenamefont {Szilner}, \citenamefont {Trotta},
  \citenamefont {Trzaska}, \citenamefont {Vakhtin}, \citenamefont {Vinodkumar},
  \citenamefont {Voskressenski},\ and\ \citenamefont {Zagrebaev}}]{itkis2004}%
  \BibitemOpen
  \bibfield  {author} {\bibinfo {author} {\bibfnamefont {M.~G.}\ \bibnamefont
  {Itkis}}, \bibinfo {author} {\bibfnamefont {J.}~\bibnamefont {\"Ayst\"o}},
  \bibinfo {author} {\bibfnamefont {S.}~\bibnamefont {Beghini}}, \bibinfo
  {author} {\bibfnamefont {A.~A.}\ \bibnamefont {Bogachev}}, \bibinfo {author}
  {\bibfnamefont {L.}~\bibnamefont {Corradi}}, \bibinfo {author} {\bibfnamefont
  {O.}~\bibnamefont {Dorvaux}}, \bibinfo {author} {\bibfnamefont
  {A.}~\bibnamefont {Gadea}}, \bibinfo {author} {\bibfnamefont
  {G.}~\bibnamefont {Giardina}}, \bibinfo {author} {\bibfnamefont
  {F.}~\bibnamefont {Hanappe}}, \bibinfo {author} {\bibfnamefont {I.~M.}\
  \bibnamefont {Itkis}}, \bibinfo {author} {\bibfnamefont {M.}~\bibnamefont
  {Jandel}}, \bibinfo {author} {\bibfnamefont {J.}~\bibnamefont {Kliman}},
  \bibinfo {author} {\bibfnamefont {S.~V.}\ \bibnamefont {Khlebnikov}},
  \bibinfo {author} {\bibfnamefont {G.~N.}\ \bibnamefont {Kniajeva}}, \bibinfo
  {author} {\bibfnamefont {N.~A.}\ \bibnamefont {Kondratiev}}, \bibinfo
  {author} {\bibfnamefont {E.~M.}\ \bibnamefont {Kozulin}}, \bibinfo {author}
  {\bibfnamefont {L.}~\bibnamefont {Krupa}}, \bibinfo {author} {\bibfnamefont
  {A.}~\bibnamefont {Latina}}, \bibinfo {author} {\bibfnamefont
  {T.}~\bibnamefont {Materna}}, \bibinfo {author} {\bibfnamefont
  {G.}~\bibnamefont {Montagnoli}}, \bibinfo {author} {\bibfnamefont {{\relax
  Yu. Ts}.}~\bibnamefont {Oganessian}}, \bibinfo {author} {\bibfnamefont
  {I.~V.}\ \bibnamefont {Pokrovsky}}, \bibinfo {author} {\bibfnamefont {E.~V.}\
  \bibnamefont {Prokhorova}}, \bibinfo {author} {\bibfnamefont
  {N.}~\bibnamefont {Rowley}}, \bibinfo {author} {\bibfnamefont {V.~A.}\
  \bibnamefont {Rubchenya}}, \bibinfo {author} {\bibfnamefont {{\relax A.
  Ya}.}~\bibnamefont {Rusanov}}, \bibinfo {author} {\bibfnamefont {R.~N.}\
  \bibnamefont {Sagaidak}}, \bibinfo {author} {\bibfnamefont {F.}~\bibnamefont
  {Scarlassara}}, \bibinfo {author} {\bibfnamefont {A.~M.}\ \bibnamefont
  {Stefanini}}, \bibinfo {author} {\bibfnamefont {L.}~\bibnamefont {Stuttge}},
  \bibinfo {author} {\bibfnamefont {S.}~\bibnamefont {Szilner}}, \bibinfo
  {author} {\bibfnamefont {M.}~\bibnamefont {Trotta}}, \bibinfo {author}
  {\bibfnamefont {W.~H.}\ \bibnamefont {Trzaska}}, \bibinfo {author}
  {\bibfnamefont {D.~N.}\ \bibnamefont {Vakhtin}}, \bibinfo {author}
  {\bibfnamefont {A.~M.}\ \bibnamefont {Vinodkumar}}, \bibinfo {author}
  {\bibfnamefont {V.~M.}\ \bibnamefont {Voskressenski}},\ and\ \bibinfo
  {author} {\bibfnamefont {V.~I.}\ \bibnamefont {Zagrebaev}},\ }\bibfield
  {title} {\bibinfo {title} {{S}hell effects in fission and quasi-fission of
  heavy and superheavy nuclei},\ }\href
  {https://doi.org/10.1016/j.nuclphysa.2004.01.022} {\bibfield  {journal}
  {\bibinfo  {journal} {Nucl. Phys. A}\ }\textbf {\bibinfo {volume} {734}},\
  \bibinfo {pages} {136} (\bibinfo {year} {2004})}\BibitemShut {NoStop}%
\bibitem [{\citenamefont {Wakhle}\ \emph {et~al.}(2014)\citenamefont {Wakhle},
  \citenamefont {Simenel}, \citenamefont {Hinde}, \citenamefont {Dasgupta},
  \citenamefont {Evers}, \citenamefont {Luong}, \citenamefont {du~Rietz},\ and\
  \citenamefont {Williams}}]{wakhle2014}%
  \BibitemOpen
  \bibfield  {author} {\bibinfo {author} {\bibfnamefont {A.}~\bibnamefont
  {Wakhle}}, \bibinfo {author} {\bibfnamefont {C.}~\bibnamefont {Simenel}},
  \bibinfo {author} {\bibfnamefont {D.~J.}\ \bibnamefont {Hinde}}, \bibinfo
  {author} {\bibfnamefont {M.}~\bibnamefont {Dasgupta}}, \bibinfo {author}
  {\bibfnamefont {M.}~\bibnamefont {Evers}}, \bibinfo {author} {\bibfnamefont
  {D.~H.}\ \bibnamefont {Luong}}, \bibinfo {author} {\bibfnamefont
  {R.}~\bibnamefont {du~Rietz}},\ and\ \bibinfo {author} {\bibfnamefont
  {E.}~\bibnamefont {Williams}},\ }\bibfield  {title} {\bibinfo {title}
  {{I}nterplay between {Q}uantum {S}hells and {O}rientation in
  {Q}uasifission},\ }\href {https://doi.org/10.1103/PhysRevLett.113.182502}
  {\bibfield  {journal} {\bibinfo  {journal} {Phys. Rev. Lett.}\ }\textbf
  {\bibinfo {volume} {113}},\ \bibinfo {pages} {182502} (\bibinfo {year}
  {2014})}\BibitemShut {NoStop}%
\bibitem [{\citenamefont {Majka}\ \emph {et~al.}(2018)\citenamefont {Majka},
  \citenamefont {P\l{}aneta}, \citenamefont {Sosin}, \citenamefont {Wieloch},
  \citenamefont {Zelga}, \citenamefont {Adamczyk}, \citenamefont {Pelczar},
  \citenamefont {Barbui}, \citenamefont {Wuenschel}, \citenamefont {Hagel},
  \citenamefont {Cao}, \citenamefont {Kim}, \citenamefont {Natowitz},
  \citenamefont {Wada}, \citenamefont {Zheng}, \citenamefont {Giuliani},\ and\
  \citenamefont {Kowalski}}]{majka2018}%
  \BibitemOpen
  \bibfield  {author} {\bibinfo {author} {\bibfnamefont {Z.}~\bibnamefont
  {Majka}}, \bibinfo {author} {\bibfnamefont {R.}~\bibnamefont {P\l{}aneta}},
  \bibinfo {author} {\bibfnamefont {Z.}~\bibnamefont {Sosin}}, \bibinfo
  {author} {\bibfnamefont {A.}~\bibnamefont {Wieloch}}, \bibinfo {author}
  {\bibfnamefont {K.}~\bibnamefont {Zelga}}, \bibinfo {author} {\bibfnamefont
  {M.}~\bibnamefont {Adamczyk}}, \bibinfo {author} {\bibfnamefont
  {K.}~\bibnamefont {Pelczar}}, \bibinfo {author} {\bibfnamefont
  {M.}~\bibnamefont {Barbui}}, \bibinfo {author} {\bibfnamefont
  {S.}~\bibnamefont {Wuenschel}}, \bibinfo {author} {\bibfnamefont
  {K.}~\bibnamefont {Hagel}}, \bibinfo {author} {\bibfnamefont
  {X.}~\bibnamefont {Cao}}, \bibinfo {author} {\bibfnamefont {E.-J.}\
  \bibnamefont {Kim}}, \bibinfo {author} {\bibfnamefont {J.}~\bibnamefont
  {Natowitz}}, \bibinfo {author} {\bibfnamefont {R.}~\bibnamefont {Wada}},
  \bibinfo {author} {\bibfnamefont {H.}~\bibnamefont {Zheng}}, \bibinfo
  {author} {\bibfnamefont {G.}~\bibnamefont {Giuliani}},\ and\ \bibinfo
  {author} {\bibfnamefont {S.}~\bibnamefont {Kowalski}},\ }\bibfield  {title}
  {\bibinfo {title} {A novel experimental setup for rare events selection and
  its potential application to super-heavy elements search},\ }\href
  {https://doi.org/10.5506/APhysPolB.49.1801} {\bibfield  {journal} {\bibinfo
  {journal} {Acta Phys. Pol. B}\ }\textbf {\bibinfo {volume} {49}},\ \bibinfo
  {pages} {1801} (\bibinfo {year} {2018})}\BibitemShut {NoStop}%
\bibitem [{\citenamefont {Wuenschel}\ \emph {et~al.}(2018)\citenamefont
  {Wuenschel}, \citenamefont {Hagel}, \citenamefont {Barbui}, \citenamefont
  {Gauthier}, \citenamefont {Cao}, \citenamefont {Wada}, \citenamefont {Kim},
  \citenamefont {Majka}, \citenamefont {P\l{}aneta}, \citenamefont {Sosin},
  \citenamefont {Wieloch}, \citenamefont {Zelga}, \citenamefont {Kowalski},
  \citenamefont {Schmidt}, \citenamefont {Ma}, \citenamefont {Zhang},\ and\
  \citenamefont {Natowitz}}]{wuenschel2018}%
  \BibitemOpen
  \bibfield  {author} {\bibinfo {author} {\bibfnamefont {S.}~\bibnamefont
  {Wuenschel}}, \bibinfo {author} {\bibfnamefont {K.}~\bibnamefont {Hagel}},
  \bibinfo {author} {\bibfnamefont {M.}~\bibnamefont {Barbui}}, \bibinfo
  {author} {\bibfnamefont {J.}~\bibnamefont {Gauthier}}, \bibinfo {author}
  {\bibfnamefont {X.~G.}\ \bibnamefont {Cao}}, \bibinfo {author} {\bibfnamefont
  {R.}~\bibnamefont {Wada}}, \bibinfo {author} {\bibfnamefont {E.~J.}\
  \bibnamefont {Kim}}, \bibinfo {author} {\bibfnamefont {Z.}~\bibnamefont
  {Majka}}, \bibinfo {author} {\bibfnamefont {R.}~\bibnamefont {P\l{}aneta}},
  \bibinfo {author} {\bibfnamefont {Z.}~\bibnamefont {Sosin}}, \bibinfo
  {author} {\bibfnamefont {A.}~\bibnamefont {Wieloch}}, \bibinfo {author}
  {\bibfnamefont {K.}~\bibnamefont {Zelga}}, \bibinfo {author} {\bibfnamefont
  {S.}~\bibnamefont {Kowalski}}, \bibinfo {author} {\bibfnamefont
  {K.}~\bibnamefont {Schmidt}}, \bibinfo {author} {\bibfnamefont
  {C.}~\bibnamefont {Ma}}, \bibinfo {author} {\bibfnamefont {G.}~\bibnamefont
  {Zhang}},\ and\ \bibinfo {author} {\bibfnamefont {J.~B.}\ \bibnamefont
  {Natowitz}},\ }\bibfield  {title} {\bibinfo {title} {Experimental survey of
  the production of $\alpha-$decaying heavy elements in
  $^{238}\mathrm{U}+{}^{232}\mathrm{Th}$ reactions at 7.5--6.1 {MeV}/nucleon},\
  }\href {https://doi.org/10.1103/PhysRevC.97.064602} {\bibfield  {journal}
  {\bibinfo  {journal} {Phys. Rev. C}\ }\textbf {\bibinfo {volume} {97}},\
  \bibinfo {pages} {064602} (\bibinfo {year} {2018})}\BibitemShut {NoStop}%
\bibitem [{\citenamefont {Feng}\ \emph {et~al.}(2009)\citenamefont {Feng},
  \citenamefont {Jin},\ and\ \citenamefont {Li}}]{feng2009a}%
  \BibitemOpen
  \bibfield  {author} {\bibinfo {author} {\bibfnamefont {Z.-Q.}\ \bibnamefont
  {Feng}}, \bibinfo {author} {\bibfnamefont {G.-M.}\ \bibnamefont {Jin}},\ and\
  \bibinfo {author} {\bibfnamefont {J.-Q.}\ \bibnamefont {Li}},\ }\bibfield
  {title} {\bibinfo {title} {Production of heavy isotopes in transfer reactions
  by collisions of $^{238}\mathrm{U}+{}^{238}\mathrm{U}$},\ }\href
  {https://doi.org/10.1103/PhysRevC.80.067601} {\bibfield  {journal} {\bibinfo
  {journal} {Phys. Rev. C}\ }\textbf {\bibinfo {volume} {80}},\ \bibinfo
  {pages} {067601} (\bibinfo {year} {2009})}\BibitemShut {NoStop}%
\bibitem [{\citenamefont {Gupta}\ \emph {et~al.}(2007)\citenamefont {Gupta},
  \citenamefont {Patra}, \citenamefont {Stevenson},\ and\ \citenamefont
  {Greiner}}]{gupta2007b}%
  \BibitemOpen
  \bibfield  {author} {\bibinfo {author} {\bibfnamefont {R.~K.}\ \bibnamefont
  {Gupta}}, \bibinfo {author} {\bibfnamefont {S.~K.}\ \bibnamefont {Patra}},
  \bibinfo {author} {\bibfnamefont {P.~D.}\ \bibnamefont {Stevenson}},\ and\
  \bibinfo {author} {\bibfnamefont {W.}~\bibnamefont {Greiner}},\ }\bibfield
  {title} {\bibinfo {title} {A highly neutron--rich cluster and/or a superheavy
  nucleus in the compound nucleus $^{238}\mathrm{U}+{}^{238}\mathrm{U}$: {A}
  mean field study},\ }\href {https://doi.org/10.1142/S0218301307006137}
  {\bibfield  {journal} {\bibinfo  {journal} {Intl. J. Mod. Phys. E}\ }\textbf
  {\bibinfo {volume} {16}},\ \bibinfo {pages} {1721} (\bibinfo {year}
  {2007})}\BibitemShut {NoStop}%
\bibitem [{\citenamefont {Sargsyan}\ \emph {et~al.}(2009)\citenamefont
  {Sargsyan}, \citenamefont {Kanokov}, \citenamefont {Adamian}, \citenamefont
  {Antonenko},\ and\ \citenamefont {Scheid}}]{sargsyan2009}%
  \BibitemOpen
  \bibfield  {author} {\bibinfo {author} {\bibfnamefont {V.~V.}\ \bibnamefont
  {Sargsyan}}, \bibinfo {author} {\bibfnamefont {Z.}~\bibnamefont {Kanokov}},
  \bibinfo {author} {\bibfnamefont {G.~G.}\ \bibnamefont {Adamian}}, \bibinfo
  {author} {\bibfnamefont {N.~V.}\ \bibnamefont {Antonenko}},\ and\ \bibinfo
  {author} {\bibfnamefont {W.}~\bibnamefont {Scheid}},\ }\bibfield  {title}
  {\bibinfo {title} {Interaction times in the
  $^{136}\mathrm{Xe}+{}^{136}\mathrm{Xe}$ and
  $^{238}\mathrm{U}+{}^{238}\mathrm{U}$ reactions with a quantum master
  equation},\ }\href {https://doi.org/10.1103/PhysRevC.80.047603} {\bibfield
  {journal} {\bibinfo  {journal} {Phys. Rev. C}\ }\textbf {\bibinfo {volume}
  {80}},\ \bibinfo {pages} {047603} (\bibinfo {year} {2009})}\BibitemShut
  {NoStop}%
\bibitem [{\citenamefont {Zhao}\ \emph {et~al.}(2009)\citenamefont {Zhao},
  \citenamefont {Wu},\ and\ \citenamefont {Li}}]{zhao2009}%
  \BibitemOpen
  \bibfield  {author} {\bibinfo {author} {\bibfnamefont {K.}~\bibnamefont
  {Zhao}}, \bibinfo {author} {\bibfnamefont {X.}~\bibnamefont {Wu}},\ and\
  \bibinfo {author} {\bibfnamefont {Z.}~\bibnamefont {Li}},\ }\bibfield
  {title} {\bibinfo {title} {{Quantum molecular dynamics study of the mass
  distribution of products in $7.0{A}$\,{M}eV
  $^{238}\mathrm{U}+{}^{238}\mathrm{U}$ collisions}},\ }\href
  {https://doi.org/10.1103/PhysRevC.80.054607} {\bibfield  {journal} {\bibinfo
  {journal} {Phys. Rev. C}\ }\textbf {\bibinfo {volume} {80}},\ \bibinfo
  {pages} {054607} (\bibinfo {year} {2009})}\BibitemShut {NoStop}%
\bibitem [{\citenamefont {{Junlong Tian}}\ \emph {et~al.}(2008)\citenamefont
  {{Junlong Tian}}, \citenamefont {{Xizhen Wu}}, \citenamefont {{Kai Zhao}},
  \citenamefont {{Yingxun Zhang}},\ and\ \citenamefont {{Zhuxia
  Li}}}]{tian2008}%
  \BibitemOpen
  \bibfield  {author} {\bibinfo {author} {\bibnamefont {{Junlong Tian}}},
  \bibinfo {author} {\bibnamefont {{Xizhen Wu}}}, \bibinfo {author}
  {\bibnamefont {{Kai Zhao}}}, \bibinfo {author} {\bibnamefont {{Yingxun
  Zhang}}},\ and\ \bibinfo {author} {\bibnamefont {{Zhuxia Li}}},\ }\bibfield
  {title} {\bibinfo {title} {{P}roperties of the composite systems formed in
  the reactions $^{238}\mathrm{U}+{}^{238}\mathrm{U}$ and
  $^{232}\mathrm{Th}+{}^{250}\mathrm{Cf}$},\ }\href
  {https://doi.org/10.1103/PhysRevC.77.064603} {\bibfield  {journal} {\bibinfo
  {journal} {Phys. Rev. C}\ }\textbf {\bibinfo {volume} {77}},\ \bibinfo
  {pages} {064603} (\bibinfo {year} {2008})}\BibitemShut {NoStop}%
\bibitem [{\citenamefont {Cusson}\ \emph {et~al.}(1980)\citenamefont {Cusson},
  \citenamefont {Maruhn},\ and\ \citenamefont {St\"ocker}}]{cusson1980}%
  \BibitemOpen
  \bibfield  {author} {\bibinfo {author} {\bibfnamefont {R.~Y.}\ \bibnamefont
  {Cusson}}, \bibinfo {author} {\bibfnamefont {J.~A.}\ \bibnamefont {Maruhn}},\
  and\ \bibinfo {author} {\bibfnamefont {H.}~\bibnamefont {St\"ocker}},\
  }\bibfield  {title} {\bibinfo {title} {Collision of
  $^{238}\mathrm{U}+{}^{238}\mathrm{U}$ using a three--dimensional {TDHF--BCS}
  model},\ }\href {https://doi.org/10.1007/BF01438163} {\bibfield  {journal}
  {\bibinfo  {journal} {Z. Phys. A}\ }\textbf {\bibinfo {volume} {294}},\
  \bibinfo {pages} {257} (\bibinfo {year} {1980})}\BibitemShut {NoStop}%
\bibitem [{\citenamefont {Umar}\ \emph {et~al.}(2017)\citenamefont {Umar},
  \citenamefont {Simenel},\ and\ \citenamefont {Ye}}]{umar2017}%
  \BibitemOpen
  \bibfield  {author} {\bibinfo {author} {\bibfnamefont {A.~S.}\ \bibnamefont
  {Umar}}, \bibinfo {author} {\bibfnamefont {C.}~\bibnamefont {Simenel}},\ and\
  \bibinfo {author} {\bibfnamefont {W.}~\bibnamefont {Ye}},\ }\bibfield
  {title} {\bibinfo {title} {Transport properties of isospin asymmetric nuclear
  matter using the time-dependent {H}artree--{F}ock method},\ }\href
  {https://doi.org/10.1103/PhysRevC.96.024625} {\bibfield  {journal} {\bibinfo
  {journal} {Phys. Rev. C}\ }\textbf {\bibinfo {volume} {96}},\ \bibinfo
  {pages} {024625} (\bibinfo {year} {2017})}\BibitemShut {NoStop}%
\bibitem [{\citenamefont {Oberacker}\ \emph {et~al.}(2014)\citenamefont
  {Oberacker}, \citenamefont {Umar},\ and\ \citenamefont
  {Simenel}}]{oberacker2014}%
  \BibitemOpen
  \bibfield  {author} {\bibinfo {author} {\bibfnamefont {V.~E.}\ \bibnamefont
  {Oberacker}}, \bibinfo {author} {\bibfnamefont {A.~S.}\ \bibnamefont
  {Umar}},\ and\ \bibinfo {author} {\bibfnamefont {C.}~\bibnamefont
  {Simenel}},\ }\bibfield  {title} {\bibinfo {title} {{D}issipative dynamics in
  quasifission},\ }\href {https://doi.org/10.1103/PhysRevC.90.054605}
  {\bibfield  {journal} {\bibinfo  {journal} {Phys. Rev. C}\ }\textbf {\bibinfo
  {volume} {90}},\ \bibinfo {pages} {054605} (\bibinfo {year}
  {2014})}\BibitemShut {NoStop}%
\bibitem [{\citenamefont {Hammerton}\ \emph {et~al.}(2015)\citenamefont
  {Hammerton}, \citenamefont {Kohley}, \citenamefont {Hinde}, \citenamefont
  {Dasgupta}, \citenamefont {Wakhle}, \citenamefont {Williams}, \citenamefont
  {Oberacker}, \citenamefont {Umar}, \citenamefont {Carter}, \citenamefont
  {Cook}, \citenamefont {Greene}, \citenamefont {Jeung}, \citenamefont {Luong},
  \citenamefont {{McNeil}}, \citenamefont {Palshetkar}, \citenamefont
  {Rafferty}, \citenamefont {Simenel},\ and\ \citenamefont
  {Stiefel}}]{hammerton2015}%
  \BibitemOpen
  \bibfield  {author} {\bibinfo {author} {\bibfnamefont {K.}~\bibnamefont
  {Hammerton}}, \bibinfo {author} {\bibfnamefont {Z.}~\bibnamefont {Kohley}},
  \bibinfo {author} {\bibfnamefont {D.~J.}\ \bibnamefont {Hinde}}, \bibinfo
  {author} {\bibfnamefont {M.}~\bibnamefont {Dasgupta}}, \bibinfo {author}
  {\bibfnamefont {A.}~\bibnamefont {Wakhle}}, \bibinfo {author} {\bibfnamefont
  {E.}~\bibnamefont {Williams}}, \bibinfo {author} {\bibfnamefont {V.~E.}\
  \bibnamefont {Oberacker}}, \bibinfo {author} {\bibfnamefont {A.~S.}\
  \bibnamefont {Umar}}, \bibinfo {author} {\bibfnamefont {I.~P.}\ \bibnamefont
  {Carter}}, \bibinfo {author} {\bibfnamefont {K.~J.}\ \bibnamefont {Cook}},
  \bibinfo {author} {\bibfnamefont {J.}~\bibnamefont {Greene}}, \bibinfo
  {author} {\bibfnamefont {D.~Y.}\ \bibnamefont {Jeung}}, \bibinfo {author}
  {\bibfnamefont {D.~H.}\ \bibnamefont {Luong}}, \bibinfo {author}
  {\bibfnamefont {S.~D.}\ \bibnamefont {{McNeil}}}, \bibinfo {author}
  {\bibfnamefont {C.~S.}\ \bibnamefont {Palshetkar}}, \bibinfo {author}
  {\bibfnamefont {D.~C.}\ \bibnamefont {Rafferty}}, \bibinfo {author}
  {\bibfnamefont {C.}~\bibnamefont {Simenel}},\ and\ \bibinfo {author}
  {\bibfnamefont {K.}~\bibnamefont {Stiefel}},\ }\bibfield  {title} {\bibinfo
  {title} {{R}educed quasifission competition in fusion reactions forming
  neutron-rich heavy elements},\ }\href
  {https://doi.org/10.1103/PhysRevC.91.041602} {\bibfield  {journal} {\bibinfo
  {journal} {Phys. Rev. C}\ }\textbf {\bibinfo {volume} {91}},\ \bibinfo
  {pages} {041602} (\bibinfo {year} {2015})}\BibitemShut {NoStop}%
\bibitem [{\citenamefont {Umar}\ and\ \citenamefont
  {Oberacker}(2015)}]{umar2015c}%
  \BibitemOpen
  \bibfield  {author} {\bibinfo {author} {\bibfnamefont {A.~S.}\ \bibnamefont
  {Umar}}\ and\ \bibinfo {author} {\bibfnamefont {V.~E.}\ \bibnamefont
  {Oberacker}},\ }\bibfield  {title} {\bibinfo {title} {{T}ime-dependent {HF}
  approach to {SHE} dynamics},\ }\href
  {https://doi.org/10.1016/j.nuclphysa.2015.02.011} {\bibfield  {journal}
  {\bibinfo  {journal} {Nucl. Phys. A}\ }\textbf {\bibinfo {volume} {944}},\
  \bibinfo {pages} {238} (\bibinfo {year} {2015})}\BibitemShut {NoStop}%
\bibitem [{\citenamefont {Umar}\ \emph {et~al.}(2016)\citenamefont {Umar},
  \citenamefont {Oberacker},\ and\ \citenamefont {Simenel}}]{umar2016}%
  \BibitemOpen
  \bibfield  {author} {\bibinfo {author} {\bibfnamefont {A.~S.}\ \bibnamefont
  {Umar}}, \bibinfo {author} {\bibfnamefont {V.~E.}\ \bibnamefont
  {Oberacker}},\ and\ \bibinfo {author} {\bibfnamefont {C.}~\bibnamefont
  {Simenel}},\ }\bibfield  {title} {\bibinfo {title} {Fusion and quasifission
  dynamics in the reactions $^{48}\mathrm{Ca}+{}^{249}\mathrm{Bk}$ and
  $^{50}\mathrm{Ti}+{}^{249}\mathrm{Bk}$ using a time-dependent
  {H}artree-{F}ock approach},\ }\href
  {https://doi.org/10.1103/PhysRevC.94.024605} {\bibfield  {journal} {\bibinfo
  {journal} {Phys. Rev. C}\ }\textbf {\bibinfo {volume} {94}},\ \bibinfo
  {pages} {024605} (\bibinfo {year} {2016})}\BibitemShut {NoStop}%
\bibitem [{\citenamefont {Godbey}\ \emph {et~al.}(2019)\citenamefont {Godbey},
  \citenamefont {Umar},\ and\ \citenamefont {Simenel}}]{godbey2019}%
  \BibitemOpen
  \bibfield  {author} {\bibinfo {author} {\bibfnamefont {K.}~\bibnamefont
  {Godbey}}, \bibinfo {author} {\bibfnamefont {A.~S.}\ \bibnamefont {Umar}},\
  and\ \bibinfo {author} {\bibfnamefont {C.}~\bibnamefont {Simenel}},\
  }\bibfield  {title} {\bibinfo {title} {Deformed shell effects in
  ${}^{48}\mathrm{Ca}+{}^{249}\mathrm{Bk}$ quasifission fragments},\ }\href
  {https://doi.org/10.1103/PhysRevC.100.024610} {\bibfield  {journal} {\bibinfo
   {journal} {Phys. Rev. C}\ }\textbf {\bibinfo {volume} {100}},\ \bibinfo
  {pages} {024610} (\bibinfo {year} {2019})}\BibitemShut {NoStop}%
\bibitem [{\citenamefont {Simenel}\ and\ \citenamefont
  {Umar}(2014)}]{simenel2014a}%
  \BibitemOpen
  \bibfield  {author} {\bibinfo {author} {\bibfnamefont {C.}~\bibnamefont
  {Simenel}}\ and\ \bibinfo {author} {\bibfnamefont {A.~S.}\ \bibnamefont
  {Umar}},\ }\bibfield  {title} {\bibinfo {title} {{F}ormation and dynamics of
  fission fragments},\ }\href {https://doi.org/10.1103/PhysRevC.89.031601}
  {\bibfield  {journal} {\bibinfo  {journal} {Phys. Rev. C}\ }\textbf {\bibinfo
  {volume} {89}},\ \bibinfo {pages} {031601} (\bibinfo {year}
  {2014})}\BibitemShut {NoStop}%
\bibitem [{\citenamefont {Scamps}\ \emph {et~al.}(2015)\citenamefont {Scamps},
  \citenamefont {Simenel},\ and\ \citenamefont {Lacroix}}]{scamps2015a}%
  \BibitemOpen
  \bibfield  {author} {\bibinfo {author} {\bibfnamefont {G.}~\bibnamefont
  {Scamps}}, \bibinfo {author} {\bibfnamefont {C.}~\bibnamefont {Simenel}},\
  and\ \bibinfo {author} {\bibfnamefont {D.}~\bibnamefont {Lacroix}},\
  }\bibfield  {title} {\bibinfo {title} {{S}uperfluid dynamics of
  $^{258}\mathrm{Fm}$ fission},\ }\href
  {https://doi.org/10.1103/PhysRevC.92.011602} {\bibfield  {journal} {\bibinfo
  {journal} {Phys. Rev. C}\ }\textbf {\bibinfo {volume} {92}},\ \bibinfo
  {pages} {011602} (\bibinfo {year} {2015})}\BibitemShut {NoStop}%
\bibitem [{\citenamefont {Goddard}\ \emph {et~al.}(2015)\citenamefont
  {Goddard}, \citenamefont {Stevenson},\ and\ \citenamefont
  {Rios}}]{goddard2015}%
  \BibitemOpen
  \bibfield  {author} {\bibinfo {author} {\bibfnamefont {P.~M.}\ \bibnamefont
  {Goddard}}, \bibinfo {author} {\bibfnamefont {P.~D.}\ \bibnamefont
  {Stevenson}},\ and\ \bibinfo {author} {\bibfnamefont {A.}~\bibnamefont
  {Rios}},\ }\bibfield  {title} {\bibinfo {title} {{F}ission dynamics within
  time-dependent {H}artree-{F}ock: deformation-induced fission},\ }\href
  {https://doi.org/10.1103/PhysRevC.92.054610} {\bibfield  {journal} {\bibinfo
  {journal} {Phys. Rev. C}\ }\textbf {\bibinfo {volume} {92}},\ \bibinfo
  {pages} {054610} (\bibinfo {year} {2015})}\BibitemShut {NoStop}%
\bibitem [{\citenamefont {Tanimura}\ \emph {et~al.}(2015)\citenamefont
  {Tanimura}, \citenamefont {Lacroix},\ and\ \citenamefont
  {Scamps}}]{tanimura2015}%
  \BibitemOpen
  \bibfield  {author} {\bibinfo {author} {\bibfnamefont {Y.}~\bibnamefont
  {Tanimura}}, \bibinfo {author} {\bibfnamefont {D.}~\bibnamefont {Lacroix}},\
  and\ \bibinfo {author} {\bibfnamefont {G.}~\bibnamefont {Scamps}},\
  }\bibfield  {title} {\bibinfo {title} {Collective aspects deduced from
  time-dependent microscopic mean-field with pairing: {A}pplication to the
  fission process},\ }\href {https://doi.org/10.1103/PhysRevC.92.034601}
  {\bibfield  {journal} {\bibinfo  {journal} {Phys. Rev. C}\ }\textbf {\bibinfo
  {volume} {92}},\ \bibinfo {pages} {034601} (\bibinfo {year}
  {2015})}\BibitemShut {NoStop}%
\bibitem [{\citenamefont {Goddard}\ \emph {et~al.}(2016)\citenamefont
  {Goddard}, \citenamefont {Stevenson},\ and\ \citenamefont
  {Rios}}]{goddard2016}%
  \BibitemOpen
  \bibfield  {author} {\bibinfo {author} {\bibfnamefont {P.~M.}\ \bibnamefont
  {Goddard}}, \bibinfo {author} {\bibfnamefont {P.~D.}\ \bibnamefont
  {Stevenson}},\ and\ \bibinfo {author} {\bibfnamefont {A.}~\bibnamefont
  {Rios}},\ }\bibfield  {title} {\bibinfo {title} {Fission dynamics within
  time--dependent {H}artree--{F}ock. {II}. {B}oost-induced fission},\ }\href
  {https://doi.org/10.1103/PhysRevC.93.014620} {\bibfield  {journal} {\bibinfo
  {journal} {Phys. Rev. C}\ }\textbf {\bibinfo {volume} {93}},\ \bibinfo
  {pages} {014620} (\bibinfo {year} {2016})}\BibitemShut {NoStop}%
\bibitem [{\citenamefont {{Aurel Bulgac}}\ \emph {et~al.}(2016)\citenamefont
  {{Aurel Bulgac}}, \citenamefont {{Piotr Magierski}}, \citenamefont {{Kenneth
  J. Roche}},\ and\ \citenamefont {{Ionel Stetcu}}}]{bulgac2016}%
  \BibitemOpen
  \bibfield  {author} {\bibinfo {author} {\bibnamefont {{Aurel Bulgac}}},
  \bibinfo {author} {\bibnamefont {{Piotr Magierski}}}, \bibinfo {author}
  {\bibnamefont {{Kenneth J. Roche}}},\ and\ \bibinfo {author} {\bibnamefont
  {{Ionel Stetcu}}},\ }\bibfield  {title} {\bibinfo {title} {{I}nduced
  {F}ission of $^{240}${P}u within a {R}eal-{T}ime {M}icroscopic {F}ramework},\
  }\href {https://doi.org/10.1103/physrevlett.116.122504} {\bibfield  {journal}
  {\bibinfo  {journal} {Phys. Rev. Lett.}\ }\textbf {\bibinfo {volume} {116}},\
  \bibinfo {pages} {122504} (\bibinfo {year} {2016})}\BibitemShut {NoStop}%
\bibitem [{\citenamefont {Tanimura}\ \emph {et~al.}(2017)\citenamefont
  {Tanimura}, \citenamefont {Lacroix},\ and\ \citenamefont
  {Ayik}}]{tanimura2017}%
  \BibitemOpen
  \bibfield  {author} {\bibinfo {author} {\bibfnamefont {Y.}~\bibnamefont
  {Tanimura}}, \bibinfo {author} {\bibfnamefont {D.}~\bibnamefont {Lacroix}},\
  and\ \bibinfo {author} {\bibfnamefont {S.}~\bibnamefont {Ayik}},\ }\bibfield
  {title} {\bibinfo {title} {Microscopic {P}hase--{S}pace {E}xploration
  {M}odeling of $^{258}\mathrm{Fm}$ {S}pontaneous {F}ission},\ }\href
  {https://doi.org/10.1103/PhysRevLett.118.152501} {\bibfield  {journal}
  {\bibinfo  {journal} {Phys. Rev. Lett.}\ }\textbf {\bibinfo {volume} {118}},\
  \bibinfo {pages} {152501} (\bibinfo {year} {2017})}\BibitemShut {NoStop}%
\bibitem [{\citenamefont {Scamps}\ and\ \citenamefont
  {Simenel}(2018)}]{scamps2018}%
  \BibitemOpen
  \bibfield  {author} {\bibinfo {author} {\bibfnamefont {G.}~\bibnamefont
  {Scamps}}\ and\ \bibinfo {author} {\bibfnamefont {C.}~\bibnamefont
  {Simenel}},\ }\bibfield  {title} {\bibinfo {title} {Impact of pear-shaped
  fission fragments on mass-asymmetric fission in actinides},\ }\href
  {https://doi.org/10.1038/s41586-018-0780-0} {\bibfield  {journal} {\bibinfo
  {journal} {Nature}\ }\textbf {\bibinfo {volume} {564}},\ \bibinfo {pages}
  {382} (\bibinfo {year} {2018})}\BibitemShut {NoStop}%
\bibitem [{\citenamefont {{Aurel Bulgac}}\ \emph {et~al.}(2018)\citenamefont
  {{Aurel Bulgac}}, \citenamefont {{Michael McNeil Forbes}}, \citenamefont
  {{Shi Jin}}, \citenamefont {{Rodrigo Navarro Perez}},\ and\ \citenamefont
  {{Nicolas Schunck}}}]{bulgac2018}%
  \BibitemOpen
  \bibfield  {author} {\bibinfo {author} {\bibnamefont {{Aurel Bulgac}}},
  \bibinfo {author} {\bibnamefont {{Michael McNeil Forbes}}}, \bibinfo {author}
  {\bibnamefont {{Shi Jin}}}, \bibinfo {author} {\bibnamefont {{Rodrigo Navarro
  Perez}}},\ and\ \bibinfo {author} {\bibnamefont {{Nicolas Schunck}}},\
  }\bibfield  {title} {\bibinfo {title} {Minimal nuclear energy density
  functional},\ }\href {https://doi.org/10.1103/PhysRevC.97.044313} {\bibfield
  {journal} {\bibinfo  {journal} {Phys. Rev. C}\ }\textbf {\bibinfo {volume}
  {97}},\ \bibinfo {pages} {044313} (\bibinfo {year} {2018})}\BibitemShut
  {NoStop}%
\bibitem [{\citenamefont {Scamps}\ and\ \citenamefont
  {Simenel}(2019)}]{scamps2019}%
  \BibitemOpen
  \bibfield  {author} {\bibinfo {author} {\bibfnamefont {G.}~\bibnamefont
  {Scamps}}\ and\ \bibinfo {author} {\bibfnamefont {C.}~\bibnamefont
  {Simenel}},\ }\bibfield  {title} {\bibinfo {title} {Effect of shell structure
  on the fission of sub--lead nuclei},\ }\href
  {https://doi.org/10.1103/PhysRevC.100.041602} {\bibfield  {journal} {\bibinfo
   {journal} {Phys. Rev. C}\ }\textbf {\bibinfo {volume} {100}},\ \bibinfo
  {pages} {041602} (\bibinfo {year} {2019})}\BibitemShut {NoStop}%
\bibitem [{\citenamefont {Simenel}\ and\ \citenamefont
  {Umar}(2018)}]{simenel2018}%
  \BibitemOpen
  \bibfield  {author} {\bibinfo {author} {\bibfnamefont {C.}~\bibnamefont
  {Simenel}}\ and\ \bibinfo {author} {\bibfnamefont {A.~S.}\ \bibnamefont
  {Umar}},\ }\bibfield  {title} {\bibinfo {title} {Heavy-ion collisions and
  fission dynamics with the time--dependent {H}artree-{F}ock theory and its
  extensions},\ }\href {https://doi.org/10.1016/j.ppnp.2018.07.002} {\bibfield
  {journal} {\bibinfo  {journal} {Prog. Part. Nucl. Phys.}\ }\textbf {\bibinfo
  {volume} {103}},\ \bibinfo {pages} {19} (\bibinfo {year} {2018})}\BibitemShut
  {NoStop}%
\bibitem [{\citenamefont {{Kazuyuki Sekizawa}}(2019)}]{sekizawa2019}%
  \BibitemOpen
  \bibfield  {author} {\bibinfo {author} {\bibnamefont {{Kazuyuki Sekizawa}}},\
  }\bibfield  {title} {\bibinfo {title} {{TDHF} {T}heory and {I}ts {E}xtensions
  for the {M}ultinucleon {T}ransfer {R}eaction: {A} {M}ini {R}eview},\ }\href
  {https://doi.org/10.3389/fphy.2019.00020} {\bibfield  {journal} {\bibinfo
  {journal} {Front. Phys.}\ }\textbf {\bibinfo {volume} {7}},\ \bibinfo {pages}
  {20} (\bibinfo {year} {2019})}\BibitemShut {NoStop}%
\bibitem [{\citenamefont {{Yu. Ts. Oganessian}}(2018)}]{priv_oganessian}%
  \BibitemOpen
  \bibfield  {author} {\bibinfo {author} {\bibnamefont {{Yu. Ts.
  Oganessian}}},\ }\href@noop {} {}\bibinfo {howpublished} {{Private
  Communication}} (\bibinfo {year} {2018})\BibitemShut {NoStop}%
\bibitem [{\citenamefont {Negele}(1982)}]{negele1982}%
  \BibitemOpen
  \bibfield  {author} {\bibinfo {author} {\bibfnamefont {J.~W.}\ \bibnamefont
  {Negele}},\ }\bibfield  {title} {\bibinfo {title} {{T}he mean-field theory of
  nuclear-structure and dynamics},\ }\href
  {https://doi.org/10.1103/RevModPhys.54.913} {\bibfield  {journal} {\bibinfo
  {journal} {Rev. Mod. Phys.}\ }\textbf {\bibinfo {volume} {54}},\ \bibinfo
  {pages} {913} (\bibinfo {year} {1982})}\BibitemShut {NoStop}%
\bibitem [{\citenamefont {Simenel}(2012)}]{simenel2012}%
  \BibitemOpen
  \bibfield  {author} {\bibinfo {author} {\bibfnamefont {C.}~\bibnamefont
  {Simenel}},\ }\bibfield  {title} {\bibinfo {title} {{N}uclear quantum
  many-body dynamics},\ }\href {https://doi.org/10.1140/epja/i2012-12152-0}
  {\bibfield  {journal} {\bibinfo  {journal} {Eur. Phys. J. A}\ }\textbf
  {\bibinfo {volume} {48}},\ \bibinfo {pages} {152} (\bibinfo {year}
  {2012})}\BibitemShut {NoStop}%
\bibitem [{\citenamefont {Bonche}\ \emph {et~al.}(1978)\citenamefont {Bonche},
  \citenamefont {Grammaticos},\ and\ \citenamefont {Koonin}}]{bonche1978}%
  \BibitemOpen
  \bibfield  {author} {\bibinfo {author} {\bibfnamefont {P.}~\bibnamefont
  {Bonche}}, \bibinfo {author} {\bibfnamefont {B.}~\bibnamefont
  {Grammaticos}},\ and\ \bibinfo {author} {\bibfnamefont {S.}~\bibnamefont
  {Koonin}},\ }\bibfield  {title} {\bibinfo {title} {{T}hree-dimensional
  time-dependent {H}artree-{F}ock calculations of
  $^{16}\mathrm{O}+{}^{16}\mathrm{O}$ and $^{40}\mathrm{Ca}+{}^{40}\mathrm{Ca}$
  fusion cross sections},\ }\href {https://doi.org/10.1103/PhysRevC.17.1700}
  {\bibfield  {journal} {\bibinfo  {journal} {Phys. Rev. C}\ }\textbf {\bibinfo
  {volume} {17}},\ \bibinfo {pages} {1700} (\bibinfo {year}
  {1978})}\BibitemShut {NoStop}%
\bibitem [{\citenamefont {Flocard}\ \emph {et~al.}(1978)\citenamefont
  {Flocard}, \citenamefont {Koonin},\ and\ \citenamefont
  {Weiss}}]{flocard1978}%
  \BibitemOpen
  \bibfield  {author} {\bibinfo {author} {\bibfnamefont {H.}~\bibnamefont
  {Flocard}}, \bibinfo {author} {\bibfnamefont {S.~E.}\ \bibnamefont
  {Koonin}},\ and\ \bibinfo {author} {\bibfnamefont {M.~S.}\ \bibnamefont
  {Weiss}},\ }\bibfield  {title} {\bibinfo {title} {{T}hree-dimensional
  time-dependent {H}artree-{F}ock calculations: {A}pplication to
  $^{16}\mathrm{O}+{}^{16}\mathrm{O}$ collisions},\ }\href
  {https://doi.org/10.1103/PhysRevC.17.1682} {\bibfield  {journal} {\bibinfo
  {journal} {Phys. Rev. C}\ }\textbf {\bibinfo {volume} {17}},\ \bibinfo
  {pages} {1682} (\bibinfo {year} {1978})}\BibitemShut {NoStop}%
\bibitem [{\citenamefont {Simenel}\ \emph {et~al.}(2001)\citenamefont
  {Simenel}, \citenamefont {Chomaz},\ and\ \citenamefont {{de
  France}}}]{simenel2001}%
  \BibitemOpen
  \bibfield  {author} {\bibinfo {author} {\bibfnamefont {C.}~\bibnamefont
  {Simenel}}, \bibinfo {author} {\bibfnamefont {P.}~\bibnamefont {Chomaz}},\
  and\ \bibinfo {author} {\bibfnamefont {G.}~\bibnamefont {{de France}}},\
  }\bibfield  {title} {\bibinfo {title} {{Q}uantum {C}alculation of the
  {D}ipole {E}xcitation in {F}usion {R}eactions},\ }\href
  {https://doi.org/10.1103/PhysRevLett.86.2971} {\bibfield  {journal} {\bibinfo
   {journal} {Phys. Rev. Lett.}\ }\textbf {\bibinfo {volume} {86}},\ \bibinfo
  {pages} {2971} (\bibinfo {year} {2001})}\BibitemShut {NoStop}%
\bibitem [{\citenamefont {Umar}\ and\ \citenamefont
  {Oberacker}(2006{\natexlab{a}})}]{umar2006d}%
  \BibitemOpen
  \bibfield  {author} {\bibinfo {author} {\bibfnamefont {A.~S.}\ \bibnamefont
  {Umar}}\ and\ \bibinfo {author} {\bibfnamefont {V.~E.}\ \bibnamefont
  {Oberacker}},\ }\bibfield  {title} {\bibinfo {title} {{T}ime dependent
  {H}artree-{F}ock fusion calculations for spherical, deformed systems},\
  }\href {https://doi.org/10.1103/PhysRevC.74.024606} {\bibfield  {journal}
  {\bibinfo  {journal} {Phys. Rev. C}\ }\textbf {\bibinfo {volume} {74}},\
  \bibinfo {pages} {024606} (\bibinfo {year} {2006}{\natexlab{a}})}\BibitemShut
  {NoStop}%
\bibitem [{\citenamefont {{Kouhei Washiyama}}\ and\ \citenamefont {{Denis
  Lacroix}}(2008)}]{washiyama2008}%
  \BibitemOpen
  \bibfield  {author} {\bibinfo {author} {\bibnamefont {{Kouhei Washiyama}}}\
  and\ \bibinfo {author} {\bibnamefont {{Denis Lacroix}}},\ }\bibfield  {title}
  {\bibinfo {title} {{E}nergy dependence of the nucleus-nucleus potential close
  to the {C}oulomb barrier},\ }\href
  {https://doi.org/10.1103/PhysRevC.78.024610} {\bibfield  {journal} {\bibinfo
  {journal} {Phys. Rev. C}\ }\textbf {\bibinfo {volume} {78}},\ \bibinfo
  {pages} {024610} (\bibinfo {year} {2008})}\BibitemShut {NoStop}%
\bibitem [{\citenamefont {Umar}\ \emph {et~al.}(2010)\citenamefont {Umar},
  \citenamefont {Oberacker}, \citenamefont {Maruhn},\ and\ \citenamefont
  {Reinhard}}]{umar2010a}%
  \BibitemOpen
  \bibfield  {author} {\bibinfo {author} {\bibfnamefont {A.~S.}\ \bibnamefont
  {Umar}}, \bibinfo {author} {\bibfnamefont {V.~E.}\ \bibnamefont {Oberacker}},
  \bibinfo {author} {\bibfnamefont {J.~A.}\ \bibnamefont {Maruhn}},\ and\
  \bibinfo {author} {\bibfnamefont {P.-G.}\ \bibnamefont {Reinhard}},\
  }\bibfield  {title} {\bibinfo {title} {{E}ntrance channel dynamics of hot and
  cold fusion reactions leading to superheavy elements},\ }\href
  {https://doi.org/10.1103/PhysRevC.81.064607} {\bibfield  {journal} {\bibinfo
  {journal} {Phys. Rev. C}\ }\textbf {\bibinfo {volume} {81}},\ \bibinfo
  {pages} {064607} (\bibinfo {year} {2010})}\BibitemShut {NoStop}%
\bibitem [{\citenamefont {Umar}\ \emph {et~al.}(2009)\citenamefont {Umar},
  \citenamefont {Oberacker}, \citenamefont {Maruhn},\ and\ \citenamefont
  {Reinhard}}]{umar2009a}%
  \BibitemOpen
  \bibfield  {author} {\bibinfo {author} {\bibfnamefont {A.~S.}\ \bibnamefont
  {Umar}}, \bibinfo {author} {\bibfnamefont {V.~E.}\ \bibnamefont {Oberacker}},
  \bibinfo {author} {\bibfnamefont {J.~A.}\ \bibnamefont {Maruhn}},\ and\
  \bibinfo {author} {\bibfnamefont {P.-G.}\ \bibnamefont {Reinhard}},\
  }\bibfield  {title} {\bibinfo {title} {{M}icroscopic calculation of
  precompound excitation energies for heavy-ion collisions},\ }\href
  {https://doi.org/10.1103/PhysRevC.80.041601} {\bibfield  {journal} {\bibinfo
  {journal} {Phys. Rev. C}\ }\textbf {\bibinfo {volume} {80}},\ \bibinfo
  {pages} {041601} (\bibinfo {year} {2009})}\BibitemShut {NoStop}%
\bibitem [{\citenamefont {{Lu Guo}}\ and\ \citenamefont {{Takashi
  Nakatsukasa}}(2012)}]{guo2012}%
  \BibitemOpen
  \bibfield  {author} {\bibinfo {author} {\bibnamefont {{Lu Guo}}}\ and\
  \bibinfo {author} {\bibnamefont {{Takashi Nakatsukasa}}},\ }\bibfield
  {title} {\bibinfo {title} {{T}ime-dependent {H}artree-{F}ock studies of the
  dynamical fusion threshold},\ }\href
  {https://doi.org/10.1051/epjconf/20123809003} {\bibfield  {journal} {\bibinfo
   {journal} {{EPJ} {W}eb {C}onf.}\ }\textbf {\bibinfo {volume} {38}},\
  \bibinfo {pages} {09003} (\bibinfo {year} {2012})}\BibitemShut {NoStop}%
\bibitem [{\citenamefont {Keser}\ \emph {et~al.}(2012)\citenamefont {Keser},
  \citenamefont {Umar},\ and\ \citenamefont {Oberacker}}]{keser2012}%
  \BibitemOpen
  \bibfield  {author} {\bibinfo {author} {\bibfnamefont {R.}~\bibnamefont
  {Keser}}, \bibinfo {author} {\bibfnamefont {A.~S.}\ \bibnamefont {Umar}},\
  and\ \bibinfo {author} {\bibfnamefont {V.~E.}\ \bibnamefont {Oberacker}},\
  }\bibfield  {title} {\bibinfo {title} {{M}icroscopic study of {C}a$+${C}a
  fusion},\ }\href {https://doi.org/10.1103/PhysRevC.85.044606} {\bibfield
  {journal} {\bibinfo  {journal} {Phys. Rev. C}\ }\textbf {\bibinfo {volume}
  {85}},\ \bibinfo {pages} {044606} (\bibinfo {year} {2012})}\BibitemShut
  {NoStop}%
\bibitem [{\citenamefont {Simenel}\ \emph
  {et~al.}(2013{\natexlab{a}})\citenamefont {Simenel}, \citenamefont {Keser},
  \citenamefont {Umar},\ and\ \citenamefont {Oberacker}}]{simenel2013a}%
  \BibitemOpen
  \bibfield  {author} {\bibinfo {author} {\bibfnamefont {C.}~\bibnamefont
  {Simenel}}, \bibinfo {author} {\bibfnamefont {R.}~\bibnamefont {Keser}},
  \bibinfo {author} {\bibfnamefont {A.~S.}\ \bibnamefont {Umar}},\ and\
  \bibinfo {author} {\bibfnamefont {V.~E.}\ \bibnamefont {Oberacker}},\
  }\bibfield  {title} {\bibinfo {title} {Microscopic study of
  ${}^{16}\mathrm{O}+{}^{16}\mathrm{O}$ fusion},\ }\href
  {https://doi.org/10.1103/PhysRevC.88.024617} {\bibfield  {journal} {\bibinfo
  {journal} {Phys. Rev. C}\ }\textbf {\bibinfo {volume} {88}},\ \bibinfo
  {pages} {024617} (\bibinfo {year} {2013}{\natexlab{a}})}\BibitemShut
  {NoStop}%
\bibitem [{\citenamefont {Oberacker}\ \emph {et~al.}(2012)\citenamefont
  {Oberacker}, \citenamefont {Umar}, \citenamefont {Maruhn},\ and\
  \citenamefont {Reinhard}}]{oberacker2012}%
  \BibitemOpen
  \bibfield  {author} {\bibinfo {author} {\bibfnamefont {V.~E.}\ \bibnamefont
  {Oberacker}}, \bibinfo {author} {\bibfnamefont {A.~S.}\ \bibnamefont {Umar}},
  \bibinfo {author} {\bibfnamefont {J.~A.}\ \bibnamefont {Maruhn}},\ and\
  \bibinfo {author} {\bibfnamefont {P.-G.}\ \bibnamefont {Reinhard}},\
  }\bibfield  {title} {\bibinfo {title} {{D}ynamic microscopic study of
  pre-equilibrium giant resonance excitation and fusion in the reactions
  ${}^{132}${S}n $+$ ${}^{48}${C}a and ${}^{124}${S}n $+$ ${}^{40}${C}a},\
  }\href {https://doi.org/10.1103/PhysRevC.85.034609} {\bibfield  {journal}
  {\bibinfo  {journal} {Phys. Rev. C}\ }\textbf {\bibinfo {volume} {85}},\
  \bibinfo {pages} {034609} (\bibinfo {year} {2012})}\BibitemShut {NoStop}%
\bibitem [{\citenamefont {Oberacker}\ \emph {et~al.}(2010)\citenamefont
  {Oberacker}, \citenamefont {Umar}, \citenamefont {Maruhn},\ and\
  \citenamefont {Reinhard}}]{oberacker2010}%
  \BibitemOpen
  \bibfield  {author} {\bibinfo {author} {\bibfnamefont {V.~E.}\ \bibnamefont
  {Oberacker}}, \bibinfo {author} {\bibfnamefont {A.~S.}\ \bibnamefont {Umar}},
  \bibinfo {author} {\bibfnamefont {J.~A.}\ \bibnamefont {Maruhn}},\ and\
  \bibinfo {author} {\bibfnamefont {P.}~\bibnamefont {Reinhard}},\ }\bibfield
  {title} {\bibinfo {title} {{M}icroscopic study of the
  $^{132,124}\mathrm{Sn}+{}^{96}\mathrm{Zr}$ reactions: {D}ynamic excitation
  energy, energy-dependent heavy-ion potential, and capture cross section},\
  }\href {https://doi.org/10.1103/PhysRevC.82.034603} {\bibfield  {journal}
  {\bibinfo  {journal} {Phys. Rev. C}\ }\textbf {\bibinfo {volume} {82}},\
  \bibinfo {pages} {034603} (\bibinfo {year} {2010})}\BibitemShut {NoStop}%
\bibitem [{\citenamefont {Umar}\ \emph {et~al.}(2012)\citenamefont {Umar},
  \citenamefont {Oberacker},\ and\ \citenamefont {Horowitz}}]{umar2012a}%
  \BibitemOpen
  \bibfield  {author} {\bibinfo {author} {\bibfnamefont {A.~S.}\ \bibnamefont
  {Umar}}, \bibinfo {author} {\bibfnamefont {V.~E.}\ \bibnamefont
  {Oberacker}},\ and\ \bibinfo {author} {\bibfnamefont {C.~J.}\ \bibnamefont
  {Horowitz}},\ }\bibfield  {title} {\bibinfo {title} {{M}icroscopic
  sub-barrier fusion calculations for the neutron star crust},\ }\href
  {https://doi.org/10.1103/PhysRevC.85.055801} {\bibfield  {journal} {\bibinfo
  {journal} {Phys. Rev. C}\ }\textbf {\bibinfo {volume} {85}},\ \bibinfo
  {pages} {055801} (\bibinfo {year} {2012})}\BibitemShut {NoStop}%
\bibitem [{\citenamefont {Simenel}\ \emph
  {et~al.}(2013{\natexlab{b}})\citenamefont {Simenel}, \citenamefont
  {Dasgupta}, \citenamefont {Hinde},\ and\ \citenamefont
  {Williams}}]{simenel2013b}%
  \BibitemOpen
  \bibfield  {author} {\bibinfo {author} {\bibfnamefont {C.}~\bibnamefont
  {Simenel}}, \bibinfo {author} {\bibfnamefont {M.}~\bibnamefont {Dasgupta}},
  \bibinfo {author} {\bibfnamefont {D.~J.}\ \bibnamefont {Hinde}},\ and\
  \bibinfo {author} {\bibfnamefont {E.}~\bibnamefont {Williams}},\ }\bibfield
  {title} {\bibinfo {title} {{M}icroscopic approach to coupled-channels effects
  on fusion},\ }\href {https://doi.org/10.1103/PhysRevC.88.064604} {\bibfield
  {journal} {\bibinfo  {journal} {Phys. Rev. C}\ }\textbf {\bibinfo {volume}
  {88}},\ \bibinfo {pages} {064604} (\bibinfo {year}
  {2013}{\natexlab{b}})}\BibitemShut {NoStop}%
\bibitem [{\citenamefont {Umar}\ \emph {et~al.}(2014)\citenamefont {Umar},
  \citenamefont {Simenel},\ and\ \citenamefont {Oberacker}}]{umar2014a}%
  \BibitemOpen
  \bibfield  {author} {\bibinfo {author} {\bibfnamefont {A.~S.}\ \bibnamefont
  {Umar}}, \bibinfo {author} {\bibfnamefont {C.}~\bibnamefont {Simenel}},\ and\
  \bibinfo {author} {\bibfnamefont {V.~E.}\ \bibnamefont {Oberacker}},\
  }\bibfield  {title} {\bibinfo {title} {{E}nergy dependence of potential
  barriers and its effect on fusion cross sections},\ }\href
  {https://doi.org/10.1103/PhysRevC.89.034611} {\bibfield  {journal} {\bibinfo
  {journal} {Phys. Rev. C}\ }\textbf {\bibinfo {volume} {89}},\ \bibinfo
  {pages} {034611} (\bibinfo {year} {2014})}\BibitemShut {NoStop}%
\bibitem [{\citenamefont {Jiang}\ \emph {et~al.}(2014)\citenamefont {Jiang},
  \citenamefont {Maruhn},\ and\ \citenamefont {Yan}}]{jiang2014}%
  \BibitemOpen
  \bibfield  {author} {\bibinfo {author} {\bibfnamefont {X.}~\bibnamefont
  {Jiang}}, \bibinfo {author} {\bibfnamefont {J.~A.}\ \bibnamefont {Maruhn}},\
  and\ \bibinfo {author} {\bibfnamefont {S.}~\bibnamefont {Yan}},\ }\bibfield
  {title} {\bibinfo {title} {{M}icroscopic study of noncentral effects in
  heavy-ion fusion reactions with spherical nuclei},\ }\href
  {https://doi.org/10.1103/PhysRevC.90.064618} {\bibfield  {journal} {\bibinfo
  {journal} {Phys. Rev. C}\ }\textbf {\bibinfo {volume} {90}},\ \bibinfo
  {pages} {064618} (\bibinfo {year} {2014})}\BibitemShut {NoStop}%
\bibitem [{\citenamefont {Koonin}\ \emph {et~al.}(1977)\citenamefont {Koonin},
  \citenamefont {Davies}, \citenamefont {Maruhn-Rezwani}, \citenamefont
  {Feldmeier}, \citenamefont {Krieger},\ and\ \citenamefont
  {Negele}}]{koonin1977}%
  \BibitemOpen
  \bibfield  {author} {\bibinfo {author} {\bibfnamefont {S.~E.}\ \bibnamefont
  {Koonin}}, \bibinfo {author} {\bibfnamefont {K.~T.~R.}\ \bibnamefont
  {Davies}}, \bibinfo {author} {\bibfnamefont {V.}~\bibnamefont
  {Maruhn-Rezwani}}, \bibinfo {author} {\bibfnamefont {H.}~\bibnamefont
  {Feldmeier}}, \bibinfo {author} {\bibfnamefont {S.~J.}\ \bibnamefont
  {Krieger}},\ and\ \bibinfo {author} {\bibfnamefont {J.~W.}\ \bibnamefont
  {Negele}},\ }\bibfield  {title} {\bibinfo {title} {{T}ime-dependent
  {H}artree-{F}ock calculations for $^{16}${O} $+$ $^{16}${O} and $^{40}${C}a
  $+$ $^{40}${C}a reactions},\ }\href
  {https://doi.org/10.1103/PhysRevC.15.1359} {\bibfield  {journal} {\bibinfo
  {journal} {Phys. Rev. C}\ }\textbf {\bibinfo {volume} {15}},\ \bibinfo
  {pages} {1359} (\bibinfo {year} {1977})}\BibitemShut {NoStop}%
\bibitem [{\citenamefont {Simenel}(2010)}]{simenel2010}%
  \BibitemOpen
  \bibfield  {author} {\bibinfo {author} {\bibfnamefont {C.}~\bibnamefont
  {Simenel}},\ }\bibfield  {title} {\bibinfo {title} {{P}article {T}ransfer
  {R}eactions with the {T}ime-{D}ependent {H}artree-{F}ock {T}heory {U}sing a
  {P}article {N}umber {P}rojection {T}echnique},\ }\href
  {https://doi.org/10.1103/PhysRevLett.105.192701} {\bibfield  {journal}
  {\bibinfo  {journal} {Phys. Rev. Lett.}\ }\textbf {\bibinfo {volume} {105}},\
  \bibinfo {pages} {192701} (\bibinfo {year} {2010})}\BibitemShut {NoStop}%
\bibitem [{\citenamefont {Simenel}(2011)}]{simenel2011}%
  \BibitemOpen
  \bibfield  {author} {\bibinfo {author} {\bibfnamefont {C.}~\bibnamefont
  {Simenel}},\ }\bibfield  {title} {\bibinfo {title} {{P}article-{N}umber
  {F}luctuations and {C}orrelations in {T}ransfer {R}eactions {O}btained
  {U}sing the {B}alian-{V}\'en\'eroni {V}ariational {P}rinciple},\ }\href
  {https://doi.org/10.1103/PhysRevLett.106.112502} {\bibfield  {journal}
  {\bibinfo  {journal} {Phys. Rev. Lett.}\ }\textbf {\bibinfo {volume} {106}},\
  \bibinfo {pages} {112502} (\bibinfo {year} {2011})}\BibitemShut {NoStop}%
\bibitem [{\citenamefont {{Kazuyuki Sekizawa}}\ and\ \citenamefont {{Kazuhiro
  Yabana}}(2013)}]{sekizawa2013}%
  \BibitemOpen
  \bibfield  {author} {\bibinfo {author} {\bibnamefont {{Kazuyuki Sekizawa}}}\
  and\ \bibinfo {author} {\bibnamefont {{Kazuhiro Yabana}}},\ }\bibfield
  {title} {\bibinfo {title} {{T}ime-dependent {H}artree-{F}ock calculations for
  multinucleon transfer processes in $^{40,48}\mathrm{Ca}+{}^{124}\mathrm{Sn}$,
  $^{40}\mathrm{Ca}+{}^{208}\mathrm{Pb}$, and
  $^{58}\mathrm{Ni}+{}^{208}\mathrm{Pb}$ reactions},\ }\href
  {https://doi.org/10.1103/PhysRevC.88.014614} {\bibfield  {journal} {\bibinfo
  {journal} {Phys. Rev. C}\ }\textbf {\bibinfo {volume} {88}},\ \bibinfo
  {pages} {014614} (\bibinfo {year} {2013})}\BibitemShut {NoStop}%
\bibitem [{\citenamefont {Scamps}\ and\ \citenamefont
  {Lacroix}(2013)}]{scamps2013a}%
  \BibitemOpen
  \bibfield  {author} {\bibinfo {author} {\bibfnamefont {G.}~\bibnamefont
  {Scamps}}\ and\ \bibinfo {author} {\bibfnamefont {D.}~\bibnamefont
  {Lacroix}},\ }\bibfield  {title} {\bibinfo {title} {{E}ffect of pairing on
  one- and two-nucleon transfer below the {C}oulomb barrier: {A} time-dependent
  microscopic description},\ }\href
  {https://doi.org/10.1103/PhysRevC.87.014605} {\bibfield  {journal} {\bibinfo
  {journal} {Phys. Rev. C}\ }\textbf {\bibinfo {volume} {87}},\ \bibinfo
  {pages} {014605} (\bibinfo {year} {2013})}\BibitemShut {NoStop}%
\bibitem [{\citenamefont {Sekizawa}\ and\ \citenamefont
  {Yabana}(2014)}]{sekizawa2014}%
  \BibitemOpen
  \bibfield  {author} {\bibinfo {author} {\bibfnamefont {K.}~\bibnamefont
  {Sekizawa}}\ and\ \bibinfo {author} {\bibfnamefont {K.}~\bibnamefont
  {Yabana}},\ }\bibfield  {title} {\bibinfo {title} {{P}article-number
  projection method in time-dependent {H}artree-{F}ock theory: {P}roperties of
  reaction products},\ }\href {https://doi.org/10.1103/PhysRevC.90.064614}
  {\bibfield  {journal} {\bibinfo  {journal} {Phys. Rev. C}\ }\textbf {\bibinfo
  {volume} {90}},\ \bibinfo {pages} {064614} (\bibinfo {year}
  {2014})}\BibitemShut {NoStop}%
\bibitem [{\citenamefont {Bourgin}\ \emph {et~al.}(2016)\citenamefont
  {Bourgin}, \citenamefont {Simenel}, \citenamefont {Courtin},\ and\
  \citenamefont {Haas}}]{bourgin2016}%
  \BibitemOpen
  \bibfield  {author} {\bibinfo {author} {\bibfnamefont {D.}~\bibnamefont
  {Bourgin}}, \bibinfo {author} {\bibfnamefont {C.}~\bibnamefont {Simenel}},
  \bibinfo {author} {\bibfnamefont {S.}~\bibnamefont {Courtin}},\ and\ \bibinfo
  {author} {\bibfnamefont {F.}~\bibnamefont {Haas}},\ }\bibfield  {title}
  {\bibinfo {title} {{M}icroscopic study of $^{40}${C}a $+$ $^{58,64}${N}i
  fusion reactions},\ }\href {https://doi.org/10.1103/PhysRevC.93.034604}
  {\bibfield  {journal} {\bibinfo  {journal} {Phys. Rev. C}\ }\textbf {\bibinfo
  {volume} {93}},\ \bibinfo {pages} {034604} (\bibinfo {year}
  {2016})}\BibitemShut {NoStop}%
\bibitem [{\citenamefont {Bottcher}\ \emph {et~al.}(1989)\citenamefont
  {Bottcher}, \citenamefont {Strayer}, \citenamefont {Umar},\ and\
  \citenamefont {Reinhard}}]{bottcher1989}%
  \BibitemOpen
  \bibfield  {author} {\bibinfo {author} {\bibfnamefont {C.}~\bibnamefont
  {Bottcher}}, \bibinfo {author} {\bibfnamefont {M.~R.}\ \bibnamefont
  {Strayer}}, \bibinfo {author} {\bibfnamefont {A.~S.}\ \bibnamefont {Umar}},\
  and\ \bibinfo {author} {\bibfnamefont {P.-G.}\ \bibnamefont {Reinhard}},\
  }\bibfield  {title} {\bibinfo {title} {{D}amped relaxation techniques to
  calculate relativistic bound-states},\ }\href
  {https://doi.org/10.1103/PhysRevA.40.4182} {\bibfield  {journal} {\bibinfo
  {journal} {Phys. Rev. A}\ }\textbf {\bibinfo {volume} {40}},\ \bibinfo
  {pages} {4182} (\bibinfo {year} {1989})}\BibitemShut {NoStop}%
\bibitem [{\citenamefont {Umar}\ and\ \citenamefont
  {Oberacker}(2006{\natexlab{b}})}]{umar2006c}%
  \BibitemOpen
  \bibfield  {author} {\bibinfo {author} {\bibfnamefont {A.~S.}\ \bibnamefont
  {Umar}}\ and\ \bibinfo {author} {\bibfnamefont {V.~E.}\ \bibnamefont
  {Oberacker}},\ }\bibfield  {title} {\bibinfo {title} {{T}hree-dimensional
  unrestricted time-dependent {H}artree-{F}ock fusion calculations using the
  full {S}kyrme interaction},\ }\href
  {https://doi.org/10.1103/PhysRevC.73.054607} {\bibfield  {journal} {\bibinfo
  {journal} {Phys. Rev. C}\ }\textbf {\bibinfo {volume} {73}},\ \bibinfo
  {pages} {054607} (\bibinfo {year} {2006}{\natexlab{b}})}\BibitemShut
  {NoStop}%
\bibitem [{\citenamefont {Maruhn}\ \emph {et~al.}(2014)\citenamefont {Maruhn},
  \citenamefont {Reinhard}, \citenamefont {Stevenson},\ and\ \citenamefont
  {Umar}}]{maruhn2014}%
  \BibitemOpen
  \bibfield  {author} {\bibinfo {author} {\bibfnamefont {J.~A.}\ \bibnamefont
  {Maruhn}}, \bibinfo {author} {\bibfnamefont {P.-G.}\ \bibnamefont
  {Reinhard}}, \bibinfo {author} {\bibfnamefont {P.~D.}\ \bibnamefont
  {Stevenson}},\ and\ \bibinfo {author} {\bibfnamefont {A.~S.}\ \bibnamefont
  {Umar}},\ }\bibfield  {title} {\bibinfo {title} {{T}he {TDHF C}ode
  {S}ky3{D}},\ }\href {https://doi.org/10.1016/j.cpc.2014.04.008} {\bibfield
  {journal} {\bibinfo  {journal} {Comput. Phys. Commun.}\ }\textbf {\bibinfo
  {volume} {185}},\ \bibinfo {pages} {2195} (\bibinfo {year}
  {2014})}\BibitemShut {NoStop}%
\bibitem [{\citenamefont {{Roger Balian}}\ and\ \citenamefont {{Marcel
  V\'en\'eroni}}(1981)}]{balian1981}%
  \BibitemOpen
  \bibfield  {author} {\bibinfo {author} {\bibnamefont {{Roger Balian}}}\ and\
  \bibinfo {author} {\bibnamefont {{Marcel V\'en\'eroni}}},\ }\bibfield
  {title} {\bibinfo {title} {{T}ime-{D}ependent {V}ariational {P}rinciple for
  {P}redicting the {E}xpectation {V}alue of an {O}bservable},\ }\href
  {https://doi.org/10.1103/PhysRevLett.47.1353} {\bibfield  {journal} {\bibinfo
   {journal} {Phys. Rev. Lett.}\ }\textbf {\bibinfo {volume} {47}},\ \bibinfo
  {pages} {1353} (\bibinfo {year} {1981})}\BibitemShut {NoStop}%
\bibitem [{\citenamefont {Dasso}\ \emph {et~al.}(1979)\citenamefont {Dasso},
  \citenamefont {Dossing},\ and\ \citenamefont {Pauli}}]{dasso1979}%
  \BibitemOpen
  \bibfield  {author} {\bibinfo {author} {\bibfnamefont {C.~H.}\ \bibnamefont
  {Dasso}}, \bibinfo {author} {\bibfnamefont {T.}~\bibnamefont {Dossing}},\
  and\ \bibinfo {author} {\bibfnamefont {H.~C.}\ \bibnamefont {Pauli}},\
  }\bibfield  {title} {\bibinfo {title} {{O}n the mass distribution in
  {T}ime-{D}ependent {H}artree-{F}ock calculations of heavy-ion collisions},\
  }\href {https://doi.org/10.1007/BF01409391} {\bibfield  {journal} {\bibinfo
  {journal} {Z. Phys. A}\ }\textbf {\bibinfo {volume} {289}},\ \bibinfo {pages}
  {395} (\bibinfo {year} {1979})}\BibitemShut {NoStop}%
\bibitem [{\citenamefont {Ayik}(2008)}]{ayik2008}%
  \BibitemOpen
  \bibfield  {author} {\bibinfo {author} {\bibfnamefont {S.}~\bibnamefont
  {Ayik}},\ }\bibfield  {title} {\bibinfo {title} {A stochastic mean-field
  approach for nuclear dynamics},\ }\href
  {https://doi.org/10.1016/j.physletb.2007.09.072} {\bibfield  {journal}
  {\bibinfo  {journal} {Phys. Lett. B}\ }\textbf {\bibinfo {volume} {658}},\
  \bibinfo {pages} {174} (\bibinfo {year} {2008})}\BibitemShut {NoStop}%
\bibitem [{\citenamefont {Lacroix}\ and\ \citenamefont
  {Ayik}(2014)}]{lacroix2014}%
  \BibitemOpen
  \bibfield  {author} {\bibinfo {author} {\bibfnamefont {D.}~\bibnamefont
  {Lacroix}}\ and\ \bibinfo {author} {\bibfnamefont {S.}~\bibnamefont {Ayik}},\
  }\bibfield  {title} {\bibinfo {title} {{S}tochastic quantum dynamics beyond
  mean field},\ }\href {https://doi.org/10.1140/epja/i2014-14095-8} {\bibfield
  {journal} {\bibinfo  {journal} {Eur. Phys. J. A}\ }\textbf {\bibinfo {volume}
  {50}},\ \bibinfo {pages} {95} (\bibinfo {year} {2014})}\BibitemShut {NoStop}%
\bibitem [{\citenamefont {{Roger Balian}}\ and\ \citenamefont {{Marcel
  V\'en\'eroni}}(1984)}]{balian1984}%
  \BibitemOpen
  \bibfield  {author} {\bibinfo {author} {\bibnamefont {{Roger Balian}}}\ and\
  \bibinfo {author} {\bibnamefont {{Marcel V\'en\'eroni}}},\ }\bibfield
  {title} {\bibinfo {title} {{F}luctuations in a time-dependent mean-field
  approach},\ }\href {https://doi.org/10.1016/0370-2693(84)92008-2} {\bibfield
  {journal} {\bibinfo  {journal} {Phys. Lett. B}\ }\textbf {\bibinfo {volume}
  {136}},\ \bibinfo {pages} {301} (\bibinfo {year} {1984})}\BibitemShut
  {NoStop}%
\bibitem [{\citenamefont {Broomfield}(2009)}]{broomfield2009}%
  \BibitemOpen
  \bibfield  {author} {\bibinfo {author} {\bibfnamefont {J.~M.~A.}\
  \bibnamefont {Broomfield}},\ }\emph {\bibinfo {title} {Calculations of {M}ass
  {D}istributions using the {B}alian-{V}\'en\'eroni {V}ariational
  {A}pproach}},\ \href@noop {} {Ph.D. thesis},\ \bibinfo  {school} {University
  of Surrey}, \bibinfo {address} {Guildford, United Kingdom} (\bibinfo {year}
  {2009})\BibitemShut {NoStop}%
\bibitem [{\citenamefont {Williams}\ \emph {et~al.}(2018)\citenamefont
  {Williams}, \citenamefont {Sekizawa}, \citenamefont {Hinde}, \citenamefont
  {Simenel}, \citenamefont {Dasgupta}, \citenamefont {Carter}, \citenamefont
  {Cook}, \citenamefont {Jeung}, \citenamefont {McNeil}, \citenamefont
  {Palshetkar}, \citenamefont {Rafferty}, \citenamefont {Ramachandran},\ and\
  \citenamefont {Wakhle}}]{williams2018}%
  \BibitemOpen
  \bibfield  {author} {\bibinfo {author} {\bibfnamefont {E.}~\bibnamefont
  {Williams}}, \bibinfo {author} {\bibfnamefont {K.}~\bibnamefont {Sekizawa}},
  \bibinfo {author} {\bibfnamefont {D.~J.}\ \bibnamefont {Hinde}}, \bibinfo
  {author} {\bibfnamefont {C.}~\bibnamefont {Simenel}}, \bibinfo {author}
  {\bibfnamefont {M.}~\bibnamefont {Dasgupta}}, \bibinfo {author}
  {\bibfnamefont {I.~P.}\ \bibnamefont {Carter}}, \bibinfo {author}
  {\bibfnamefont {K.~J.}\ \bibnamefont {Cook}}, \bibinfo {author}
  {\bibfnamefont {D.~Y.}\ \bibnamefont {Jeung}}, \bibinfo {author}
  {\bibfnamefont {S.~D.}\ \bibnamefont {McNeil}}, \bibinfo {author}
  {\bibfnamefont {C.~S.}\ \bibnamefont {Palshetkar}}, \bibinfo {author}
  {\bibfnamefont {D.~C.}\ \bibnamefont {Rafferty}}, \bibinfo {author}
  {\bibfnamefont {K.}~\bibnamefont {Ramachandran}},\ and\ \bibinfo {author}
  {\bibfnamefont {A.}~\bibnamefont {Wakhle}},\ }\bibfield  {title} {\bibinfo
  {title} {Exploring {Z}eptosecond {Q}uantum {E}quilibration {D}ynamics: {F}rom
  {D}eep-{I}nelastic to {F}usion-{F}ission {O}utcomes in
  $^{58}\mathrm{Ni}+{}^{60}\mathrm{Ni}$ {R}eactions},\ }\href
  {https://doi.org/10.1103/PhysRevLett.120.022501} {\bibfield  {journal}
  {\bibinfo  {journal} {Phys. Rev. Lett.}\ }\textbf {\bibinfo {volume} {120}},\
  \bibinfo {pages} {022501} (\bibinfo {year} {2018})}\BibitemShut {NoStop}%
\bibitem [{\citenamefont {Bonche}\ and\ \citenamefont
  {Flocard}(1985)}]{bonche1985}%
  \BibitemOpen
  \bibfield  {author} {\bibinfo {author} {\bibfnamefont {P.}~\bibnamefont
  {Bonche}}\ and\ \bibinfo {author} {\bibfnamefont {H.}~\bibnamefont
  {Flocard}},\ }\bibfield  {title} {\bibinfo {title} {{D}ispersion of one-body
  operators with the {B}alian-{V}eneroni variational principle},\ }\href
  {https://doi.org/10.1016/0375-9474(85)90232-5} {\bibfield  {journal}
  {\bibinfo  {journal} {Nucl. Phys. A}\ }\textbf {\bibinfo {volume} {437}},\
  \bibinfo {pages} {189} (\bibinfo {year} {1985})}\BibitemShut {NoStop}%
\bibitem [{\citenamefont {{Ka--Hae Kim}}\ \emph {et~al.}(1997)\citenamefont
  {{Ka--Hae Kim}}, \citenamefont {{Takaharu Otsuka}},\ and\ \citenamefont
  {{Paul Bonche}}}]{kim1997}%
  \BibitemOpen
  \bibfield  {author} {\bibinfo {author} {\bibnamefont {{Ka--Hae Kim}}},
  \bibinfo {author} {\bibnamefont {{Takaharu Otsuka}}},\ and\ \bibinfo {author}
  {\bibnamefont {{Paul Bonche}}},\ }\bibfield  {title} {\bibinfo {title}
  {{T}hree-dimensional {TDHF} calculations for reactions of unstable nuclei},\
  }\href {https://doi.org/10.1088/0954-3899/23/10/014} {\bibfield  {journal}
  {\bibinfo  {journal} {J. Phys. G}\ }\textbf {\bibinfo {volume} {23}},\
  \bibinfo {pages} {1267} (\bibinfo {year} {1997})}\BibitemShut {NoStop}%
\bibitem [{\citenamefont {Umar}\ \emph {et~al.}(1991)\citenamefont {Umar},
  \citenamefont {Strayer}, \citenamefont {Wu}, \citenamefont {Dean},\ and\
  \citenamefont {G\"u\c{c}l\"u}}]{umar1991a}%
  \BibitemOpen
  \bibfield  {author} {\bibinfo {author} {\bibfnamefont {A.~S.}\ \bibnamefont
  {Umar}}, \bibinfo {author} {\bibfnamefont {M.~R.}\ \bibnamefont {Strayer}},
  \bibinfo {author} {\bibfnamefont {J.~S.}\ \bibnamefont {Wu}}, \bibinfo
  {author} {\bibfnamefont {D.~J.}\ \bibnamefont {Dean}},\ and\ \bibinfo
  {author} {\bibfnamefont {M.~C.}\ \bibnamefont {G\"u\c{c}l\"u}},\ }\bibfield
  {title} {\bibinfo {title} {{N}uclear {H}artree-{F}ock calculations with
  splines},\ }\href {https://doi.org/10.1103/PhysRevC.44.2512} {\bibfield
  {journal} {\bibinfo  {journal} {Phys. Rev. C}\ }\textbf {\bibinfo {volume}
  {44}},\ \bibinfo {pages} {2512} (\bibinfo {year} {1991})}\BibitemShut
  {NoStop}%
\bibitem [{\citenamefont {Fu}\ \emph {et~al.}(2018)\citenamefont {Fu},
  \citenamefont {Tong}, \citenamefont {Wang}, \citenamefont {Wang},
  \citenamefont {Wang}, \citenamefont {Wang},\ and\ \citenamefont
  {Yao}}]{fu2018}%
  \BibitemOpen
  \bibfield  {author} {\bibinfo {author} {\bibfnamefont {Y.}~\bibnamefont
  {Fu}}, \bibinfo {author} {\bibfnamefont {H.}~\bibnamefont {Tong}}, \bibinfo
  {author} {\bibfnamefont {X.~F.}\ \bibnamefont {Wang}}, \bibinfo {author}
  {\bibfnamefont {H.}~\bibnamefont {Wang}}, \bibinfo {author} {\bibfnamefont
  {D.~Q.}\ \bibnamefont {Wang}}, \bibinfo {author} {\bibfnamefont {X.~Y.}\
  \bibnamefont {Wang}},\ and\ \bibinfo {author} {\bibfnamefont {J.~M.}\
  \bibnamefont {Yao}},\ }\bibfield  {title} {\bibinfo {title} {Microscopic
  analysis of shape transition in neutron-deficient {Y}b isotopes},\ }\href
  {https://doi.org/10.1103/PhysRevC.97.014311} {\bibfield  {journal} {\bibinfo
  {journal} {Phys. Rev. C}\ }\textbf {\bibinfo {volume} {97}},\ \bibinfo
  {pages} {014311} (\bibinfo {year} {2018})}\BibitemShut {NoStop}%
\bibitem [{\citenamefont {Nomura}\ \emph {et~al.}(2011)\citenamefont {Nomura},
  \citenamefont {Otsuka}, \citenamefont {Rodr\'{\i}guez-Guzm\'an},
  \citenamefont {Robledo},\ and\ \citenamefont {Sarriguren}}]{nomura2011}%
  \BibitemOpen
  \bibfield  {author} {\bibinfo {author} {\bibfnamefont {K.}~\bibnamefont
  {Nomura}}, \bibinfo {author} {\bibfnamefont {T.}~\bibnamefont {Otsuka}},
  \bibinfo {author} {\bibfnamefont {R.}~\bibnamefont
  {Rodr\'{\i}guez-Guzm\'an}}, \bibinfo {author} {\bibfnamefont {L.~M.}\
  \bibnamefont {Robledo}},\ and\ \bibinfo {author} {\bibfnamefont
  {P.}~\bibnamefont {Sarriguren}},\ }\bibfield  {title} {\bibinfo {title}
  {Collective structural evolution in neutron-rich {Y}b, {H}f, {W}, {O}s, and
  {P}t isotopes},\ }\href {https://doi.org/10.1103/PhysRevC.84.054316}
  {\bibfield  {journal} {\bibinfo  {journal} {Phys. Rev. C}\ }\textbf {\bibinfo
  {volume} {84}},\ \bibinfo {pages} {054316} (\bibinfo {year}
  {2011})}\BibitemShut {NoStop}%
\bibitem [{\citenamefont {Robledo}\ \emph {et~al.}(2009)\citenamefont
  {Robledo}, \citenamefont {Rodr{\'{\i}}guez-Guzm{\'{a}}n},\ and\ \citenamefont
  {Sarriguren}}]{robledo2009}%
  \BibitemOpen
  \bibfield  {author} {\bibinfo {author} {\bibfnamefont {L.~M.}\ \bibnamefont
  {Robledo}}, \bibinfo {author} {\bibfnamefont {R.}~\bibnamefont
  {Rodr{\'{\i}}guez-Guzm{\'{a}}n}},\ and\ \bibinfo {author} {\bibfnamefont
  {P.}~\bibnamefont {Sarriguren}},\ }\bibfield  {title} {\bibinfo {title} {Role
  of triaxiality in the ground-state shape of neutron-rich {Y}b, {H}f, {W},
  {O}s and {P}t isotopes},\ }\href
  {https://doi.org/10.1088/0954-3899/36/11/115104} {\bibfield  {journal}
  {\bibinfo  {journal} {J. Phys. G}\ }\textbf {\bibinfo {volume} {36}},\
  \bibinfo {pages} {115104} (\bibinfo {year} {2009})}\BibitemShut {NoStop}%
\bibitem [{\citenamefont {Sarriguren}\ \emph {et~al.}(2008)\citenamefont
  {Sarriguren}, \citenamefont {Rodr\'{\i}guez-Guzm\'an},\ and\ \citenamefont
  {Robledo}}]{sarriguren2008}%
  \BibitemOpen
  \bibfield  {author} {\bibinfo {author} {\bibfnamefont {P.}~\bibnamefont
  {Sarriguren}}, \bibinfo {author} {\bibfnamefont {R.}~\bibnamefont
  {Rodr\'{\i}guez-Guzm\'an}},\ and\ \bibinfo {author} {\bibfnamefont {L.~M.}\
  \bibnamefont {Robledo}},\ }\bibfield  {title} {\bibinfo {title} {Shape
  transitions in neutron-rich {Y}b, {H}f, {W}, {O}s, and {P}t isotopes within a
  {S}kyrme {H}artree-{F}ock + {BCS} approach},\ }\href
  {https://doi.org/10.1103/PhysRevC.77.064322} {\bibfield  {journal} {\bibinfo
  {journal} {Phys. Rev. C}\ }\textbf {\bibinfo {volume} {77}},\ \bibinfo
  {pages} {064322} (\bibinfo {year} {2008})}\BibitemShut {NoStop}%
\bibitem [{\citenamefont {Xu}\ \emph {et~al.}(2011)\citenamefont {Xu},
  \citenamefont {Hua}, \citenamefont {Li}, \citenamefont {Meng}, \citenamefont
  {Li}, \citenamefont {Xu}, \citenamefont {Shi}, \citenamefont {Liu},
  \citenamefont {Zhang}, \citenamefont {Li}, \citenamefont {Zhu}, \citenamefont
  {Wu}, \citenamefont {Li}, \citenamefont {He}, \citenamefont {Zhou},
  \citenamefont {Wang}, \citenamefont {Ye}, \citenamefont {Jiang},
  \citenamefont {Zheng}, \citenamefont {Lou}, \citenamefont {Ma}, \citenamefont
  {Wang}, \citenamefont {Cheng},\ and\ \citenamefont {He}}]{xu2011}%
  \BibitemOpen
  \bibfield  {author} {\bibinfo {author} {\bibfnamefont {C.}~\bibnamefont
  {Xu}}, \bibinfo {author} {\bibfnamefont {H.}~\bibnamefont {Hua}}, \bibinfo
  {author} {\bibfnamefont {X.~Q.}\ \bibnamefont {Li}}, \bibinfo {author}
  {\bibfnamefont {J.}~\bibnamefont {Meng}}, \bibinfo {author} {\bibfnamefont
  {Z.~H.}\ \bibnamefont {Li}}, \bibinfo {author} {\bibfnamefont {F.~R.}\
  \bibnamefont {Xu}}, \bibinfo {author} {\bibfnamefont {Y.}~\bibnamefont
  {Shi}}, \bibinfo {author} {\bibfnamefont {H.~L.}\ \bibnamefont {Liu}},
  \bibinfo {author} {\bibfnamefont {S.~Q.}\ \bibnamefont {Zhang}}, \bibinfo
  {author} {\bibfnamefont {Z.~Y.}\ \bibnamefont {Li}}, \bibinfo {author}
  {\bibfnamefont {L.~H.}\ \bibnamefont {Zhu}}, \bibinfo {author} {\bibfnamefont
  {X.~G.}\ \bibnamefont {Wu}}, \bibinfo {author} {\bibfnamefont {G.~S.}\
  \bibnamefont {Li}}, \bibinfo {author} {\bibfnamefont {C.~Y.}\ \bibnamefont
  {He}}, \bibinfo {author} {\bibfnamefont {S.~G.}\ \bibnamefont {Zhou}},
  \bibinfo {author} {\bibfnamefont {S.~Y.}\ \bibnamefont {Wang}}, \bibinfo
  {author} {\bibfnamefont {Y.~L.}\ \bibnamefont {Ye}}, \bibinfo {author}
  {\bibfnamefont {D.~X.}\ \bibnamefont {Jiang}}, \bibinfo {author}
  {\bibfnamefont {T.}~\bibnamefont {Zheng}}, \bibinfo {author} {\bibfnamefont
  {J.~L.}\ \bibnamefont {Lou}}, \bibinfo {author} {\bibfnamefont {L.~Y.}\
  \bibnamefont {Ma}}, \bibinfo {author} {\bibfnamefont {E.~H.}\ \bibnamefont
  {Wang}}, \bibinfo {author} {\bibfnamefont {Y.~Y.}\ \bibnamefont {Cheng}},\
  and\ \bibinfo {author} {\bibfnamefont {C.}~\bibnamefont {He}},\ }\bibfield
  {title} {\bibinfo {title} {New insight into the shape coexistence and shape
  evolution of $^{157}\mathrm{Yb}$},\ }\href
  {https://doi.org/10.1103/PhysRevC.83.014318} {\bibfield  {journal} {\bibinfo
  {journal} {Phys. Rev. C}\ }\textbf {\bibinfo {volume} {83}},\ \bibinfo
  {pages} {014318} (\bibinfo {year} {2011})}\BibitemShut {NoStop}%
\bibitem [{\citenamefont {{C\'edric Simenel}}\ and\ \citenamefont {{Benoit
  Avez}}(2008)}]{simenel2008}%
  \BibitemOpen
  \bibfield  {author} {\bibinfo {author} {\bibnamefont {{C\'edric Simenel}}}\
  and\ \bibinfo {author} {\bibnamefont {{Benoit Avez}}},\ }\bibfield  {title}
  {\bibinfo {title} {Time--dependent {H}artree--{F}ock description of heavy
  {io}ns fusion},\ }\href {https://doi.org/10.1142/S0218301308009525}
  {\bibfield  {journal} {\bibinfo  {journal} {Intl. J. Mod. Phys. E}\ }\textbf
  {\bibinfo {volume} {17}},\ \bibinfo {pages} {31} (\bibinfo {year}
  {2008})}\BibitemShut {NoStop}%
\bibitem [{\citenamefont {Simenel}\ \emph {et~al.}(2017)\citenamefont
  {Simenel}, \citenamefont {Umar}, \citenamefont {Godbey}, \citenamefont
  {Dasgupta},\ and\ \citenamefont {Hinde}}]{simenel2017}%
  \BibitemOpen
  \bibfield  {author} {\bibinfo {author} {\bibfnamefont {C.}~\bibnamefont
  {Simenel}}, \bibinfo {author} {\bibfnamefont {A.~S.}\ \bibnamefont {Umar}},
  \bibinfo {author} {\bibfnamefont {K.}~\bibnamefont {Godbey}}, \bibinfo
  {author} {\bibfnamefont {M.}~\bibnamefont {Dasgupta}},\ and\ \bibinfo
  {author} {\bibfnamefont {D.~J.}\ \bibnamefont {Hinde}},\ }\bibfield  {title}
  {\bibinfo {title} {How the {P}auli exclusion principle affects fusion of
  atomic nuclei},\ }\href {https://doi.org/10.1103/physrevc.95.031601}
  {\bibfield  {journal} {\bibinfo  {journal} {Phys. Rev. C}\ }\textbf {\bibinfo
  {volume} {95}},\ \bibinfo {pages} {031601} (\bibinfo {year}
  {2017})}\BibitemShut {NoStop}%
\bibitem [{\citenamefont {Shen}\ \emph {et~al.}(1987)\citenamefont {Shen},
  \citenamefont {Albinski}, \citenamefont {Gobbi}, \citenamefont {Gralla},
  \citenamefont {Hildenbrand}, \citenamefont {Herrmann}, \citenamefont
  {Kuzminski}, \citenamefont {M\"uller}, \citenamefont {Stelzer}, \citenamefont
  {T{\~{o}}ke}, \citenamefont {Back}, \citenamefont {Bj\o{}rnholm},\ and\
  \citenamefont {S\o{}rensen}}]{shen1987}%
  \BibitemOpen
  \bibfield  {author} {\bibinfo {author} {\bibfnamefont {W.~Q.}\ \bibnamefont
  {Shen}}, \bibinfo {author} {\bibfnamefont {J.}~\bibnamefont {Albinski}},
  \bibinfo {author} {\bibfnamefont {A.}~\bibnamefont {Gobbi}}, \bibinfo
  {author} {\bibfnamefont {S.}~\bibnamefont {Gralla}}, \bibinfo {author}
  {\bibfnamefont {K.~D.}\ \bibnamefont {Hildenbrand}}, \bibinfo {author}
  {\bibfnamefont {N.}~\bibnamefont {Herrmann}}, \bibinfo {author}
  {\bibfnamefont {J.}~\bibnamefont {Kuzminski}}, \bibinfo {author}
  {\bibfnamefont {W.~F.~J.}\ \bibnamefont {M\"uller}}, \bibinfo {author}
  {\bibfnamefont {H.}~\bibnamefont {Stelzer}}, \bibinfo {author} {\bibfnamefont
  {J.}~\bibnamefont {T{\~{o}}ke}}, \bibinfo {author} {\bibfnamefont {B.~B.}\
  \bibnamefont {Back}}, \bibinfo {author} {\bibfnamefont {S.}~\bibnamefont
  {Bj\o{}rnholm}},\ and\ \bibinfo {author} {\bibfnamefont {S.~P.}\ \bibnamefont
  {S\o{}rensen}},\ }\bibfield  {title} {\bibinfo {title} {{F}ission and
  quasifission in {U}-induced reactions},\ }\href
  {https://doi.org/10.1103/PhysRevC.36.115} {\bibfield  {journal} {\bibinfo
  {journal} {Phys. Rev. C}\ }\textbf {\bibinfo {volume} {36}},\ \bibinfo
  {pages} {115} (\bibinfo {year} {1987})}\BibitemShut {NoStop}%
\bibitem [{\citenamefont {Hinde}\ \emph {et~al.}(2008)\citenamefont {Hinde},
  \citenamefont {Thomas}, \citenamefont {du~Rietz}, \citenamefont
  {Diaz-Torres}, \citenamefont {Dasgupta}, \citenamefont {Brown}, \citenamefont
  {Evers}, \citenamefont {Gasques}, \citenamefont {Rafiei},\ and\ \citenamefont
  {Rodriguez}}]{hinde2008}%
  \BibitemOpen
  \bibfield  {author} {\bibinfo {author} {\bibfnamefont {D.~J.}\ \bibnamefont
  {Hinde}}, \bibinfo {author} {\bibfnamefont {R.~G.}\ \bibnamefont {Thomas}},
  \bibinfo {author} {\bibfnamefont {R.}~\bibnamefont {du~Rietz}}, \bibinfo
  {author} {\bibfnamefont {A.}~\bibnamefont {Diaz-Torres}}, \bibinfo {author}
  {\bibfnamefont {M.}~\bibnamefont {Dasgupta}}, \bibinfo {author}
  {\bibfnamefont {M.~L.}\ \bibnamefont {Brown}}, \bibinfo {author}
  {\bibfnamefont {M.}~\bibnamefont {Evers}}, \bibinfo {author} {\bibfnamefont
  {L.~R.}\ \bibnamefont {Gasques}}, \bibinfo {author} {\bibfnamefont
  {R.}~\bibnamefont {Rafiei}},\ and\ \bibinfo {author} {\bibfnamefont {M.~D.}\
  \bibnamefont {Rodriguez}},\ }\bibfield  {title} {\bibinfo {title}
  {{D}isentangling {E}ffects of {N}uclear {S}tructure in {H}eavy {E}lement
  {F}ormation},\ }\href {https://doi.org/10.1103/PhysRevLett.100.202701}
  {\bibfield  {journal} {\bibinfo  {journal} {Phys. Rev. Lett.}\ }\textbf
  {\bibinfo {volume} {100}},\ \bibinfo {pages} {202701} (\bibinfo {year}
  {2008})}\BibitemShut {NoStop}%
\bibitem [{\citenamefont {Simenel}\ \emph {et~al.}(2012)\citenamefont
  {Simenel}, \citenamefont {Hinde}, \citenamefont {{du Rietz}}, \citenamefont
  {Dasgupta}, \citenamefont {Evers}, \citenamefont {Lin}, \citenamefont
  {Luong},\ and\ \citenamefont {Wakhle}}]{simenel2012b}%
  \BibitemOpen
  \bibfield  {author} {\bibinfo {author} {\bibfnamefont {C.}~\bibnamefont
  {Simenel}}, \bibinfo {author} {\bibfnamefont {D.~J.}\ \bibnamefont {Hinde}},
  \bibinfo {author} {\bibfnamefont {R.}~\bibnamefont {{du Rietz}}}, \bibinfo
  {author} {\bibfnamefont {M.}~\bibnamefont {Dasgupta}}, \bibinfo {author}
  {\bibfnamefont {M.}~\bibnamefont {Evers}}, \bibinfo {author} {\bibfnamefont
  {C.~J.}\ \bibnamefont {Lin}}, \bibinfo {author} {\bibfnamefont {D.~H.}\
  \bibnamefont {Luong}},\ and\ \bibinfo {author} {\bibfnamefont
  {A.}~\bibnamefont {Wakhle}},\ }\bibfield  {title} {\bibinfo {title}
  {{I}nfluence of entrance-channel magicity and isospin on quasi-fission},\
  }\href {https://doi.org/10.1016/j.physletb.2012.03.063} {\bibfield  {journal}
  {\bibinfo  {journal} {Phys. Lett. B}\ }\textbf {\bibinfo {volume} {710}},\
  \bibinfo {pages} {607} (\bibinfo {year} {2012})}\BibitemShut {NoStop}%
\bibitem [{\citenamefont {du~Rietz}\ \emph {et~al.}(2013)\citenamefont
  {du~Rietz}, \citenamefont {Williams}, \citenamefont {Hinde}, \citenamefont
  {Dasgupta}, \citenamefont {Evers}, \citenamefont {Lin}, \citenamefont
  {Luong}, \citenamefont {Simenel},\ and\ \citenamefont
  {Wakhle}}]{durietz2013}%
  \BibitemOpen
  \bibfield  {author} {\bibinfo {author} {\bibfnamefont {R.}~\bibnamefont
  {du~Rietz}}, \bibinfo {author} {\bibfnamefont {E.}~\bibnamefont {Williams}},
  \bibinfo {author} {\bibfnamefont {D.~J.}\ \bibnamefont {Hinde}}, \bibinfo
  {author} {\bibfnamefont {M.}~\bibnamefont {Dasgupta}}, \bibinfo {author}
  {\bibfnamefont {M.}~\bibnamefont {Evers}}, \bibinfo {author} {\bibfnamefont
  {C.~J.}\ \bibnamefont {Lin}}, \bibinfo {author} {\bibfnamefont {D.~H.}\
  \bibnamefont {Luong}}, \bibinfo {author} {\bibfnamefont {C.}~\bibnamefont
  {Simenel}},\ and\ \bibinfo {author} {\bibfnamefont {A.}~\bibnamefont
  {Wakhle}},\ }\bibfield  {title} {\bibinfo {title} {Mapping quasifission
  characteristics and timescales in heavy element formation reactions},\ }\href
  {https://doi.org/10.1103/PhysRevC.88.054618} {\bibfield  {journal} {\bibinfo
  {journal} {Phys. Rev. C}\ }\textbf {\bibinfo {volume} {88}},\ \bibinfo
  {pages} {054618} (\bibinfo {year} {2013})}\BibitemShut {NoStop}%
\bibitem [{\citenamefont {Morjean}\ \emph {et~al.}(2017)\citenamefont
  {Morjean}, \citenamefont {Hinde}, \citenamefont {Simenel}, \citenamefont
  {Jeung}, \citenamefont {Airiau}, \citenamefont {Cook}, \citenamefont
  {Dasgupta}, \citenamefont {Drouart}, \citenamefont {Jacquet}, \citenamefont
  {Kalkal}, \citenamefont {Palshetkar}, \citenamefont {Prasad}, \citenamefont
  {Rafferty}, \citenamefont {Simpson}, \citenamefont {Tassan-Got},
  \citenamefont {Vo-Phuoc},\ and\ \citenamefont {Williams}}]{morjean2017}%
  \BibitemOpen
  \bibfield  {author} {\bibinfo {author} {\bibfnamefont {M.}~\bibnamefont
  {Morjean}}, \bibinfo {author} {\bibfnamefont {D.~J.}\ \bibnamefont {Hinde}},
  \bibinfo {author} {\bibfnamefont {C.}~\bibnamefont {Simenel}}, \bibinfo
  {author} {\bibfnamefont {D.~Y.}\ \bibnamefont {Jeung}}, \bibinfo {author}
  {\bibfnamefont {M.}~\bibnamefont {Airiau}}, \bibinfo {author} {\bibfnamefont
  {K.~J.}\ \bibnamefont {Cook}}, \bibinfo {author} {\bibfnamefont
  {M.}~\bibnamefont {Dasgupta}}, \bibinfo {author} {\bibfnamefont
  {A.}~\bibnamefont {Drouart}}, \bibinfo {author} {\bibfnamefont
  {D.}~\bibnamefont {Jacquet}}, \bibinfo {author} {\bibfnamefont
  {S.}~\bibnamefont {Kalkal}}, \bibinfo {author} {\bibfnamefont {C.~S.}\
  \bibnamefont {Palshetkar}}, \bibinfo {author} {\bibfnamefont
  {E.}~\bibnamefont {Prasad}}, \bibinfo {author} {\bibfnamefont
  {D.}~\bibnamefont {Rafferty}}, \bibinfo {author} {\bibfnamefont {E.~C.}\
  \bibnamefont {Simpson}}, \bibinfo {author} {\bibfnamefont {L.}~\bibnamefont
  {Tassan-Got}}, \bibinfo {author} {\bibfnamefont {K.}~\bibnamefont
  {Vo-Phuoc}},\ and\ \bibinfo {author} {\bibfnamefont {E.}~\bibnamefont
  {Williams}},\ }\bibfield  {title} {\bibinfo {title} {Evidence for the {R}ole
  of {P}roton {S}hell {C}losure in {Q}uasifission {R}eactions from {X--Ray}
  {F}luorescence of {M}ass--{I}dentified {F}ragments},\ }\href
  {https://doi.org/10.1103/PhysRevLett.119.222502} {\bibfield  {journal}
  {\bibinfo  {journal} {Phys. Rev. Lett.}\ }\textbf {\bibinfo {volume} {119}},\
  \bibinfo {pages} {222502} (\bibinfo {year} {2017})}\BibitemShut {NoStop}%
\bibitem [{\citenamefont {Mohanto}\ \emph {et~al.}(2018)\citenamefont
  {Mohanto}, \citenamefont {Hinde}, \citenamefont {Banerjee}, \citenamefont
  {Dasgupta}, \citenamefont {Jeung}, \citenamefont {Simenel}, \citenamefont
  {Simpson}, \citenamefont {Wakhle}, \citenamefont {Williams}, \citenamefont
  {Carter}, \citenamefont {Cook}, \citenamefont {Luong}, \citenamefont
  {Palshetkar},\ and\ \citenamefont {Rafferty}}]{mohanto2018}%
  \BibitemOpen
  \bibfield  {author} {\bibinfo {author} {\bibfnamefont {G.}~\bibnamefont
  {Mohanto}}, \bibinfo {author} {\bibfnamefont {D.~J.}\ \bibnamefont {Hinde}},
  \bibinfo {author} {\bibfnamefont {K.}~\bibnamefont {Banerjee}}, \bibinfo
  {author} {\bibfnamefont {M.}~\bibnamefont {Dasgupta}}, \bibinfo {author}
  {\bibfnamefont {D.~Y.}\ \bibnamefont {Jeung}}, \bibinfo {author}
  {\bibfnamefont {C.}~\bibnamefont {Simenel}}, \bibinfo {author} {\bibfnamefont
  {E.~C.}\ \bibnamefont {Simpson}}, \bibinfo {author} {\bibfnamefont
  {A.}~\bibnamefont {Wakhle}}, \bibinfo {author} {\bibfnamefont
  {E.}~\bibnamefont {Williams}}, \bibinfo {author} {\bibfnamefont {I.~P.}\
  \bibnamefont {Carter}}, \bibinfo {author} {\bibfnamefont {K.~J.}\
  \bibnamefont {Cook}}, \bibinfo {author} {\bibfnamefont {D.~H.}\ \bibnamefont
  {Luong}}, \bibinfo {author} {\bibfnamefont {C.~S.}\ \bibnamefont
  {Palshetkar}},\ and\ \bibinfo {author} {\bibfnamefont {D.~C.}\ \bibnamefont
  {Rafferty}},\ }\bibfield  {title} {\bibinfo {title} {Interplay of spherical
  closed shells and {$N/Z$} asymmetry in quasifission dynamics},\ }\href
  {https://doi.org/10.1103/PhysRevC.97.054603} {\bibfield  {journal} {\bibinfo
  {journal} {Phys. Rev. C}\ }\textbf {\bibinfo {volume} {97}},\ \bibinfo
  {pages} {054603} (\bibinfo {year} {2018})}\BibitemShut {NoStop}%
\bibitem [{\citenamefont {Hinde}\ \emph {et~al.}(2018)\citenamefont {Hinde},
  \citenamefont {Jeung}, \citenamefont {Prasad}, \citenamefont {Wakhle},
  \citenamefont {Dasgupta}, \citenamefont {Evers}, \citenamefont {Luong},
  \citenamefont {du~Rietz}, \citenamefont {Simenel}, \citenamefont {Simpson},\
  and\ \citenamefont {Williams}}]{hinde2018}%
  \BibitemOpen
  \bibfield  {author} {\bibinfo {author} {\bibfnamefont {D.~J.}\ \bibnamefont
  {Hinde}}, \bibinfo {author} {\bibfnamefont {D.~Y.}\ \bibnamefont {Jeung}},
  \bibinfo {author} {\bibfnamefont {E.}~\bibnamefont {Prasad}}, \bibinfo
  {author} {\bibfnamefont {A.}~\bibnamefont {Wakhle}}, \bibinfo {author}
  {\bibfnamefont {M.}~\bibnamefont {Dasgupta}}, \bibinfo {author}
  {\bibfnamefont {M.}~\bibnamefont {Evers}}, \bibinfo {author} {\bibfnamefont
  {D.~H.}\ \bibnamefont {Luong}}, \bibinfo {author} {\bibfnamefont
  {R.}~\bibnamefont {du~Rietz}}, \bibinfo {author} {\bibfnamefont
  {C.}~\bibnamefont {Simenel}}, \bibinfo {author} {\bibfnamefont {E.~C.}\
  \bibnamefont {Simpson}},\ and\ \bibinfo {author} {\bibfnamefont
  {E.}~\bibnamefont {Williams}},\ }\bibfield  {title} {\bibinfo {title}
  {Sub--barrier quasifission in heavy element formation reactions with deformed
  actinide target nuclei},\ }\href {https://doi.org/10.1103/PhysRevC.97.024616}
  {\bibfield  {journal} {\bibinfo  {journal} {Phys. Rev. C}\ }\textbf {\bibinfo
  {volume} {97}},\ \bibinfo {pages} {024616} (\bibinfo {year}
  {2018})}\BibitemShut {NoStop}%
\bibitem [{\citenamefont {Sekizawa}\ and\ \citenamefont
  {Yabana}(2016)}]{sekizawa2016}%
  \BibitemOpen
  \bibfield  {author} {\bibinfo {author} {\bibfnamefont {K.}~\bibnamefont
  {Sekizawa}}\ and\ \bibinfo {author} {\bibfnamefont {K.}~\bibnamefont
  {Yabana}},\ }\bibfield  {title} {\bibinfo {title} {{T}ime-dependent
  {H}artree-{F}ock calculations for multinucleon transfer and quasifission
  processes in the $^{64}\text{Ni}+{}^{238}\text{U}$ reaction},\ }\href
  {https://doi.org/10.1103/PhysRevC.93.054616} {\bibfield  {journal} {\bibinfo
  {journal} {Phys. Rev. C}\ }\textbf {\bibinfo {volume} {93}},\ \bibinfo
  {pages} {054616} (\bibinfo {year} {2016})}\BibitemShut {NoStop}%
\bibitem [{\citenamefont {{Edward Simpson}}(2019)}]{anu_chart}%
  \BibitemOpen
  \bibfield  {author} {\bibinfo {author} {\bibnamefont {{Edward Simpson}}},\
  }\href {https://people.physics.anu.edu.au/~ecs103/chart/} {\emph {\bibinfo
  {title} {{The} {C}olourful {N}uclide {C}hart}}},\ \bibinfo {type} {Tech.
  Rep.}\ (\bibinfo  {institution} {Australian National University},\ \bibinfo
  {year} {2019})\BibitemShut {NoStop}%
\bibitem [{\citenamefont {Itkis}\ \emph {et~al.}(2011)\citenamefont {Itkis},
  \citenamefont {Kozulin}, \citenamefont {Itkis}, \citenamefont {Knyazheva},
  \citenamefont {Bogachev}, \citenamefont {Chernysheva}, \citenamefont {Krupa},
  \citenamefont {Oganessian}, \citenamefont {Zagrebaev}, \citenamefont
  {Rusanov}, \citenamefont {Goennenwein}, \citenamefont {Dorvaux},
  \citenamefont {Stuttg\'e}, \citenamefont {Hanappe}, \citenamefont {Vardaci},\
  and\ \citenamefont {{Go\'es de Brennand}}}]{itkis2011}%
  \BibitemOpen
  \bibfield  {author} {\bibinfo {author} {\bibfnamefont {I.~M.}\ \bibnamefont
  {Itkis}}, \bibinfo {author} {\bibfnamefont {E.~M.}\ \bibnamefont {Kozulin}},
  \bibinfo {author} {\bibfnamefont {M.~G.}\ \bibnamefont {Itkis}}, \bibinfo
  {author} {\bibfnamefont {G.~N.}\ \bibnamefont {Knyazheva}}, \bibinfo {author}
  {\bibfnamefont {A.~A.}\ \bibnamefont {Bogachev}}, \bibinfo {author}
  {\bibfnamefont {E.~V.}\ \bibnamefont {Chernysheva}}, \bibinfo {author}
  {\bibfnamefont {L.}~\bibnamefont {Krupa}}, \bibinfo {author} {\bibfnamefont
  {{\relax Yu. Ts}.}~\bibnamefont {Oganessian}}, \bibinfo {author}
  {\bibfnamefont {V.~I.}\ \bibnamefont {Zagrebaev}}, \bibinfo {author}
  {\bibfnamefont {{\relax A. Ya}.}~\bibnamefont {Rusanov}}, \bibinfo {author}
  {\bibfnamefont {F.}~\bibnamefont {Goennenwein}}, \bibinfo {author}
  {\bibfnamefont {O.}~\bibnamefont {Dorvaux}}, \bibinfo {author} {\bibfnamefont
  {L.}~\bibnamefont {Stuttg\'e}}, \bibinfo {author} {\bibfnamefont
  {F.}~\bibnamefont {Hanappe}}, \bibinfo {author} {\bibfnamefont
  {E.}~\bibnamefont {Vardaci}},\ and\ \bibinfo {author} {\bibfnamefont
  {E.}~\bibnamefont {{Go\'es de Brennand}}},\ }\bibfield  {title} {\bibinfo
  {title} {{F}ission and quasifission modes in heavy-ion-induced reactions
  leading to the formation of {H}s$^{*}$},\ }\href
  {https://doi.org/10.1103/PhysRevC.83.064613} {\bibfield  {journal} {\bibinfo
  {journal} {Phys. Rev. C}\ }\textbf {\bibinfo {volume} {83}},\ \bibinfo
  {pages} {064613} (\bibinfo {year} {2011})}\BibitemShut {NoStop}%
\bibitem [{\citenamefont {Itkis}\ \emph {et~al.}(2015)\citenamefont {Itkis},
  \citenamefont {Vardaci}, \citenamefont {Itkis}, \citenamefont {Knyazheva},\
  and\ \citenamefont {Kozulin}}]{itkis2015}%
  \BibitemOpen
  \bibfield  {author} {\bibinfo {author} {\bibfnamefont {M.~G.}\ \bibnamefont
  {Itkis}}, \bibinfo {author} {\bibfnamefont {E.}~\bibnamefont {Vardaci}},
  \bibinfo {author} {\bibfnamefont {I.~M.}\ \bibnamefont {Itkis}}, \bibinfo
  {author} {\bibfnamefont {G.~N.}\ \bibnamefont {Knyazheva}},\ and\ \bibinfo
  {author} {\bibfnamefont {E.~M.}\ \bibnamefont {Kozulin}},\ }\bibfield
  {title} {\bibinfo {title} {Fusion and fission of heavy and superheavy nuclei
  (experiment)},\ }\href {https://doi.org/10.1016/j.nuclphysa.2015.09.007}
  {\bibfield  {journal} {\bibinfo  {journal} {Nucl. Phys. A}\ }\textbf
  {\bibinfo {volume} {944}},\ \bibinfo {pages} {204} (\bibinfo {year}
  {2015})}\BibitemShut {NoStop}%
\bibitem [{\citenamefont {Kozulin}\ \emph {et~al.}(2019)\citenamefont
  {Kozulin}, \citenamefont {Knyazheva}, \citenamefont {Ghosh}, \citenamefont
  {Sen}, \citenamefont {Itkis}, \citenamefont {Itkis}, \citenamefont {Novikov},
  \citenamefont {Diatlov}, \citenamefont {Pchelintsev}, \citenamefont
  {Bhattacharya}, \citenamefont {Bhattacharya}, \citenamefont {Banerjee},
  \citenamefont {Saveleva},\ and\ \citenamefont {Vorobiev}}]{kozulin2019}%
  \BibitemOpen
  \bibfield  {author} {\bibinfo {author} {\bibfnamefont {E.~M.}\ \bibnamefont
  {Kozulin}}, \bibinfo {author} {\bibfnamefont {G.~N.}\ \bibnamefont
  {Knyazheva}}, \bibinfo {author} {\bibfnamefont {T.~K.}\ \bibnamefont
  {Ghosh}}, \bibinfo {author} {\bibfnamefont {A.}~\bibnamefont {Sen}}, \bibinfo
  {author} {\bibfnamefont {I.~M.}\ \bibnamefont {Itkis}}, \bibinfo {author}
  {\bibfnamefont {M.~G.}\ \bibnamefont {Itkis}}, \bibinfo {author}
  {\bibfnamefont {K.~V.}\ \bibnamefont {Novikov}}, \bibinfo {author}
  {\bibfnamefont {I.~N.}\ \bibnamefont {Diatlov}}, \bibinfo {author}
  {\bibfnamefont {I.~V.}\ \bibnamefont {Pchelintsev}}, \bibinfo {author}
  {\bibfnamefont {C.}~\bibnamefont {Bhattacharya}}, \bibinfo {author}
  {\bibfnamefont {S.}~\bibnamefont {Bhattacharya}}, \bibinfo {author}
  {\bibfnamefont {K.}~\bibnamefont {Banerjee}}, \bibinfo {author}
  {\bibfnamefont {E.~O.}\ \bibnamefont {Saveleva}},\ and\ \bibinfo {author}
  {\bibfnamefont {I.~V.}\ \bibnamefont {Vorobiev}},\ }\bibfield  {title}
  {\bibinfo {title} {Fission and quasifission of the composite system
  $\mathrm{Z}=114$ formed in heavy-ion reactions at energies near the coulomb
  barrier},\ }\href {https://doi.org/10.1103/PhysRevC.99.014616} {\bibfield
  {journal} {\bibinfo  {journal} {Phys. Rev. C}\ }\textbf {\bibinfo {volume}
  {99}},\ \bibinfo {pages} {014616} (\bibinfo {year} {2019})}\BibitemShut
  {NoStop}%
\bibitem [{\citenamefont {Banerjee}\ \emph {et~al.}(2019)\citenamefont
  {Banerjee}, \citenamefont {Hinde}, \citenamefont {Dasgupta}, \citenamefont
  {Simpson}, \citenamefont {Jeung}, \citenamefont {Simenel}, \citenamefont
  {{Swinton-Bland}}, \citenamefont {Williams}, \citenamefont {Carter},
  \citenamefont {Cook}, \citenamefont {David}, \citenamefont {D\"ullmann},
  \citenamefont {Khuyagbaatar}, \citenamefont {Kindler}, \citenamefont
  {Lommel}, \citenamefont {Prasad}, \citenamefont {Sengupta}, \citenamefont
  {Smith}, \citenamefont {{Vo-Phuoc}}, \citenamefont {Walshe},\ and\
  \citenamefont {Yakushev}}]{banerjee2019}%
  \BibitemOpen
  \bibfield  {author} {\bibinfo {author} {\bibfnamefont {K.}~\bibnamefont
  {Banerjee}}, \bibinfo {author} {\bibfnamefont {D.~J.}\ \bibnamefont {Hinde}},
  \bibinfo {author} {\bibfnamefont {M.}~\bibnamefont {Dasgupta}}, \bibinfo
  {author} {\bibfnamefont {E.~C.}\ \bibnamefont {Simpson}}, \bibinfo {author}
  {\bibfnamefont {D.~Y.}\ \bibnamefont {Jeung}}, \bibinfo {author}
  {\bibfnamefont {C.}~\bibnamefont {Simenel}}, \bibinfo {author} {\bibfnamefont
  {B.~M.~A.}\ \bibnamefont {{Swinton-Bland}}}, \bibinfo {author} {\bibfnamefont
  {E.}~\bibnamefont {Williams}}, \bibinfo {author} {\bibfnamefont {I.~P.}\
  \bibnamefont {Carter}}, \bibinfo {author} {\bibfnamefont {K.~J.}\
  \bibnamefont {Cook}}, \bibinfo {author} {\bibfnamefont {H.~M.}\ \bibnamefont
  {David}}, \bibinfo {author} {\bibfnamefont {C.}~\bibnamefont {D\"ullmann}},
  \bibinfo {author} {\bibfnamefont {J.}~\bibnamefont {Khuyagbaatar}}, \bibinfo
  {author} {\bibfnamefont {B.}~\bibnamefont {Kindler}}, \bibinfo {author}
  {\bibfnamefont {B.}~\bibnamefont {Lommel}}, \bibinfo {author} {\bibfnamefont
  {E.}~\bibnamefont {Prasad}}, \bibinfo {author} {\bibfnamefont
  {C.}~\bibnamefont {Sengupta}}, \bibinfo {author} {\bibfnamefont {J.~F.}\
  \bibnamefont {Smith}}, \bibinfo {author} {\bibfnamefont {K.}~\bibnamefont
  {{Vo-Phuoc}}}, \bibinfo {author} {\bibfnamefont {J.}~\bibnamefont {Walshe}},\
  and\ \bibinfo {author} {\bibfnamefont {A.}~\bibnamefont {Yakushev}},\
  }\bibfield  {title} {\bibinfo {title} {Mechanisms {S}uppressing {S}uperheavy
  {E}lement {Y}ields in {C}old {F}usion {R}eactions},\ }\href
  {https://doi.org/10.1103/PhysRevLett.122.232503} {\bibfield  {journal}
  {\bibinfo  {journal} {Phys. Rev. Lett.}\ }\textbf {\bibinfo {volume} {122}},\
  \bibinfo {pages} {232503} (\bibinfo {year} {2019})}\BibitemShut {NoStop}%
\bibitem [{\citenamefont {Thomas}\ \emph {et~al.}(2008)\citenamefont {Thomas},
  \citenamefont {Hinde}, \citenamefont {Duniec}, \citenamefont {Zenke},
  \citenamefont {Dasgupta}, \citenamefont {Brown}, \citenamefont {Evers},
  \citenamefont {Gasques}, \citenamefont {Rodriguez},\ and\ \citenamefont
  {Diaz-Torres}}]{thomas2008}%
  \BibitemOpen
  \bibfield  {author} {\bibinfo {author} {\bibfnamefont {R.~G.}\ \bibnamefont
  {Thomas}}, \bibinfo {author} {\bibfnamefont {D.~J.}\ \bibnamefont {Hinde}},
  \bibinfo {author} {\bibfnamefont {D.}~\bibnamefont {Duniec}}, \bibinfo
  {author} {\bibfnamefont {F.}~\bibnamefont {Zenke}}, \bibinfo {author}
  {\bibfnamefont {M.}~\bibnamefont {Dasgupta}}, \bibinfo {author}
  {\bibfnamefont {M.~L.}\ \bibnamefont {Brown}}, \bibinfo {author}
  {\bibfnamefont {M.}~\bibnamefont {Evers}}, \bibinfo {author} {\bibfnamefont
  {L.~R.}\ \bibnamefont {Gasques}}, \bibinfo {author} {\bibfnamefont {M.~D.}\
  \bibnamefont {Rodriguez}},\ and\ \bibinfo {author} {\bibfnamefont
  {A.}~\bibnamefont {Diaz-Torres}},\ }\bibfield  {title} {\bibinfo {title}
  {{E}ntrance channel dependence of quasifission in reactions forming
  $^{220}\mathrm{Th}$},\ }\href {https://doi.org/10.1103/PhysRevC.77.034610}
  {\bibfield  {journal} {\bibinfo  {journal} {Phys. Rev. C}\ }\textbf {\bibinfo
  {volume} {77}},\ \bibinfo {pages} {034610} (\bibinfo {year}
  {2008})}\BibitemShut {NoStop}%
\bibitem [{\citenamefont {Sekizawa}(2017{\natexlab{b}})}]{sekizawa2017}%
  \BibitemOpen
  \bibfield  {author} {\bibinfo {author} {\bibfnamefont {K.}~\bibnamefont
  {Sekizawa}},\ }\bibfield  {title} {\bibinfo {title} {Microscopic description
  of production cross sections including deexcitation effects},\ }\href
  {https://doi.org/10.1103/physrevc.96.014615} {\bibfield  {journal} {\bibinfo
  {journal} {Phys. Rev. C}\ }\textbf {\bibinfo {volume} {96}},\ \bibinfo
  {pages} {014615} (\bibinfo {year} {2017}{\natexlab{b}})}\BibitemShut
  {NoStop}%
\bibitem [{\citenamefont {Bulgac}\ \emph {et~al.}(2019)\citenamefont {Bulgac},
  \citenamefont {Jin},\ and\ \citenamefont {Stetcu}}]{bulgac2019}%
  \BibitemOpen
  \bibfield  {author} {\bibinfo {author} {\bibfnamefont {A.}~\bibnamefont
  {Bulgac}}, \bibinfo {author} {\bibfnamefont {S.}~\bibnamefont {Jin}},\ and\
  \bibinfo {author} {\bibfnamefont {I.}~\bibnamefont {Stetcu}},\ }\bibfield
  {title} {\bibinfo {title} {Unitary evolution with fluctuations and
  dissipation},\ }\href {https://doi.org/10.1103/PhysRevC.100.014615}
  {\bibfield  {journal} {\bibinfo  {journal} {Phys. Rev. C}\ }\textbf {\bibinfo
  {volume} {100}},\ \bibinfo {pages} {014615} (\bibinfo {year}
  {2019})}\BibitemShut {NoStop}%
\end{thebibliography}%


\end{document}